\documentclass[journal]{IEEEtran}

\usepackage{graphicx}
\usepackage{flushend}
\usepackage{amsmath,amsfonts}
\usepackage{tabularx}
\usepackage{array}
\renewcommand{\arraystretch}{1.3}
\usepackage{verbatim}
\usepackage{graphicx}
\usepackage{float} 
\usepackage{cite}
\usepackage{amsthm}
\usepackage{amssymb}
\usepackage{xcolor}
\usepackage{url}
\usepackage{multicol}
\usepackage{placeins}
\usepackage[T1]{fontenc}
\usepackage{amsmath,bm,amsfonts,amssymb,graphicx,color,multirow,setspace,xfrac,comment,xcolor}

\usepackage[font=normalsize,labelfont=sf,textfont=sf]{subfig}

\usepackage{booktabs}

\usepackage{tikz}
\usetikzlibrary{calc,decorations.pathreplacing,arrows.meta}

\begin{document}

\title{Industrial Viewpoints on RAN Technologies for 6G}
 
\author{Mansoor Shafi, \textit{Life Fellow, IEEE}, Erik G. Larsson, \textit{Fellow, IEEE}, Xingqin Lin, \textit{Senior Member, IEEE}, Dorin Panaitopol, \textit{Senior Member, IEEE}, Stefan Parkvall, \textit{Fellow, IEEE}, Flavien Ronteix-Jacquet, and Antti Toskala
\thanks{This article has been accepted for publication in \emph{Proceedings of the IEEE}. \copyright~2026 IEEE. Personal use of this material is permitted. Permission from IEEE must be obtained for all other uses, in any current or future media, including reprinting/republishing this material for advertising or promotional purposes, creating new collective works, for resale or redistribution to servers or lists, or reuse of any copyrighted component of this work in other works.}
\thanks{Mansoor Shafi is with Spark NZ Ltd and an Adjunct Professor at the Department of Electrical and Computer Engineering, Victoria University of Wellington, and the University of Canterbury. Email: mansoor.shafi@spark.co.nz}
\thanks{Erik G. Larsson is with Link\"oping University, 
Dept. of Electrical Engineering (ISY),
 581 83 Link\"oping,
Sweden. Email: erik.g.larsson@liu.se}
\thanks{Xingqin Lin is with NVIDIA, Santa Clara, CA 95051, USA. Email: xingqinl@nvidia.com}
\thanks{Dorin Panaitopol is with THALES SIX GTS FRANCE, Gennevilliers, 92622, France. Email: dorin.panaitopol@thalesgroup.com}
\thanks{Stefan Parkvall is with Ericsson Research, 16480 Stockholm, Sweden. Email: stefan.parkvall@ericsson.com}
\thanks{Flavien Ronteix-Jacquet is with the European Commission Joint Research Center (JRC), Ispra, Italy. Email: flavien.ronteix-jacquet@ec.europa.eu}
\thanks{Antti Toskala is with Nokia Standards, Espoo, Finland. Email: antti.toskala@nokia.com}
}

\maketitle
 
\begin{abstract}
Sixth-generation (6G) mobile systems have entered the standardization phase, with commercial deployments anticipated around 2030. This article presents an industrial perspective on the performance targets and key technologies shaping the 6G radio access network (RAN). It examines advances in multiple-input multiple-output (MIMO),
artificial intelligence (AI), waveforms, channel coding, signal constellations, integrated sensing and communication
(ISAC), and the
integration of terrestrial and non-terrestrial networks (NTNs). Although detailed specification work is still ahead, the scope of the 6G studies has been defined and substantial technical progress has been made. We assess the emerging technologies and design choices from the perspectives of implementation complexity, computational and energy efficiency, interoperability, deployment feasibility, and migration from 5G. By connecting research advances with practical standardization and commercialization considerations, this article aims to inform researchers and industry practitioners developing the 6G RAN.
\end{abstract}


\vspace{-0.2cm}
\section{Introduction}

Mobile communications have become an essential part of our lives and have been pervasive since the introduction of the smartphone. Since the deployment of third-generation (3G) systems around 2000, a new generation of mobile communications has emerged roughly every ten years. The fourth generation (4G) was standardized and deployed around 2010, followed by fifth-generation (5G) New Radio (NR) around 2020, which is now extensively deployed worldwide.

Since 3G, mobile-system specifications have been developed by the 3rd Generation Partnership Project (3GPP). Radio access has evolved from code-division multiple access (CDMA) in 3G to orthogonal frequency-division multiplexing (OFDM)-based technologies in 4G and beyond. The 5G standard, first introduced in Release 15, was designed to support a diverse range of services through three usage scenarios: enhanced mobile broadband (eMBB), ultra-reliable low-latency communications (URLLC), and massive machine-type communications (mMTC). At present, however, most 5G services focus on eMBB and fixed wireless access (FWA), while mMTC still relies largely on solutions inherited from 4G. 5G has been enhanced through several 3GPP releases, most recently Release 19, completed at the end of 2025. The key characteristics introduced across the mobile generations from 3G to 5G are summarized in Table~\ref{tab:standards}. 

\begin{table*}[t!]
  \centering
  \caption{Evolution of Mobile Standards (simplified). The 6G values are expected.} 
  \scalebox{0.78}{
  \begin{tabular}{ |p{2.5cm}|p{1.5cm}|p{1.5cm}|p{2cm}|p{2cm}|p{2cm}|p{2cm}|}
    \hline
    \textbf{Defining Features} &\textbf{3G (IMT 2000)} & \textbf{3.5G} & \textbf{4G (IMT 2010)} & \textbf{5G (IMT 2020)} & \textbf{5G-Advanced} & \textbf{6G (IMT 2030)} \\
    \hline
    \midrule\hline
    \textbf{Starting 3GPP release} & Rel 99 & Rel 5 & Rel 8 & Rel 15 & Rel 18 & Rel 21 \\ 
    \hline
    \textbf{Radio access standard name} & WCDMA & HSPA & LTE / LTE-A (from Rel 10) & NR & NR & 6GR \\ 
    \hline
    \textbf{KPI references}  & \cite{iturm1225}& \cite{iturm1225}&\cite{ITURM2134}  & \cite{ITURM2410} & \cite{ITURM2410} & \cite{ITURM2030} \\
    \hline
    \textbf{Carrier} & \multicolumn{2}{|c|}{< 3 GHz} & \multicolumn{1}{|c|}{< 6 GHz} & \multicolumn{2}{|c|}{FR1, FR2 for TN} & \multicolumn{1}{|c|}{ext. FR1, FR2} \\ 
    \textbf{frequency} & \multicolumn{2}{|c|}{} & \multicolumn{1}{|c|}{} & \multicolumn{2}{|c|}{FR1-NTN, FR2-NTN for NTN} & \multicolumn{1}{|c|}{} \\
    \hline
    \textbf{Channel bandwidth } & \multicolumn{2}{|c|}{5 MHz} & \multicolumn{1}{|c|}{$\le$ 20 MHz} & \multicolumn{2}{|c|}{$\le$ 100 MHz (FR1) } & $\le$ 400/200 MHz (FR1) \\
    \textbf{DL/UL}             & \multicolumn{2}{|c|}{}      & \multicolumn{1}{|c|}{}             & \multicolumn{2}{|c|}{$\le$ 400 MHz (FR2) } & $\le$ 400 MHz (FR2) \\
    \hline
    \textbf{Waveform} & \multicolumn{2}{|c|}{CDMA} & \multicolumn{1}{|c|}{CP-OFDM (DL) }   & \multicolumn{3}{|c|}{CP-OFDM (DL)} \\
                      & \multicolumn{2}{|c|}{}     & \multicolumn{1}{|c|}{DFT-s-OFDM (UL) } & \multicolumn{3}{|c|}{CP-OFDM, DFT-s-OFDM (UL)} \\ 
    \hline
    \textbf{Subcarrier spacing} & \multicolumn{2}{|c|}{N/A} & \multicolumn{1}{|c|}{15 kHz} & \multicolumn{3}{|c|}{15--120 (960) kHz} \\
    \hline
    \textbf{Modulation (max) } & \multicolumn{1}{|c|}{QPSK} & \multicolumn{1}{|c|}{16-QAM} & 256-QAM & \multicolumn{3}{|c|}{1024-QAM} 
    \\
    \hline
    \textbf{Duplexing} & FDD  & \multicolumn{2}{c|}{FDD, TDD} & FDD, TDD, dyn-TDD & \multicolumn{2}{|c|}{FDD, TDD, dyn-TDD, SBFD} \\
    \hline
    \textbf{MIMO} &\multicolumn{2}{|c|}{no} & $8\times8$ SU-MIMO & \multicolumn{3}{|c|}{massive MIMO, MU-MIMO, D-MIMO (for 6G)} \\
    \hline
    \textbf{Coding} & \multicolumn{3}{|c|}{Turbo, convolutional} &  \multicolumn{3}{|c|}{LDPC, polar, enhanced LDPC for 6G} \\
    \hline
    \textbf{Beamforming} & \multicolumn{2}{|c|}{no} & limited & \multicolumn{3}{|c|}{advanced} \\
    \hline
    \textbf{Latency} & \multicolumn{2}{|c|}{20--50 ms} & \multicolumn{1}{|c|}{5--10 ms} & \multicolumn{3}{|c|}{$\sim$1 ms} \\
    \hline
    \textbf{Key services} & CS voice & \multicolumn{5}{|c|}{data} \\
    \hline
    \textbf{UEs} & phone & smartphone & {smartphone, IoT} & \multicolumn{3}{|c|}{smartphone, IoT, wearables, FWA, ...} \\ 
    \hline
    \textbf{Core network} & \multicolumn{2}{|c|}{3GC} & EPC & EPC (NSA), 5GC (SA) & 5GC (SA) & 5GC evo \\
    \hline
    \textbf{Circuit/packet switched} & \multicolumn{2}{|c|}{CS/PS} & \multicolumn{4}{|c|}{PS only} \\
    \hline
    \textbf{Architecture} & \multicolumn{6}{|c|}{See Fig.~\ref{fig:architecture}} \\
    \hline
    \textbf{NTN support} & \multicolumn{3}{|c|}{no} & \multicolumn{2}{|l|}{NTN starting with Rel 17 (5G NR-based)}  & native \\
    & \multicolumn{3}{|c|}{} & \multicolumn{2}{|l|}{NTN starting with Rel 18 (LTE-based)}  &  \\
    \hline
    \textbf{AI/ML} & \multicolumn{3}{|c|}{ N/A} & not standardized & selected use cases & AI-native \\
    \hline 
    \bottomrule
  \end{tabular}
  }
  \label{tab:standards}
\end{table*}

5G Release 19 is expected to serve as a precursor to 6G \cite{Commag1}. It already supports several technologies relevant to 6G, including extended reality (XR), artificial intelligence/machine learning (AI/ML), and non-terrestrial networks (NTNs). The academic community has also embraced 6G research, as reflected in the large number of keynote talks at flagship conferences and IEEE journal papers on 6G architecture, use cases, and physical-layer technologies. IEEE journals have published several special issues on 6G; in particular, the \emph{Proceedings of the IEEE} has published two such issues \cite{Proceedings1,Proceedings2}---see also the references cited therein. A comprehensive review of this literature is beyond the scope of this article. Table~\ref{Tab:summary} summarizes the selected surveys and highlights how our focus differs from theirs. We also discuss key papers throughout the article and direct readers to the references cited therein.

\begin{figure}[t!] %
\centering
\includegraphics[width=0.6\columnwidth]{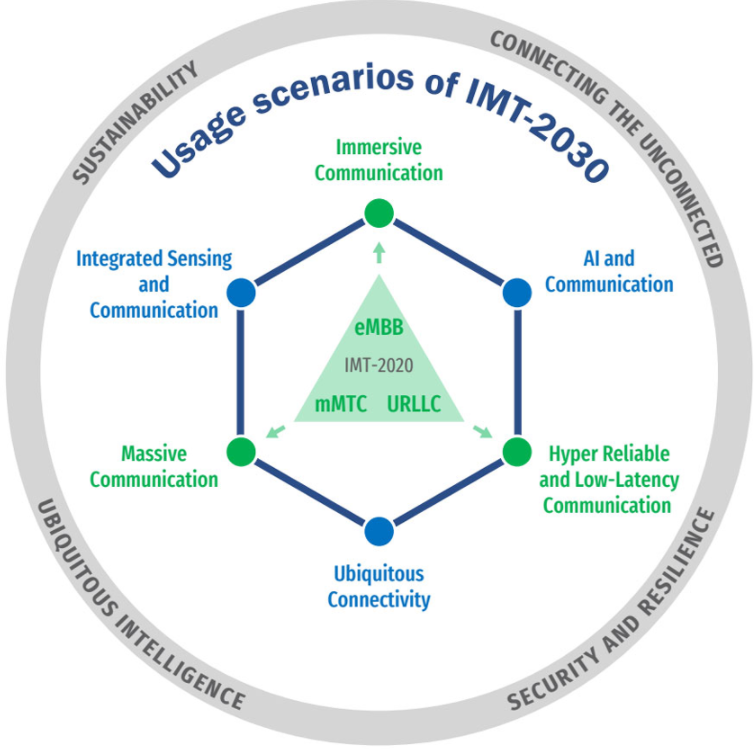} %
\begingroup
\captionsetup{font=footnotesize}
\caption{IMT-2030 usage scenarios and their evolution from IMT-2020, within the International Mobile Telecommunications (IMT) framework. The inner triangle shows the three IMT-2020 scenarios: enhanced mobile broadband (eMBB), massive machine-type communications (mMTC), and ultra-reliable low-latency communications (URLLC). Green arrows indicate their evolution toward immersive communication, massive communication, and hyper-reliable and low-latency communication. The outer hexagon additionally includes ubiquitous connectivity, artificial intelligence (AI) and communication, and integrated sensing and communication. The surrounding ring identifies four overarching themes: sustainability, connecting the unconnected, ubiquitous intelligence, and security and resilience. Reproduced from \cite{ITUFRamework}.}
\label{fig:ITUdiagram}
\endgroup
\vspace{-10pt}
\end{figure}

\begin{table*}[t!]
\centering
\caption{Summary of Selected 6G Surveys.} %
\scalebox{0.78}{
\begin{tabular}{|p{1.5cm}|p{9.0cm}|p{9.0cm}|}
\hline
\textbf{References} & \textbf{Main Topics} & \textbf{Differences}
\tabularnewline\hline
\midrule\hline
\textbf{\cite{10499974}} & Architecture-focused, use of THz bands. & Our focus is on standardization.
\tabularnewline\hline
\textbf{\cite{9295376}} & Summary of applications, KPIs and RAN capabilities. & Some of the RAN capabilities \cite{9295376} are not in the scope of standardization. Our focus is on standardization. 
\tabularnewline\hline
\textbf{\cite{10994811}} & 
GenAI-centric survey focused on how generative models address data scarcity and enable 6G use cases. &  We take a broader RAN-standardization and air-interface perspective, covering AI-native RAN architecture, 3GPP 5G-Advanced lessons, and practical constraints. \tabularnewline\hline
\textbf{\cite{10570308}} & An in-depth tutorial on NTN and how AI/ML and O-RAN can be leveraged for NTNs. &  NTN is a part of the 6G RAN. We present a  complete view of TN and NTN in standardization.
\tabularnewline\hline
\textbf{\cite{10329947}} & The paper advances Open RAN principles for 6G. &  The
proposed openness of interfaces requires interoperability and poses a complexity burden and operational challenge across the vendors involved. 
\tabularnewline\hline
\textbf{\cite{8766143}} & Several 6G technologies are presented, utilizing THz bands, visible light communication (VLC), supermassive MIMO, large intelligent surfaces, holographic beamforming, orbital angular momentum multiplexing, etc. & The ideas presented are far from standardization, which is the focus of our paper. \tabularnewline\hline
\textbf{\cite{8869705}} & A forward-looking vision of disruptive technologies for 6G. & The 6G KPIs and vision are far from standardization, which is the focus of our paper. 
\tabularnewline\hline
\textbf{\cite{9711524}} & The paper presents a digital-twin-based architecture for 6G. & Our focus is on the standardization of the RAN that will support  the 6G applications. 
\tabularnewline\hline
\textbf{\cite{9170653}} & Visionary technologies that could provide the step changes needed to support 6G are presented.
 & Many of the ideas are not in the scope of standards, which is the focus of our paper.
\tabularnewline\hline
\textbf{\cite{8946626}} & A survey of the convergence of NTN and TN is given, and a convergence architecture and open issues are presented. & We present a holistic view of the 6G RAN in standardization that includes NTN and TN.
\tabularnewline\hline
\textbf{\cite{DBLP:journals/corr/abs-1906-00741}} & The  
key features of 6G and the required enabling communication technologies such as the use of THz bands are presented. & The enabling technologies discussed are largely not in the scope of standardization, which is the key focus of our paper. 
 \tabularnewline\hline
\textbf{\cite{11456641}} & Tutorial for a JSAC special issue, focusing on air-interface technologies with potential for fundamental enhancements (e.g., flexible coexistence, AI-native design, spectrum and energy efficiency, and ISAC). & We take a more industry-focused perspective, with opinions from authors affiliated with operators, vendors, and academia, emphasizing practical standardization, deployment, implementation challenges, and backward compatibility with 5G. 
\tabularnewline\hline

\textbf{\cite{TATARIA1}} &  The key features of 6G and the required enabling
communication technologies such as the use of THz bands are presented. 
 & The enabling technologies discussed are largely not in the scope of standardization, which is the key focus of our paper. 
 \tabularnewline\hline
  
  \textbf{\cite{mostafa}} & The paper reviews trends in communication and possible technologies for 6G and challenges at very high frequencies, e.g., a transceiver architecture for 6G in the sub-THz bands.
 & Given the decisions on spectrum in \cite{WRC2023}, low and mid bands are preferred for 6G, as discussed in this paper. 

 \tabularnewline\hline
  \textbf{\cite{6Gandbeyond}} & 6G use cases, their KPIs, and a comparison with 5G are given. The 6G network operates in the sub-THz bands, uses novel architectures, and improves coverage, including deep-space connectivity and an internet-of-space-things, but no use cases for the internet-of-space-things are provided. 
 & Given the decisions on spectrum in \cite{WRC2023}, low and mid bands are preferred for 6G, as discussed in this paper. 
 \tabularnewline\hline
\bottomrule
\end{tabular}}
\label{Tab:summary}
\vspace{-10pt}
\end{table*}
In parallel, the ITU Radiocommunication Sector (ITU-R) has taken a leap into the future by completing a report \cite{ITUFRamework} on the vision and use cases of 6G, referred to as International Mobile Telecommunications (IMT)-2030. The scenarios of IMT-2030 are shown in Fig.~\ref{fig:ITUdiagram}. %
Many industry white papers have also addressed 6G \cite{qcom1,E3,Nokia,Samsung2}. The works in \cite{Samsung2,Ericsson,Ericsson2} envision disruptive communication technologies in response to societal and lifestyle changes, including (1) \emph{immersive communications}, in which immersive reality becomes a preferred means of communication; (2) \emph{connectivity for all things}, extending substantially beyond 5G and enabling communication among things; (3) \emph{integrated sensing and communication} (ISAC); (4) \emph{ubiquitous connectivity}; and (5) \emph{AI and communications}.

\begin{figure*}[ht]
    \centering
    \includegraphics[width=17cm]{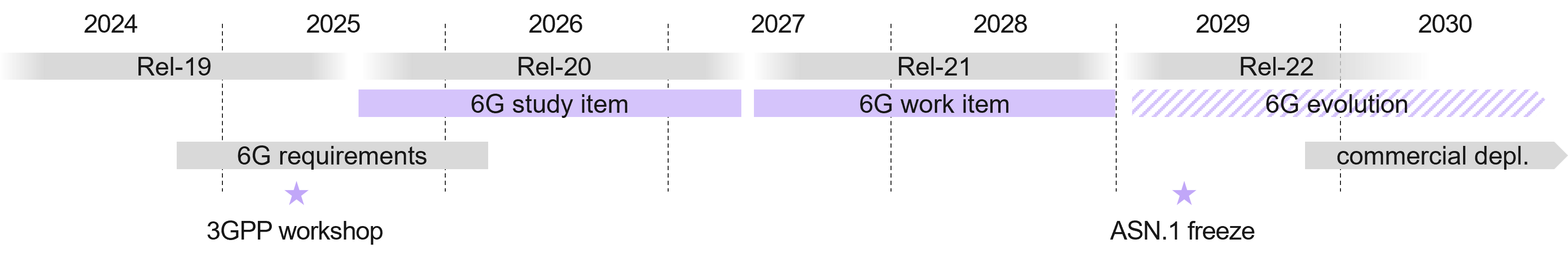} %
    \begingroup
\captionsetup{font=footnotesize}
    \caption{Indicative timeline for 6G standardization in the 3rd Generation Partnership Project (3GPP). The bars show the envisaged progression from requirements definition and the Release 20 study item to the Release 21 work item and subsequent evolution in Release 22. The stars identify the 3GPP 6G workshop and the Abstract Syntax Notation One (ASN.1) freeze, marking completion of the relevant protocol encodings. Commercial deployments are anticipated around 2030. Release boundaries and deployment timing are indicative.}
    \label{fig:timeline}
    \endgroup
    \vspace{-5pt}
\end{figure*}

The requirements and end-to-end architecture are discussed in a European research initiative \cite{hexax} and in several other industry white papers. Turning to standardization, 3GPP held a 6G workshop in March 2025 covering both the radio access network (RAN) and core network (CN) \cite{6GWS-25243}. At this workshop, operators and vendors presented their visions and key directions for 6G. The standardization work consists of two phases: a 6G study-item phase in Release 20 \cite{RP-251881}, currently underway, followed by a work-item phase in Release 21. As shown in Fig.~\ref{fig:timeline}, the complete 3GPP specifications are expected to be available in early 2029 \cite{3GPPtimeline}.

The detailed scope of the Release 20 6G study is provided in \cite{6GWS-25243}. In brief, it covers:
\begin{itemize} %
  \item A single technology framework based on a standalone (SA) architecture;
  \item Radio-interface protocol architecture and procedures (user plane, control plane, security aspects, etc.);
  \item Support for various services and functionalities (sensing, AI/ML for 6G, leveraging the 5G framework, etc.);
  \item Migration from 5G to 6G (5G--6G multi-radio-access-technology (multi-RAT) spectrum sharing, additional mechanisms if necessary, etc.);
  \item Physical-layer solutions\footnote{New physical-layer solutions may be considered, but their gains, especially those of new waveforms and channel coding, must be evaluated against the corresponding 5G NR solutions.} (multiplexing schemes, waveforms, MIMO, control signaling, etc.);
  \item Mobility and radio resource management (RRM);
  \item Performance requirements (radio-frequency (RF) aspects, RRM aspects, etc.);
  \item RAN architecture, interface protocols, and procedures. %
\end{itemize}

Many academic visions for 6G summarized in Table~\ref{Tab:summary}
are not aligned with practical industrial requirements and current
3GPP standardization priorities. This article focuses on technologies
being considered for standardization, assessed against practical
implementation and deployment constraints. For example, sub-THz
bands, frequently emphasized in academic 6G research, are outside
the scope of the current 3GPP 6G Radio study.
 The site densification required for wide-area deployment at
these frequencies could make such deployments prohibitively expensive,
as suggested by the slow uptake of millimeter-wave (mmWave) bands
\cite{shaficband}. As another example, reconfigurable intelligent surfaces (RISs) have attracted extensive
academic attention, but have gained limited industrial support for
adoption in commercial cellular networks. RIS-specific standardization
is not part of the current 3GPP 6G Radio study. Practical adoption
faces challenges including channel acquisition and control overhead,
installation and maintenance costs, and coexistence with existing
networks~\cite{2024riscellularnetworks}.

The contributions of this paper are summarized as follows:
\begin{itemize}
\item We present a holistic view of what the 6G RAN is expected to look like, covering its principal building blocks with an emphasis on standardization.
\item For each building block, we discuss implementation and deployment aspects and, where possible, identify practical challenges and limits.
\item We discuss how non-terrestrial systems will be integrated with terrestrial systems in 6G to provide ubiquitous and global coverage.
\item We discuss support for AI-native capabilities and ISAC in the
RAN and the associated practical implementation and deployment
constraints.
\item We provide a comprehensive bibliography that enables interested readers to explore these topics in greater detail.
\end{itemize}
The remainder of this paper is organized as follows. Section~\ref{sec:usecases} presents the 6G use cases, and Section~\ref{sec:KPI} describes the 6G performance requirements. Section~\ref{sec:bands} discusses the new spectrum bands for 6G, although 6G will operate across all available bands. Section~\ref{sec:arch} then examines the architecture, followed by a discussion of the RAN building blocks in Section~\ref{sec:tech}. Finally, Section~\ref{sec:conc} presents the conclusions.

\vspace{-0.2cm}
\section{6G Use Cases}
\label{sec:usecases}

The usage scenarios envisaged for 6G expand upon those of 5G (i.e., eMBB, URLLC, and mMTC, as discussed in \cite{ITU1}) to include evolved forms of these scenarios; see Fig.~\ref{fig:ITUdiagram}. In addition, 6G is expected to enable new usage scenarios based on capabilities such as AI and ISAC, which were not part of the original 5G framework. The 3GPP report in \cite{22870} summarizes the 6G use cases and service requirements. Owing to space limitations, only a brief list of these use cases is
given below:
\begin{itemize}
\item \textbf{Immersive Communication} (IC): This usage scenario covers use cases that provide users with rich, high-fidelity, interactive video experiences, including interactions with machine interfaces.
\item \textbf{Artificial Intelligence and Communication}: This scenario encompasses two complementary concepts: (1) AI for 6G systems, in which AI capabilities support networks and devices in providing services; and (2) 6G systems for AI, in which 6G system functionalities support and enable AI applications and services.
\item \textbf{Integrated Sensing and Communication}: This scenario enables new applications and services that rely on wide-area, multidimensional sensing to provide spatial information about unconnected objects and connected devices, as well as their movements and surroundings.
\item \textbf{Ubiquitous Connectivity}: This scenario aims to enhance connectivity, for example, through NTN access, and help bridge the digital divide.
\item \textbf{Massive Connectivity}: This usage scenario extends 5G mMTC and supports connectivity for massive numbers of devices or sensors across a wide range of use cases and applications, including battery-free Internet of Things (IoT) devices and devices requiring long battery lifetimes
\cite{sensorssurvey}.
\item \textbf{Home Robots}: Examples include home robots that perform household chores, support social interaction, and improve quality of life.
\item \textbf{Fixed Wireless Access}: 6G networks will continue to support FWA, as in 5G. FWA devices differ from smartphones in their traffic patterns and mobility profiles.
\item \textbf{Compute}: If 6G networks evolve into cloud-service platforms, locally available computing resources could be made available to subscribers, enabling a range of computing services over the mobile network.
\item \textbf{Non-Terrestrial Networks}:
The integration of NTN into 6G is expected to provide global, resilient, and energy-efficient coverage \cite{RP-251881}.
\end{itemize}

\vspace{-0.2cm}
\section{6G Performance Requirements}
\label{sec:KPI} %

The 6G RAN key performance indicators (KPIs) specified in 3GPP TR~38.914~\cite{38914} cover data rate, spectral efficiency, latency, reliability, connection density, mobility, bandwidth, positioning, sensing, AI-related capabilities, coverage, and composite service requirements. Table~\ref{tab:5g_6g_kpi_comparison} compares a selected set of these targets with their 5G counterparts in TR~38.913~\cite{38913}. 

\begin{table*}[t]
\caption{Selected 5G and 6G RAN performance targets from
3GPP TR~38.913 and TR~38.914
\cite{38913,38914}.
Values separated by ``/'' denote DL/UL.}
\label{tab:5g_6g_kpi_comparison}
\centering
\small
\setlength{\tabcolsep}{4pt}
\renewcommand{\arraystretch}{1.12}

\begin{tabularx}{\textwidth}{
    @{}p{0.17\textwidth}
       p{0.25\textwidth}
       p{0.28\textwidth}
       X@{}}
\toprule
\textbf{KPI} &
\textbf{5G RAN target: TR~38.913} &
\textbf{6G RAN target: TR~38.914} &
\textbf{Nominal change} \\
\midrule

Peak data rate
& 20/10~Gbit/s
& $ 36$/$ 18$~Gbit/s
& $1.8\times$ \\

Peak spectral efficiency
& 30/15~bit/s/Hz
& $ 60$/$ 30$~bit/s/Hz
& $2\times$ \\

Average spectral efficiency
& $7.8/5.4$~bit/s/Hz/TRP
  (dense urban)
& $ 23.4$/$ 16.2$~bit/s/Hz/TRP
  (dense urban)
& $3\times$ \\

$5^{\mathrm{th}}$-percentile user spectral efficiency
& $0.225/0.150$~bit/s/Hz
  (dense urban)
& $ 0.675$/$ 0.450$~bit/s/Hz
  (dense urban)
& $3\times$ \\

User-plane latency
& 4~ms (eMBB); 0.5~ms (URLLC)\textsuperscript{a}
& $4$~ms (IC); $1$~ms (HRLLC)
& Not tightened numerically \\

Control-plane latency
& 10~ms
& $ 20$~ms
& Not tightened numerically \\

Reliability
& $1-10^{-5}$ for 32 bytes within 1~ms
& $ 1-10^{-5}$ for 32 bytes L2 PDU within
  1~ms
& Not tightened numerically \\

Connection density
& $10^{6}$ devices/km$^{2}$
& $ 10^{6}$ devices/km$^{2}$
& Same headline density \\

Mobility
& 500~km/h
& 500~km/h (and up to 1000/1200~km/h for airplane)
& Higher mobility considered for airplane \\
\bottomrule
\end{tabularx}

\vspace{1mm}
\begin{minipage}{\textwidth}
\footnotesize
\textit{Note:}
\textsuperscript{a}IMT-2020 targets a user-plane latency requirement of 1~ms for URLLC, whereas 3GPP TR~38.913 sets a more stringent target of 0.5~ms.
\end{minipage}
\end{table*}

As shown in Table~\ref{tab:5g_6g_kpi_comparison}, the principal numerical improvements from 5G to 6G concern throughput and spectral efficiency. The peak data rate targets increase by a factor of 1.8, while peak spectral efficiency doubles. In the dense urban scenario, both average spectral efficiency and $5^{\mathrm{th}}$-percentile user spectral efficiency increase by a factor of three. The latter improvements are particularly relevant to network capacity and cell-edge user experience. The transition to 6G does not, however, uniformly tighten every conventional KPI. The user-plane latency targets are 4~ms for IC and 1~ms for hyper-reliable and low-latency communication (HRLLC), compared with 4~ms for eMBB and 0.5~ms for URLLC in 5G. The control-plane latency target is 20~ms for 6G, compared with 10~ms for 5G. Similarly, the reliability target remains $1-10^{-5}$ for a 32-byte packet delivered within 1~ms, while the connection density target remains $10^{6}$ devices/km$^{2}$. The maximum evaluated terrestrial mobility remains 500~km/h, although 6G additionally considers an airplane mobility class with speeds up to 1000/1200~km/h. 

Beyond the directly comparable targets in Table~\ref{tab:5g_6g_kpi_comparison}, the 6G study introduces or expands evaluation dimensions related to area traffic capacity, positioning, energy efficiency, sensing, AI, resilience, extended connectivity, coverage, and composite service performance. Thus, the 6G performance framework combines selective increases in capacity-related targets with a broader set of capabilities and deployment scenarios.

\begin{figure}
    \centering
    \includegraphics[width=0.9\columnwidth]{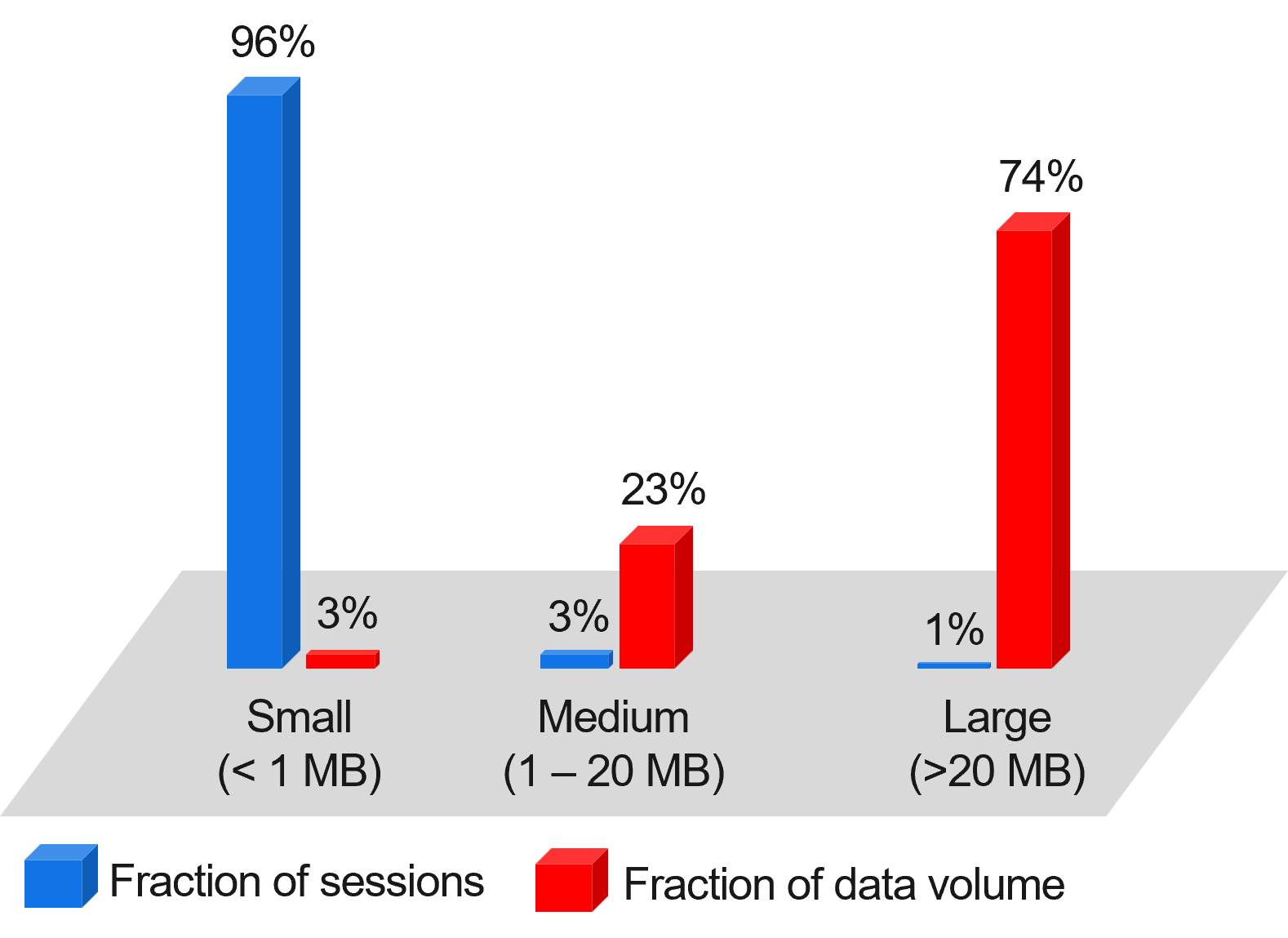} %
    \begingroup
\captionsetup{font=footnotesize}
    \caption{Measured downlink traffic in a commercial European network, grouped by session size. Small sessions (<1 MB) account for 96\% of sessions but only 3\% of data volume. Medium sessions (1--20 MB) account for 3\% of sessions and 23\% of data volume, whereas large sessions (>20 MB) account for 1\% of sessions and 74\% of data volume. The contrasting distributions illustrate why radio access network (RAN) design must efficiently support both frequent small transfers and the relatively few large transfers that dominate traffic volume.}
    \label{fig:bursty_traffic}
    \endgroup
    \vspace{-5pt}
\end{figure}

It is noted that metrics such as peak rates do not have any bearing on the end-user performance experienced in the field. This type of metric has an inherent full-buffer assumption, that is, there is an infinite amount of data to transfer, while traffic in real networks typically is very bursty. In fact, most data sessions generate only a small amount of data, often on the order of 10 kB, while a small number of sessions constitute the bulk of the traffic volume. In Fig.~\ref{fig:bursty_traffic}, measurements on a real, commercial network are shown, which clearly illustrate this aspect: most sessions (96\%) are small and a few sessions (1\%) carry most of the data (74\%). This traffic behavior needs to be accounted for in the design of 6G, which needs to handle both small and large data sessions in an efficient way. 

Low latency remains important for 6G, but the widely publicized ``1 ms 5G latency'' requires qualification. It was an IMT-2020 one-way radio network target under specific evaluation conditions, rather than a guarantee of 1 ms end-to-end latency in commercial deployments. In practice, commercial 5G has been led primarily by eMBB and FWA, while the industrial realization and uptake of URLLC have been more limited than initially anticipated. Moreover, end-to-end latency depends on transport, device support, system integration, processing, traffic load, and radio conditions. Although HRLLC is identified by ITU-R as an IMT-2030 usage scenario~\cite{ITUFRamework}, the corresponding 6G user-plane latency target in 3GPP TR~38.914 is not tighter than the 5G URLLC target. From an industrial viewpoint, this may be interpreted as placing greater emphasis on dependable end-to-end performance by jointly considering latency, reliability, and availability than on pursuing a still-lower headline latency value~\cite{38914}.

Low latency is nevertheless particularly important for small data sessions, for which connection establishment can constitute a significant fraction of the total transfer time. It also enables transport protocols such as transmission control protocol (TCP) and quick user datagram protocol internet connections (QUIC) to increase their sending rates quickly and utilize the available radio resources efficiently~\cite{TCP_slow_start}. Relevant RAN mechanisms include camping on an appropriate carrier to avoid inter-frequency handovers, rapid connection setup, and early channel state information (CSI) reporting. For the larger data sessions, spectrum-efficient transmission schemes are crucial. Carrier aggregation and/or large transmission bandwidths, advanced MIMO schemes, and accurate CSI reports are examples of technologies that can be used.

\vspace{-0.2cm}
\section{Spectrum Bands}
\label{sec:bands}

Many academic papers on 6G have treated 6G as synonymous with the use of sub-THz bands, much as 5G has often been treated as synonymous with mmWave bands.
Neither characterization is accurate. Most deployed 5G cellular systems today operate below mmWave, including in the C-band; higher-frequency bands are generally less suitable for wide-area deployment \cite{shaficband}. Providing wide-area coverage at mmWave frequencies can require substantial network densification, with short inter-site distances, to address coverage limitations \cite{shaficband}. RISs have been proposed to mitigate these limitations, but their deployment presents operational and logistical challenges \cite{TATARIA1}. RIS-specific standardization is also outside the scope of the current 3GPP 6G Radio study. These coverage and deployment challenges are likely to be more pronounced at sub-THz frequencies \cite{TATARIA1}. Against this background, mid-band spectrum is expected to be central to 6G, consistent with the candidate ranges identified in \cite{WRC2023}. Existing mobile bands are also expected to support 6G, much as they support 5G today.

The World Radiocommunication Conference 2023 (WRC-23) \cite{WRC2023} identified the following frequency ranges for study with a view to possible IMT identification at WRC-27. Their sharing and compatibility conditions will also be considered at WRC-27.
\begin{itemize}
\item 4.4--4.8 GHz, or parts thereof, in ITU-R Regions 1 and 3;
\item 7.125--8.4 GHz, or parts thereof, in ITU-R Regions 2 and 3;
\item 7.125--7.25 GHz and 7.75--8.4 GHz, or parts thereof, in ITU-R Region 1;
\item 14.8--15.35 GHz.
\end{itemize}
Current experience with mmWave indicates that higher-frequency bands, including sub-THz spectrum, are generally unsuitable for wide-area cellular deployments \cite{shaficband}. WRC-23 also identified all or part of the 6.425--7.125 GHz band for IMT in several countries and regions \cite{WRC2023}; this spectrum may therefore also support 6G.

The candidate bands identified for possible IMT use in \cite{WRC2023} are already heavily used by a variety of incumbent services. Coexistence studies between future IMT systems, including 6G, and these incumbent services are being conducted within the ITU-R. These studies will inform decisions on the spectrum that may be identified for IMT and the technical and regulatory conditions needed to protect incumbent services. Nevertheless, 6G will also operate in existing bands currently used by 4G and/or 5G. A fundamental requirement for 6G is therefore the ability to dynamically share spectrum resources with 5G, as discussed in Section~\ref{subsec:MRSS}.

Table~\ref{tab:standards} provides an overview of the frequency bands and channel bandwidths used by terrestrial mobile systems from 3G through 5G, together with the expected values for 6G \cite{TS38104}. Frequency bands for satellite systems are discussed in Section~\ref{subsec:ntn}. As shown in Table~\ref{tab:standards}, both carrier bandwidths and operating frequency ranges have increased substantially since the early generations of mobile systems. Carrier bandwidths are expected to reach 400 MHz or more in 6G. However, transmission over such wide carrier bandwidths poses challenges. In particular, the challenges of maintaining RF linearity across a wide operating bandwidth are discussed in \cite{TATARIA1}.

\vspace{-0.3cm}
\section{Architecture}
\label{sec:arch}

\subsection{RAN Architecture and Interfaces}
\label{subsec:RANarch}
\begin{figure*}[t!]
    \centering
    \includegraphics[width=2.0\columnwidth]{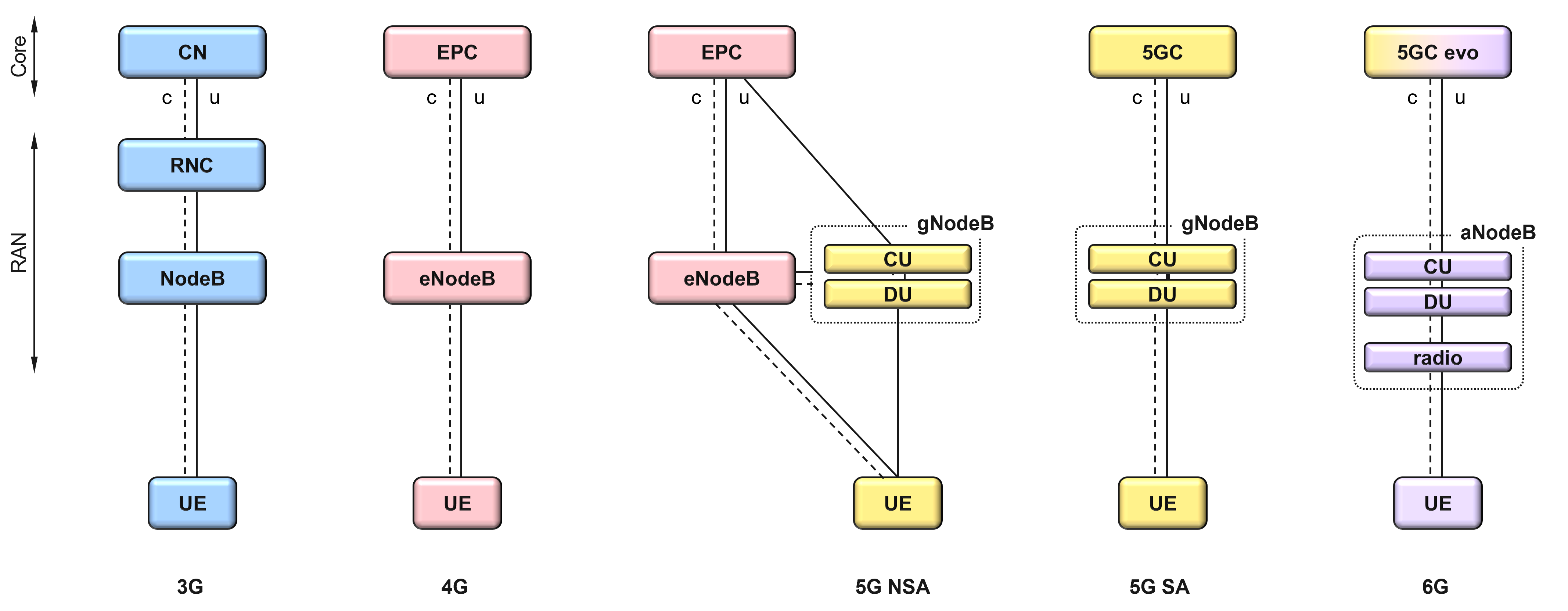}
    \begingroup
\captionsetup{font=footnotesize}
\caption{Simplified evolution of the functional radio access network (RAN) and core network (CN) architecture from 3G to a prospective 6G system. The 3G radio network controller (RNC) disappears in the flatter 4G architecture. In 5G non-standalone (NSA), the 4G evolved packet core (EPC) and eNodeB anchor connectivity, whereas 5G standalone (SA) uses the 5G core (5GC). The prospective 6G base station, labeled aNodeB, connects to an evolved 5GC and contains central-unit (CU), distributed-unit (DU), and radio functions. UE denotes user equipment; dashed ``c'' and solid ``u'' links represent control and user planes. The boxes represent logical network functions.}
    \label{fig:architecture}
    \endgroup
\end{figure*}

Over the different generations, summarized in Table~\ref{tab:standards}, the RAN architecture has evolved as illustrated (in a simplified form) in Fig.~\ref{fig:architecture}. In 3G, the RAN consisted of two nodes, the radio network controller (RNC), to which multiple NodeBs connect (``NodeB'' is the name of a logical node that can be realized as a base station). With the introduction of 4G, a flat architecture was agreed upon, the RNC was dropped, and a new core network, the evolved packet core (EPC), was defined.  When 5G was introduced, two architectures were defined: non-standalone (NSA) and standalone (SA). In the former, the user equipment (UE) is connected to both 4G and 5G, with 4G being responsible for connection establishment and 5G acting as a data rate booster. In the standalone alternative, on the other hand, a new 5G core network is used and all functionality, including connection establishment, is handled by 5G. The reason for having two options, NSA and SA, was to allow operators to gradually introduce 5G and to benefit from the wide-area 4G coverage. Although having multiple architectural options might look tempting at first sight, it does add to the overall complexity of the system \cite{Cagenius:18}. It has also fragmented the market and thereby delayed the uptake of 5G SA. Currently, the majority of the world's 5G networks are NSA, while SA is required to get the full 5G benefits. 

With this in mind, 3GPP aims at an SA architecture as the baseline \cite{RP-253876}. Although not finally decided, the major trend is toward evolving the 5G core to also handle 6G rather than developing a new 6G core network. There are two main arguments for using an evolved 5G CN. First, the service-based 5G CN architecture is very flexible \cite{Rommer:25}; new functionality to support 6G services such as ISAC can be easily added. Second, as many operators are just starting to migrate from 5G NSA to 5G SA, investments in a completely new CN in parallel (or shortly after) can be financially difficult to motivate. The structure of the RAN-CN interface is currently being discussed in RAN3; a likely scenario is a point-to-point interface between logical nodes (as in 5G), evolved to be more cloud-friendly. 

One purpose of standardization is to provide a healthy ecosystem with open multi-vendor interfaces defined when motivated by business reasons. Clearly, the UE-RAN interface (known as Uu in 3GPP) is a prime example of a business-motivated multi-vendor interface, as are the RAN-CN interface and a RAN-RAN interface for mobility between 6G base stations (aNodeBs). With the emergence of Open RAN, the lower-layer split (LLS), that is, an interface between the radio unit and the remaining part of the base station processing, has emerged as an increasingly important open interface. There is a strong push in the industry to align the overall architecture work in 3GPP with the LLS work in the O-RAN Alliance, as illustrated by the summary from the joint 3GPP and O-RAN Alliance workshop in April 2025 \cite{3ORW-250035}. 

As shown in Fig.~\ref{fig:architecture}, the 5G architecture allows the gNodeB to be split into two parts, a central unit (CU) and a distributed unit (DU), with a standardized F1 interface between the two. A relevant question is whether this split should be kept in 6G or not. Originally intended as an open interface, the F1 interface in 5G has, with a few exceptions, not been used in a multi-vendor setting. Splitting the control of the RAN into a CU and a DU complicates the overall management of the system and may unnecessarily limit the overall performance. However, migration of existing 5G deployments relying on the CU-DU split also needs to be taken into account. Allowing a CU-DU split also in 6G but not designing the RAN2 protocols on the assumption of this split being present could be a way to strike a balance between supporting deployments based on this split and still enabling performance optimizations.

Migration from 5G to 6G is crucial for an operator, especially given that very little, if any, new spectrum is available in the lower frequency range (FR) bands (i.e., FR1), which are important for coverage. The key mechanism for 5G-to-6G migration identified by 3GPP \cite{RP-253876} is multi-RAT spectrum sharing (MRSS), described in Section~\ref{subsec:MRSS}. Whether any complementary mechanisms to support the 5G-to-6G migration are needed or not, for example, dual connectivity, is to be studied by 3GPP until September 2026 \cite{RP-253876}. 

Security is an area of growing importance. 6G will build on the already strong security mechanisms built into 5G. Ciphering schemes will be evolved to be resistant to future quantum-computer-based attacks (``post-quantum ciphers''). 3GPP is also looking into whether some of the RAN-internal control information between the UE and the aNodeB needs increased protection. %

It is important to understand that the architecture discussed in the previous paragraphs is the standardization (functional) architecture, that is, the entities and interfaces defined in 3GPP. From an implementation or deployment perspective, different architectures can be envisioned, for example, locating different parts of the CN at different geographical sites, making a ``single-box'' RAN implementation, or implementing a product with both RAN and CN running on the same hardware. These implementation choices are not part of the 3GPP discussions.

\subsection{Multi-RAT Spectrum Sharing}
\label{subsec:MRSS}

MRSS is a key feature of 6G, enabling a smooth migration from 5G to 6G by dynamically sharing spectrum between the two. When designing the MRSS scheme, several aspects are important to consider. First, as an existing 5G device cannot be changed, any new 6G signals must be invisible to a 5G device. Furthermore, from an operator perspective, the changes to the configuration of the 5G part of the network, if any, and any performance impact on 5G, must be minimized. Finally, to simplify the 6G design, the same set of access procedures should preferably be used regardless of whether the device tries to access a 6G-only carrier or an MRSS carrier that supports both 5G and 6G. 

The details of the MRSS design remain to be discussed in 3GPP. Nevertheless, a few aspects can be identified. The decision to use the same waveform as in 5G (i.e., OFDM) and the same set of subcarrier spacings simplifies the MRSS design. This allows a very dynamic sharing of resources between 5G and 6G following the short-term traffic variations and avoiding any guard bands between 5G and 6G. Scheduling and other resource-allocation mechanisms can be used to dynamically share resources between 5G and 6G.

Hiding the 6G synchronization signal block (SSB) used for initial access and mobility measurements (if such a signal is defined) can, for example, use the reserved resource mechanism defined in 5G as part of the forward-compatibility design \cite{Dahlman:23}. Alternatively, the multiple SSB occasions defined in 5G as part of beam-sweeping can be used with unoccupied 5G SSB positions used for the 6G SSB. Similarly, the random-access occasions for 6G can reuse the 5G ones, albeit with different random-access preambles, thereby being ``hidden'' from the 5G part of the network. 

To achieve a very low MRSS overhead, the physical downlink control channel (PDCCH) in 6G needs to be designed such that dynamic sharing with 5G is possible. Analysis in 3GPP has shown that a total MRSS overhead on the order of a few percent is possible \cite{R1-2606927} and that 6G can boost the performance significantly compared to 5G. Thus, the real performance of an MRSS carrier is better than that of a 5G-only carrier, taking both the overhead cost and the 6G gain into account. %

\subsection{O-RAN Aspects}

Open RAN denotes the general architectural approach of
disaggregating RAN functions through softwarization and
virtualization and introducing open interfaces between them \cite{polese,Andrews_2024,bonati}.
O-RAN refers more specifically to the architecture and
specifications developed by the O-RAN Alliance. These interfaces
are intended to enable multi-vendor interoperability, although
standardization alone does not guarantee it. O-RAN builds on the 3GPP RAN architecture. While 3GPP
defines the radio interface, RAN functions, and overall 6G system
architecture, the O-RAN Alliance is expected to complement these
specifications. Close
coordination between the two organizations is therefore important,
as discussed at the joint 3GPP and O-RAN Alliance workshop on 6G
coordination~\cite{RP-250866}.

At the specification level, the 5G O-RAN architecture adopts the
3GPP CU/DU and CU control-plane/user-plane decompositions and the corresponding
E1 and F1 interfaces. It additionally specifies the radio unit (RU) and the
7-2x LLS between the DU and RU as the open
fronthaul. The open fronthaul comprises control, user, and
synchronization planes, together with a management plane for RU
management. The service management and orchestration (SMO)
framework manages O-RAN network functions over the O1 interface
and cloud infrastructure and workloads over O2
\cite{ORANArchitecture2026,ORANFronthaul2026}. The architecture also introduces complementary control loops for
RAN optimization. The non-real-time RAN intelligent controller
(non-RT RIC), located within the SMO, hosts rApps and provides
policy, enrichment information, and AI/ML model management services
to the near-RT RIC over A1. The near-RT RIC hosts xApps and uses
E2 to collect measurements from, and control exposed functions
in, E2 nodes such as the CU and DU. Non-RT control loops
typically operate over intervals of at least one second, whereas
near-RT control loops operate approximately between 10~ms and
1~s. 

For 6G, the open fronthaul between the DU and RU is of
particular interest. Wider bandwidths, larger antenna arrays,
advanced beamforming, and NTN integration
will affect the required fronthaul capacity, latency,
synchronization, and functional split. The corresponding O-RAN
interfaces and service models will therefore need to evolve in
coordination with the 3GPP 6G architecture. From an industrial viewpoint, open interfaces are necessary but
not sufficient for plug-and-play multi-vendor operation.
Commercial deployment also requires agreed interface profiles,
conformance and interoperability testing, compatible service
models and software lifecycles, end-to-end performance validation,
and appropriate security controls. Consequently,
disaggregation introduces system-integration and operational
complexity across the participating vendors.

\subsection{Cloud RAN Impacts}

Cloud-based processing with cloud RAN (CRAN) is possible and already used in 5G. Typically, CRAN implementations have considered hardware acceleration for signal decoding (to achieve better energy efficiency and a low load for cloud processing) or for inline acceleration. 

The larger bandwidths in 6G will have increased requirements for the connectivity between the cloud infrastructure and the radio sites. The resulting data rate needed for the fronthaul connection will increase as a result of the larger bandwidths, suggesting the use of an enhanced common public radio interface (eCPRI)-type approach based on the work in O-RAN for the fronthaul from the radio site toward the cloud infrastructure, as sending pure I/Q samples might be an inefficient solution. The cloud-friendliness of the RAN-CN interface will also be investigated, as mentioned in Section~\ref{subsec:RANarch}. %

\begin{figure}[htbp!]
    \centering
    \vspace{-2pt}
    \includegraphics[width=1.0\columnwidth]{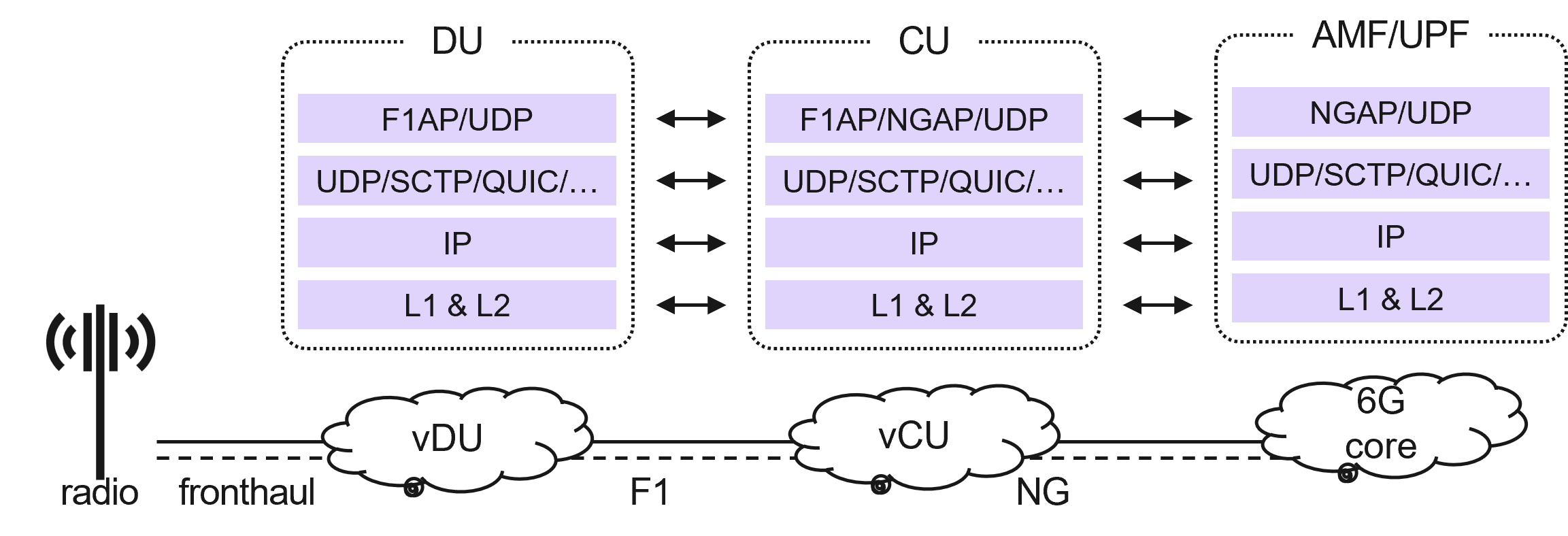}
    \begingroup
\captionsetup{font=footnotesize}
    \caption{Illustrative cloud radio access network (CRAN) implementation for 6G. The radio connects through fronthaul to a virtualized distributed unit (vDU), followed by a virtualized central unit (vCU) and the core network. F1 and NG label the DU--CU and RAN--core interfaces using familiar 5G terminology. F1AP and NGAP denote F1 Application Protocol and NG Application Protocol; UDP and IP denote User Datagram Protocol and Internet Protocol. QUIC is shown as a candidate transport, with L1 and L2 representing the physical and data-link layers. AMF and UPF denote the access and mobility management function and user plane function \cite{23501}. The stacks illustrate a possible design for 6G.}

    \label{fig:cloud_6G_architecture_v3}
    \endgroup
    \vspace{-10pt}
\end{figure}

The 6G architecture with cloud-based implementation is shown in Fig.~\ref{fig:cloud_6G_architecture_v3}. The interfaces between the CN and RAN have always been defined as  open interfaces in 3GPP. O-RAN is expected to address the fronthaul interface for 6G implementations. 

\section{Technologies for 6G RAN}
\label{sec:tech}

\subsection{Foundations of (Massive) MIMO}\label{subsec:mimofund}
\label{sec:mimo}

Massive MIMO is the dominant physical-layer technology in 5G, and conceivably in all future wireless systems \cite{larsson2014massive}.
The MIMO framework for 6G will, to a large extent, build upon the Massive MIMO 5G solutions developed from Release 15 and onward \cite{Rel18mimo}. When operating in reciprocity-based beamforming mode in time-division duplex (TDD), Massive MIMO is agnostic of the antenna array topology (though most arrays in practice use $\lambda/2$-spaced antennas, where $\lambda$ denotes wavelength), of the propagation environment, and of whether the
terminals are in the geometric far-field or near-field of the array.

Massive MIMO is built upon two fundamental insights:
\begin{itemize}
\item When there is an excess of service (base station) antennas relative to the number of users, linear signal processing (e.g., maximum-ratio combining and zero-forcing) becomes nearly optimal. This substantially simplifies the signal processing, compared to what would be required for conventional multi-user MIMO (MU-MIMO) with a number of base station antennas comparable to the number of multiplexed terminals. In fact, somewhat paradoxically, the information-theoretic analysis of MIMO simplifies in the large-number-of-antennas regime. Non-asymptotic, rigorous lower bounds on the ergodic Shannon capacity are known for imperfect CSI, both for single- and multi-cell systems \cite{marzetta2016fund}; these bounds take on a simple form and may be directly used for system optimization.

\item If the system operates in TDD, electromagnetic reciprocity can be exploited to obtain downlink CSI from uplink pilots. This makes the resources required to acquire CSI for downlink beamforming \emph{independent} of the number of service antennas.  
\end{itemize}
\paragraph{What Can Massive MIMO Offer and How Does It Work?}
A useful abstraction of the wireless channel is that the time-frequency domain can be segmented into coherence blocks, whose time duration equals the channel coherence time (say $T_c$) and whose
bandwidth equals the channel coherence bandwidth (say $B_c$).
With an appropriate degree of approximation (as made precise, for example, in \cite[Ch. 2]{marzetta2016fund}), the channel is a static linear system within a coherence block. This coherence-block approximation of the channel fading is useful to conceptualize the workings of massive MIMO, but since the channel response does vary within a block, practical signal processing algorithm implementations
require the use of interpolation methods. In the 3GPP standard,
the size of a resource block is comparable to that of a coherence block.
At a fixed mobility (in meters per second), the dimensionality ($B_c\cdot T_c$) of the coherence interval is proportional to the wavelength, because the Doppler spread is proportional to the carrier frequency.
In \cite{marzetta2016fund}, it is shown that in independent Rayleigh fading, a single isolated cell with a massive MIMO base station with $M$ antennas serving $K$ single-antenna terminals with zero-forcing decoding and max-min fairness power control (all $K$ terminals receiving the same rate: uniform quality of service (QoS)) can achieve an uplink sum rate of
\begin{align}
&  \frac{KB}{2}\cdot \left( 1-\frac{K}{B_cT_c}\right) \nonumber \\
  \cdot & \log_2\left( 1 + \frac{(M-K)\cdot \mbox{SNR}}{\min_k \frac{1}{\gamma_k} +
    \mbox{SNR}\cdot \left( \sum_{k=1}^K \frac{\beta_k}{\gamma_k} -1 \right) } \right). \label{eq:e1}
  \end{align}
In (\ref{eq:e1}), $K$, $M$, $B_c$, and $T_c$ are as defined earlier; $B$ is the system bandwidth; SNR is the signal-to-noise ratio; $\beta_k$ is the path loss with the $k$th terminal; and $\gamma_k$ is the mean-square channel estimate. One always has $\gamma_k\le\beta_k$, and in the high-SNR region, $\gamma_k\approx \beta_k$. The ratio $\gamma_k/\beta_k$ quantifies the noisiness of the uplink pilots.
The fundamental formula (\ref{eq:e1}) offers a number of important
insights concerning the operation of massive MIMO (in a single cell):
\begin{itemize}
\item Capacity scales proportionally to bandwidth $B$, assuming the SNR is fixed (which requires increasing the transmit power proportionally to $B$).
  
\item The factor that pre-multiplies the logarithm accounts for the fraction of resources that must be spent on uplink pilots for the base station to acquire CSI. This factor is maximized for $K=B_cT_c/2$. Consequently, the multiplexing gain (maximum permissible $K$) is limited by the channel coherence. Eventually, if $K$ equals $B_cT_c$, all resources are spent on pilots and none on data payload.  
  
\item The term after ``$1+$'' inside the logarithm is the effective signal-to-interference-plus-noise ratio (SINR) per terminal. It scales with $M-K$: $M$ for the array gain, but $K$ degrees of freedom are consumed by the interference nulling of the zero-forcing processing; hence $M-K$. This relation also shows that to multiplex $K$ terminals, one needs at least $K$ antennas; preferably, at least twice or thrice that, but beyond that point, the gain in rate by increasing $M$ further is only logarithmic. Hence, in view of the observation in the previous bullet, it is ultimately $B_cT_c$ that determines how many base station antennas are useful. The SINR also scales approximately inversely proportionally to $K$, reflecting that the amount of interference in the cell grows with $K$. The sole source of this interference is channel estimation errors: with perfect CSI, the zero-forcing detector would cancel all but the desired signal. 
Somewhat paradoxically, the effective SINR is approximately proportional to $M$ (if $M\gg K$), despite the fact that the number of unknown channel coefficients to estimate in every coherence block scales proportionally to $M$ too; even though the number of unknown parameters increases, performance uniformly increases with $M$.

\item Since the optimal number of terminals to multiplex, $K$, is proportional to $T_c$ \cite{marzetta2016fund}, it is also proportional to the wavelength (assuming $B_c$ does not change with frequency, which is consistent with measurements \cite{adhikary2014joint}). 
At the same time, if $M$ scales with $K$, the quantity inside the logarithm is constant. This means that the sum rate can double for every doubling of the wavelength: one MHz of bandwidth at a 100 MHz carrier has tenfold value compared to one MHz of bandwidth at a 1 GHz carrier. This shows the power of spatial multiplexing, assuming that enough terminals and data are available for such multiplexing to be meaningful, and that the physical dimensions (form factor) of the array can be tolerated.
\end{itemize}

In a multi-cell system, matters are considerably more involved, as
there will be interference among the cells. This interference comes in two forms: non-coherent and coherent \cite{marzetta2016fund}. For illustration, consider the uplink. Non-coherent interference originates from cells that use different pilots from those used in the home cell; its effect is similar to that of thermal noise: when increasing the number of base station antennas, $M$, the interference is simply averaged out.
Coherent interference, in contrast, originates from cells that use
\emph{the same} pilots as in the home cell. This coherent
interference, also known as pilot contamination, does not average out when increasing $M$. Rather, its magnitude \emph{scales with $M$}, so that the benefit of having a large antenna array is essentially lost. Much methodology has been developed to mitigate this phenomenon \cite{sanguinetti2019toward}, supplementing the uplink pilots with side information about the 
channel to enable more accurate channel estimates.

\paragraph{CSI, TDD, and FDD}
In TDD operation, terminals transmit pilots on the uplink; CSI is then obtained by the base station. By virtue of reciprocity, this CSI is valid also on the downlink. In practice, this requires the MIMO array at the base station to be reciprocity-calibrated, which can be achieved using algorithms that measure the channel response between antennas in the array \cite{vieira2017reciprocity}. The main advantage of TDD is the scalability with respect to $M$: \emph{the pilot overhead is independent of the number of base station antennas}, although this overhead scales with the number of UEs (assuming some form of uplink pilots such as the sounding reference signal (SRS)). 

In frequency-division duplex (FDD), the situation is different. While large arrays certainly can be used, the overhead required to obtain downlink CSI scales proportionally to $M$. First, one cannot rely on uplink pilots, so every antenna on the downlink has to send an orthogonal pilot; the cost of this scales with $M$. Second, the terminals must estimate the channel and feed CSI back to the base station over a control link, an activity whose cost also scales with $M$.

The fact that most new spectrum is licensed for TDD operation
testifies to the advantages of TDD operation, as predicted by the
physics and information-theoretic arguments of
\cite{marzetta2006fast,marzetta2016fund}; see \cite{Rel18mimo} for
CSI acquisition in FDD bands. 
However, it is noted that TDD also has drawbacks, for example, uplink latency (the need to wait for an uplink slot) and reduced (uplink) coverage compared to FDD.

\paragraph{Distributed MIMO}
Distributed MIMO (D-MIMO) is a technology that appears under many
different names: cell-free massive MIMO
\cite{ngo2017cell,ngo2024ultradense}, user-centric MIMO
\cite{demir2021foundations}, network MIMO \cite{venkatesan2007network},
ubiquitous MIMO, 
RadioStripes \cite{InterdonatoBNFL2019}, RadioWeaves
\cite{van2019radioweaves}, 
large intelligent surfaces
\cite{hu2018beyond}\footnote{Not to be conflated with \emph{reconfigurable}
intelligent surfaces (RISs).}, and pCell. The idea is to connect
several geographically separated MIMO arrays (antenna panels) together using a fast backhaul, and in such a way that these panels can operate phase-coherently together. Effectively, any terminal is served by a multiplicity of antenna panels in its vicinity, hence the term
``user-centric.'' Cell borders disappear, or at least have a different meaning, hence the term ``cell-free.'' The superiority of D-MIMO over small-cell deployments is well documented \cite{ngo2017cell} and is a simple consequence of the fact that D-MIMO suppresses interference, whereas densification (with small cells) makes it worse.

Two main technical challenges in D-MIMO are to distribute the data over the backhaul and to achieve phase-synchronous transmission on the downlink \cite{xu2025distributed,ngo2024ultradense}. On the uplink, pilots and data see the same channel, so phase coherence is not an issue. On the downlink, assuming TDD with reciprocity-based beamforming, the antenna panels have to be \emph{jointly reciprocity-calibrated} \cite{nissel2022correctly,larsson2023phase,ganesan2023beamsync}. This can be achieved either through the distribution of a common clock reference (e.g., via a fiber cable), or by equipping the antenna panels with independent oscillators and having them perform bidirectional over-the-air  calibration measurements on one another \cite{larsson2024massive,vieralarsson_pimrc}. In the latter case, these calibration measurements have to be integrated into the TDD flow \cite{ngo2025breaking}, and they have to be frequent enough that recalibration is accomplished as soon as the local phase references at the different panels have drifted significantly away from one another. Over-the-air reciprocity calibration in D-MIMO is an active research topic. One may alternatively involve the terminals in this process, but this is considered undesirable.

\vspace{-0.3cm}
\subsection{Industrial and 3GPP Aspects of MIMO}
\label{industrymimo}

Next, we discuss how MU-MIMO and massive MIMO are being implemented in the 3GPP standards. 

\subsubsection{MIMO in the 5G-to-6G Transition}
\label{subsubsec:6Gantennas}

While the MIMO framework for 6G will, to a large extent, build upon the corresponding 5G framework,
there are aspects that will change, based on experience from 5G in the field as well as to address new, previously unsupported deployment scenarios. In 5G Release 20, there will be work on early CSI reporting to quickly provide the base station with the necessary information to handle bursty traffic \cite{RP-251856}. Similar solutions should also be incorporated in the 6G design.

Uplink MIMO is becoming increasingly important, as
some use cases (e.g., FWA) require high data rates not only in the downlink but also in the uplink. Increasing the number of uplink transmit antennas is one possibility (see Table~\ref{tab:mimo} for some indicative numbers). Coherent uplink MIMO is also getting more attention. Although it is formally part of the 5G specifications, it has not been implemented in practice. The acceptable phase error between the transmit antennas (in 5G, the same value is used for 2, 4, and 8 transmit antennas) is one example where the 6G specifications can be improved. Support for two-layer MIMO with discrete Fourier transform spread OFDM (DFT-s-OFDM) has been agreed and can improve cell-edge data rates (in 5G, only single-layer transmission is supported for DFT-s-OFDM).

Downlink MIMO is also likely to see improvements in 6G. One clear trend is the increase in the number of antenna elements, with massive MIMO becoming even more massive, especially in the centimeter-wave range, as seen in Table~\ref{tab:mimo}. By increasing the number of antenna elements (in principle, four times as many antenna elements can be used at 7 GHz compared to 3.5 GHz without increasing the physical size of the antenna unit), the increased path loss at the higher carrier frequencies can be compensated for. This supports deploying 6G at 7~GHz on the existing 5G C-band
(around 3.5~GHz) site grid without a major impact on downlink coverage. However, integrating such a large number of antenna elements in one antenna unit is not a simple task and is challenging in terms of, e.g., cooling. 

The number of downlink MIMO layers, especially for MU-MIMO, may increase; proposals mention up to 16 layers for single-user MIMO (SU-MIMO) and 64 for MU-MIMO.
This will have an impact on, for example, the demodulation reference signal (DM-RS) design to provide up to 48 orthogonal DM-RS ports.

\begin{figure}[t!]
   \centering
   \includegraphics[width=1.0\columnwidth]{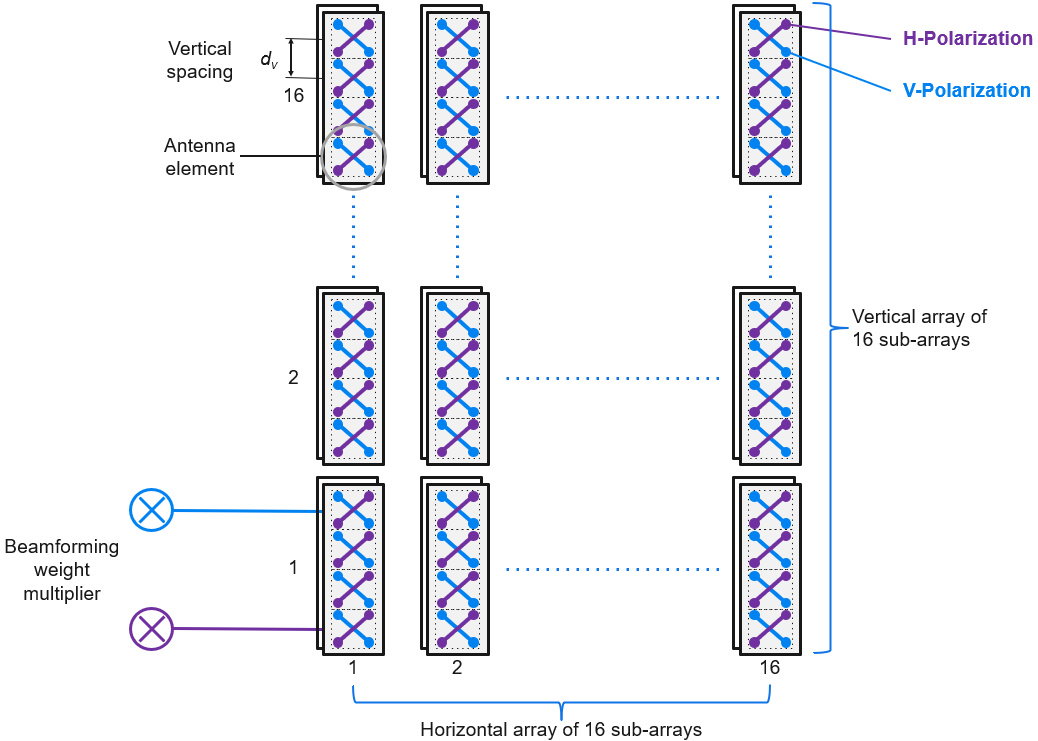}
\begingroup
\captionsetup{font=footnotesize}
\caption{Example antenna layout for a large 6G array. The array contains $16\times16$ subarrays, each comprising four vertically stacked dual-polarized antenna elements: 1024 dual-polarized elements, or 2048 polarization branches, in total. Blue and purple distinguish the two orthogonal polarizations. The vertical element spacing is denoted by $d_v$, and the circled multiplication symbols represent beamforming weights. Dotted lines indicate omitted subarrays between the illustrated rows and columns. With one channel state information (CSI) port and one power amplifier per subarray and polarization, this example has 512 CSI ports.}
    \label{fig:Antennas}
    \endgroup
    \vspace{-5pt}
\end{figure}

CSI acquisition is essential to virtually all high-performing MIMO schemes, and the CSI accuracy can often be the limiting factor \cite{marzetta2016fund}. Given the trend toward a large number of antenna elements, the number of CSI ports will also increase in 6G to 256. An example of a 6G antenna layout with 2048 antenna elements is given in Fig.~\ref{fig:Antennas}. The DM-RS has also been proposed, either as the main way of acquiring CSI or as a complement to CSI reference signal (CSI-RS)-based schemes \cite{6GWS-250071, 6GWS-250084}. This could reduce CSI-RS overhead for very large arrays.

Another aspect related to CSI reporting is the CSI codebook, that is, how to encode the channel conditions prior to reporting them to the 5G base station (gNB). In the first release of 5G, two types of codebooks (Type I and Type II) were specified, and later releases added enhancements to these \cite{Rel18mimo}. Type I assumes a beamforming matrix $\bf{W}=\bf{W}_1\bf{W}_2$, where $\bf{W}_1$ captures the long-term, frequency-independent properties of the channel and $\bf{W}_2$ the short-term, frequency-dependent properties; see \cite{Rel18mimo} for the definitions and computation of $\bf{W}_1$ and $\bf{W}_2$. The main intention with Type I is single-user MIMO with up to 8 spatial layers. 

Type II follows a similar structure of $\bf{W}$ but extends $\bf{W}_1$ such that multiple beams can be reported (up to two in Release 15; later releases support up to four), thus providing additional resolution in the spatial domain at the cost of relatively large CSI reports. In later releases, the Type II codebook has been enhanced by structuring $\bf{W}_2=\tilde{\bf{W}}_2\bf{W}^\text{H}_{f,k}$ to compress the frequency-domain information \cite{Rel18mimo}. However, Type II codebooks have seen very limited practical deployment. Defining a Type-II-like codebook as the baseline in 6G is useful \cite{6GWS-250163}. The work on AI-based CSI feedback in 5G will also be important to 6G; see Section~\ref{sec:aiai}.

SRSs are part of the 5G specifications and used for reciprocity-based schemes. However, the UE requirements  for SRS transmissions are very loose and limit the performance of reciprocity-based schemes with the existing devices. Tightening the 6G device requirements for SRS and improving SRS reception algorithms will be considered in 6G \cite{6GWS-250036, 6GWS-250153}.

\begin{table}[htbp] %
    \centering
    \caption{Summary of typical MIMO configurations (downlink/uplink) supported by the 5G specifications, typically deployed in commercial 5G networks at 3.5 GHz, and proposed for 6G (primarily for centimeter wave).}
    \scalebox{0.85}{\begin{tabular}{|l|l|l|l|}
    \hline
         &  \textbf{5G spec.} & \textbf{5G depl.} & \textbf{6G}%
        \tabularnewline\hline
        \midrule\hline
         \textbf{SU-MIMO layers} & 8/8 & 4/4 & 8/8 \\ 
         \textbf{MU-MIMO layers} & 48/48 & 16/4 & 48 \\ \hline
         \textbf{CSI ports per carrier} & 128 & 8--32 & 256 \\ \hline
             \textbf{UE antenna elements} &  2/4 Tx/Rx & 2/4 Tx/Rx &  4--8/8--16 Tx/Rx \\ \hline
         \textbf{gNB antenna elements} & 16--128  & 16--128 & 1024--4096  %
        \tabularnewline\hline
        \bottomrule
    \end{tabular}}
    \label{tab:mimo}
    \vspace{-5pt}
\end{table}

\subsubsection{Codebook Enhancements for 6G}

With the antenna numbers increasing in 6G (see Table~\ref{tab:mimo}), the number of CSI ports is also increasing, resulting in a significantly higher CSI overhead. CSI compression will also benefit from the use of AI in the RAN enhancements. Results in Fig.~\ref{fig:ai_csi} show a wide range of squared generalized cosine similarity (SGCS) gains, from 1.4\% to 21.4\%, between target CSI and reconstructed CSI. Enhancements to Type II codebooks defined for 5G can result in a spectral efficiency advantage of up to about 20\% \cite{6GWS-250068} relative to 5G. 
The design of enhancements to 5G codebooks is therefore important. These concepts and associated beam management are also discussed in
\cite{joon,dreifuerst,df1}.

Neural network principles are being studied in 6G for beam management. Reference \cite{zhang2020learningbeamcodebooksneural} proposes an artificial-neural-network-based framework for learning environment-aware beamforming codebooks for large-scale MIMO systems.

\subsubsection{Practical Challenges}
\begin{figure*}[t!] %
    \centering
    \includegraphics[width=17cm]{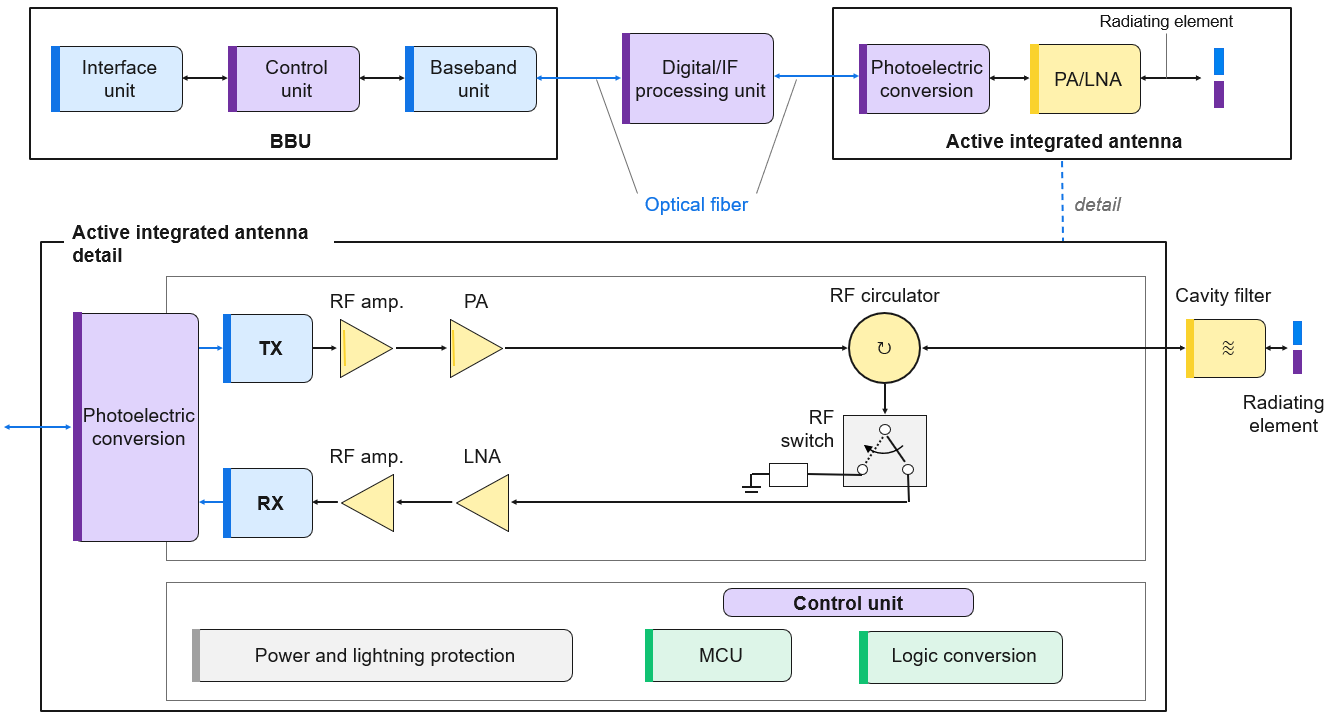}
    \begingroup
\captionsetup{font=footnotesize}
\caption{Typical base station transceiver architecture, adapted from \cite{ITURM2334,TATARIA1}. The upper diagram shows the baseband unit (BBU), digital/intermediate-frequency (IF) processing, optical-fiber connections, and active integrated antenna. The lower diagram expands the antenna-side transmit (Tx) and receive (Rx) paths, including radio-frequency (RF) amplifiers, the power amplifier (PA), low-noise amplifier (LNA), RF circulator, RF switch, and cavity filter. The supporting circuitry provides control, power and lightning protection, and logic conversion; MCU denotes microcontroller unit. Only one radiating element and its associated transceiver chain are illustrated for clarity.}
    \label{fig:AIRT}
    \endgroup
    \vspace{-10pt}
\end{figure*}

The architecture of a typical base station transceiver is depicted in Fig.~\ref{fig:AIRT}. Only one transceiver is shown for simplicity, but their number will equal the number of Tx/Rx modules in Table~\ref{tab:mimo}. The Tx and Rx perform mixing and de-mixing, cascaded amplifiers provide power gain, and bandpass filters are required to meet out-of-band emission requirements. While the operating bandwidths of some bands are quite large (several thousand MHz; see Table~\ref{tab:standards}), the carrier bandwidths are limited to a maximum of 400 MHz. This is because the noise floor rise caused by larger carrier bandwidths will result in SNR degradation and reduced coverage. However, with a large operating bandwidth, the radio performance at the lower and upper band edges is expected to be quite different from the band center unless all the RF circuitry is flat across the operating bandwidth, which in turn is a significant design challenge.

There is a good discussion on RF challenges with respect to frequency in \cite{tr38820}. Some examples follow. The noise figures (NFs) of both the base station and UE will rise with frequency. Contributions to the NF do not come from the low-noise amplifier (LNA) alone but also from the bandwidth, linearity, and dynamic range of the amplifier. All elements in the full RF receiver chain all the way up to the radiating elements will contribute to the overall receiver performance, including the switch (for TDD), analog-to-digital converters (ADCs), routing and filter losses, etc. It might be possible to reduce the noise contribution to the NF from the ADC by using more bits, but this would result in increased power consumption and heat dissipation, as a single added ADC bit would result in four times higher power consumption, something that is very relevant for UEs.

The power amplifier (PA) efficiency trend is shown in Figs.~\ref{fig:PAE1} and~\ref{fig:PAE2}, and saturated power levels are given in \cite{tr38820}. Silicon technologies are limited to a maximum peak power of 2 W. With gallium nitride (GaN) technology, the peak output power can be increased to a maximum of 20 W (at 24 GHz).

Antenna arrays in 5G and 6G base stations will be made of arrays of subarrays, where a subarray is an array of vertically stacked slant-polarized elements (see the antenna layout figures and the discussion in \cite{Rel18mimo}). Here, a PA is used to feed a subarray in each polarization. A PA driving a multi-element subarray must be capable of producing a higher level of RF power to drive the subarray. GaN and gate-all-around (GAA)-based technologies provide up to an order of magnitude higher power levels compared to silicon-based technologies. While higher PA output powers are desirable, the PA output power is also chosen to meet the adjacent channel leakage ratio (ACLR) requirements that must be met for out-of-band emission and coexistence conditions \cite{ITUr2070}. PA nonlinearity could result in spatially sensitive ACLR and cause the antenna array radiation to also fall in unintended directions \cite{mollen1,mollen2}.
Each transceiver chain also has a bandpass filter, as shown in Fig.~\ref{fig:AIRT}.
These filters have insertion losses, and increasing the number of Tx chains will therefore result in increased insertion losses. The above issues have to be kept in mind when using large antenna arrays.

\subsubsection {Has the MIMO Contribution to the Spectral Efficiency of the Physical Layer Reached Saturation?}

It might be thought, after decades of research and development, that physical-layer technology has come close to the theoretical limits. Certainly, it is widely agreed that phase-coherent D-MIMO with reciprocity-based CSI acquisition (which scales favorably with the number of service antennas) is the ultimate physical-layer solution. However, implementation efforts to date remain quite far from what is possible in the long run. The challenges of distributing backhaul and achieving phase-synchronous downlink operation are the main hurdles, in addition to obtaining additional sites. For these reasons, industry seems slow, or even reluctant, to adopt D-MIMO. 
Initial trials of D-MIMO (referred to as multiple transmission reception points (multi-TRP)),  as discussed in \cite{Rel18mimo}, indicated
 only marginal gains in spectral efficiency. However,  theoretical studies \cite{ganesan2020radioweaves,demir2021foundations}  as well as field measurements in relevant conditions  (e.g., \cite{xu2024spatial} indoors) demonstrate that the potential spectral efficiency of fully coherent D-MIMO with aggressive spatial  multiplexing of many terminals simultaneously is huge.
 
 In what ways could one improve cellular massive MIMO without going all the way to D-MIMO? While
 cellular MIMO is a very good technology, coverage holes and difficulties in sending multiple streams to multi-antenna users because of insufficient channel rank remain issues. One option is RIS, but RISs have large form factors, require copious amounts of training and control overhead, and lack band selectivity. In fact, the lack of band selectivity may be the single most important obstacle to the use of RIS in practical networks, as a non-selective RIS would cause coexistence problems between operators in neighboring bands. The deployment of RIS nodes poses logistical and operational difficulties that outweigh any benefits \cite{2024riscellularnetworks}.
 
 A different, radically new approach is to deploy swarms (large numbers) of small, inexpensive
  wireless repeaters (full-duplex relays) \cite{willhammar2025achieving}. Such repeaters can be made reciprocal \cite{larsson2024reciprocity} and therefore function as \emph{active scatterers} in the channel, like ordinary channel scatterers but with amplification. In a TDD MU-MIMO system, these repeaters would become completely transparent to the network and to the users.
  In \cite{willhammar2025achieving}, it is demonstrated that while swarm-repeater-assisted MIMO in TDD cannot achieve the same performance as that of D-MIMO, it can come quite close. This is possible without any backhauling or phase synchronization; all that is required is a low-rate control channel to the repeaters to configure their gain, to align them with the TDD pattern, and to enable reciprocity calibration.
  
  Another direction for the physical layer is the integration of \emph{computing and communication}. For example, simply by virtue of the superposition principle for wave propagation, addition (and in fact, any mathematical operation representable as a nomographic function) can take place over the air (OTA) by several devices transmitting simultaneously in the same time-frequency resource \cite{chen2023over,csahin2023survey}.
  In principle, this renders the computation completely scalable with respect to the number of devices involved.
Such OTA computation can be accomplished through either coherent or non-coherent signal combination. A challenge with coherent combination is phase noise (simultaneously transmitting devices must be aligned in phase) \cite{dahl2024over}. Non-coherent combination does not require such phase alignment, and works even in fully decentralized (multipoint device-to-device) settings \cite{michelusi2024non}. 
The integration of both coherent and non-coherent OTA computation into future systems remains an open field. It is unlikely to become a part of 6G, at least initially, but it is a good topic for basic research with relevance for future generations of systems, perhaps beyond 6G.

\vspace{-0.3cm}
\subsection{AI/ML-Native Wireless Network}
\label{subsec:ai}

Recent breakthroughs in AI/ML have demonstrated remarkable success in domains such as image recognition and natural language processing. This success has motivated researchers and developers to explore their applicability to wireless communications, including 6G RAN \cite{o2017introduction, chen2019artificial, letaief2019roadmap}.

\subsubsection{Why and What Is AI-Native Wireless?}

Numerous research studies and industry initiatives have demonstrated that AI/ML presents a powerful tool for handling complex problems in wireless networks \cite{zappone2019wireless, zhang2019deep}. One key role of AI/ML is to tackle hard-to-model problems by determining appropriate representations for complex wireless phenomena that are difficult to capture with traditional models. This includes addressing nonlinear effects, interference, and the environmental variability intrinsic to real-world wireless channels \cite{zappone2019wireless, simeone2018very}. By modeling these nonlinear functions more accurately than conventional linear approximations, AI/ML can provide an adaptable framework that better reflects the actual behavior of wireless environments.

In addition to improved modeling, AI/ML methods excel in finding near-optimal solutions that are computationally feasible. Many optimization problems in wireless communications are computationally intractable when seeking optimal solutions using classical approaches. AI/ML algorithms, however, can approximate these optimal solutions efficiently, thereby overcoming the limitations imposed by the computational infeasibility of traditional methods \cite{zappone2019wireless, simeone2018very}. This capability is particularly beneficial for tasks such as parameter optimization, where AI-enhanced methods can continuously and intelligently adjust parameters to improve system performance without being constrained by rigid, predefined configurations.

The vision of AI-native wireless marks a shift away from the traditional ``one-size-fits-all'' approach toward AI-driven, site-specific optimization. Core AI/ML engines are first trained offline in controlled lab environments to develop robust baseline models. Once deployed, these models enter an on-site adaptation phase in which models continuously fine-tune parameters in real time, responding to local propagation conditions, user behaviors, and interference dynamics. This dual-stage AI lifecycle enables each cell site to evolve autonomously, delivering tailored performance improvements in spectral efficiency, energy use, and user experience.

When evaluating AI/ML solutions for 6G, it is crucial to benchmark them against state-of-the-art classical algorithms rather than against overly simplistic textbook examples. This ensures that any claimed gains truly reflect advances over what is already deployed in high-performance systems, rather than improvements that only appear large when compared to naive baselines. By rigorously comparing AI-driven approaches to the mature, optimized methods, we can more accurately assess where AI/ML offers genuine advantages in terms of spectral efficiency, robustness, complexity, and energy efficiency.

Overall, the role of AI/ML in 6G wireless is to provide innovative solutions where traditional techniques fall short \cite{letaief2019roadmap, yang2020artificial}. ``AI native'' is a trending concept that has been extensively used in the 6G discussions. While there is no universally agreed-upon formal definition of ``AI native'' in wireless communications, the term broadly refers to a design and operational paradigm in which AI/ML is integrated across every layer of the system \cite{itu2022future}. In particular, AI/ML functionalities are envisioned to be present from the very start of 6G. The implication is that, from initial development, the network leverages AI/ML algorithms not only for specific optimization tasks but also as a core design principle in managing resources, orchestrating network behaviors, and delivering services. This early integration ensures that AI/ML is a fundamental building block rather than an afterthought \cite{lin2023embracing, soldati2023approaching}.

Standardization flexibility further characterizes an AI-native approach. Rather than adhering to strict standards that specify fixed parameters for network operations, this approach may define flexible guidelines that enable adaptive behaviors, context-aware configurations, and continuous improvements driven by real-world data \cite{hoydis2021toward}. This flexibility allows networks to dynamically tailor their performance characteristics, ensuring optimal user experiences and efficient spectrum utilization without being overly constrained by static protocols.

\begin{figure}[t!] %
\centering
\includegraphics[width=1.0\columnwidth]{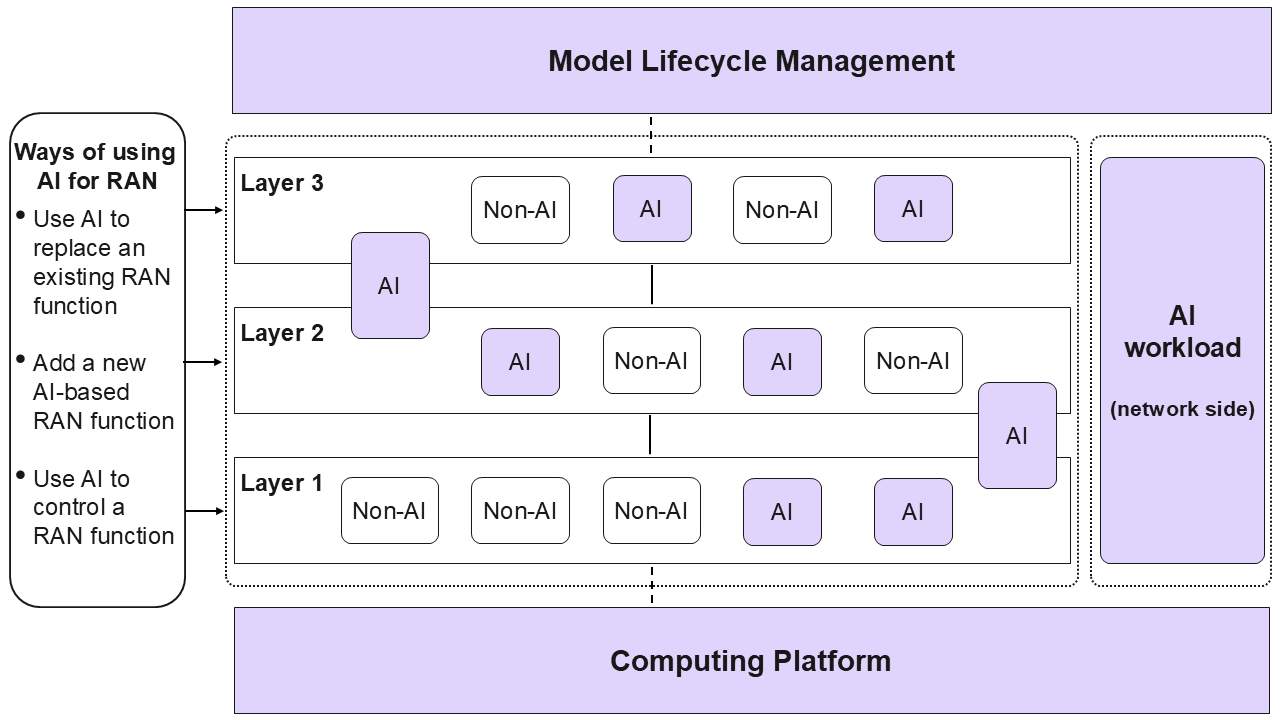}
\begingroup
\captionsetup{font=footnotesize}
\caption{Conceptual framework for an artificial-intelligence-native (AI-native) 6G radio access network (RAN). Layers~1--3 contain AI-based functions (purple blocks) alongside conventional functions (white blocks). AI can replace an existing RAN function, introduce a new function, or control an existing function; blocks spanning layers illustrate possible cross-layer processing. Model lifecycle management covers model development, deployment, performance monitoring, and updating. The computing platform supports the radio processing functions and can also host network-side AI workloads beyond radio protocol processing, shown on the right. The allocation of functions is illustrative.}
\label{fig:ai-framework}
\endgroup
\vspace{-8pt}
\end{figure}

In summary, while ``AI native'' remains an evolving and somewhat loosely defined concept, its core idea centers on the seamless, intrinsic integration of AI technologies into the network's architecture and operations from the outset. In Fig.~\ref{fig:ai-framework}, we present a pragmatic conceptual framework for AI-native 6G RAN. The radio protocol stack consists of three layers, each of which may contain AI-based modules, conventional (non-AI) modules, or a combination of both. Fig.~\ref{fig:ai-framework} illustrates that AI can be utilized in one of three ways: (1) as a replacement for an existing RAN function, (2) as a newly introduced AI-based function, or (3) as a control mechanism over an existing RAN function \cite{ericsson2023defining}. Above the radio protocol layers is a model lifecycle management block responsible for developing, deploying, and updating AI models used in the network, covering everything from initial training to ongoing retraining and performance monitoring. At the bottom is the computing platform, providing the underlying hardware resources (e.g., central processing units (CPUs), graphics processing units (GPUs), and specialized accelerators) necessary to run the computing workloads efficiently. On the right side, an AI workload area indicates that RAN-independent AI workloads may be processed within the network infrastructure, reflecting the trend toward converged compute-communication platforms wherein AI workloads can run alongside RAN workloads. In short, Fig.~\ref{fig:ai-framework} shows that we can incrementally integrate AI where it offers the most benefit while maintaining compatibility with conventional RAN functions. In this way, Fig.~\ref{fig:ai-framework} captures the essence of a hybrid approach to designing 6G RAN, one where AI and classical techniques work together. In the following subsections, we discuss more concrete design aspects to better understand the conceptual framework illustrated in Fig.~\ref{fig:ai-framework}.

\subsubsection{AI-Native Air Interface}
\label{sec:aiai}

An AI-native air interface represents a radical departure from conventional radio-interface design \cite{hoydis2021toward}. Traditionally, air-interface design has relied on predetermined pilot signals, fixed modulation and coding schemes, and standardized feedback protocols for CSI. In contrast, an AI-native design leverages data-driven models that learn and adapt to the underlying channel and network conditions \cite{dorner2017deep}. The integration of AI/ML techniques in the 6G air interface will span all layers of the radio protocol stack, from the physical layer through the medium access control (MAC) layer and up to the radio resource control (RRC) layer \cite{itu2022future}. At the physical layer, AI/ML is primarily used to optimize signal processing functions that were traditionally addressed with fixed algorithms \cite{o2017introduction, qin2019deep}. In 5G-Advanced, AI/ML techniques have been explored to optimize the physical layer by enhancing CSI feedback, predicting beam directions, and improving positioning accuracy \cite{lin2022overview}:
\vspace{-0.03cm}
\begin{itemize}
\item \textbf{CSI feedback}: AI/ML is applied to CSI feedback through two distinct cases. The first case involves CSI compression, where the UE employs an AI-based encoder to compress the CSI into a more compact form \cite{guo2022overview}. The gNB then uses a corresponding AI-based decoder to reconstruct the CSI, thereby reducing the feedback payload while preserving the CSI accuracy. The second case focuses on time-domain CSI prediction with a UE-sided model. In this approach, the UE uses historical CSI measurements to predict future channel conditions, effectively mitigating the adverse effects of channel aging that are particularly significant in high-mobility scenarios \cite{luo2018channel}.
\item \textbf{Beam management}: Traditional beam management techniques typically involve exhaustive measurements across a large set of beams, resulting in significant overhead and latency. AI-based approaches streamline this process by predicting the optimal beam selection using reduced measurements \cite{alrabeiah2020deep, khan2023machine}. In the spatial-domain downlink beam prediction use case, the system leverages a designated subset of beam measurements to determine the best current beam, thereby reducing the need for exhaustive scanning. In the time-domain variant, the model forecasts the best beam for future time instances based on past measurements. These predictive methods not only reduce signaling overhead but also improve the speed and accuracy of beam selection, which is particularly beneficial in the challenging propagation environments encountered in mmWave systems.
\item \textbf{Positioning}: Accurate positioning is vital for a range of applications, from location-based services to industrial automation. AI/ML has been applied to enhance positioning accuracy in two key directions. Direct AI-based positioning employs AI/ML models such as convolutional neural networks that take inputs like channel impulse responses or power delay profiles to directly estimate a UE's location \cite{soltani2020more}. Alternatively, AI-assisted positioning uses AI/ML models to generate intermediate statistics, such as probabilities of line-of-sight conditions, angle-of-arrival estimates, or time-of-arrival data, which are then used to estimate the UE's position \cite{liu2023machine}. These AI-enabled approaches have demonstrated improvements, especially in challenging environments where traditional geometry-based methods struggle due to heavy non-line-of-sight conditions or dense multipath propagation.
\end{itemize}

In 3GPP Release 18, extensive evaluation results presented by many industry players in 3GPP demonstrated tangible benefits from AI-based beam management and AI-based positioning \cite{3gpp2024aitr}. 
In Release 19, 3GPP specified the necessary signaling and mechanisms to enable AI-based beam management and AI-based positioning \cite{lin2025bridge}. For time-domain CSI prediction, Release 18 yielded promising concepts but lacked comprehensive comparisons against classical, non-AI methods and raised questions about computational cost. Release 19 therefore extended the investigations, further quantifying performance gains and evaluating the complexity tradeoff \cite{3gpp2024aitr}. Following the extended study, 3GPP specified the necessary signaling and mechanisms to enable AI-based CSI prediction in Release 19. 

\begin{figure}[t!] %
\centering
\includegraphics[width=1.0\columnwidth]{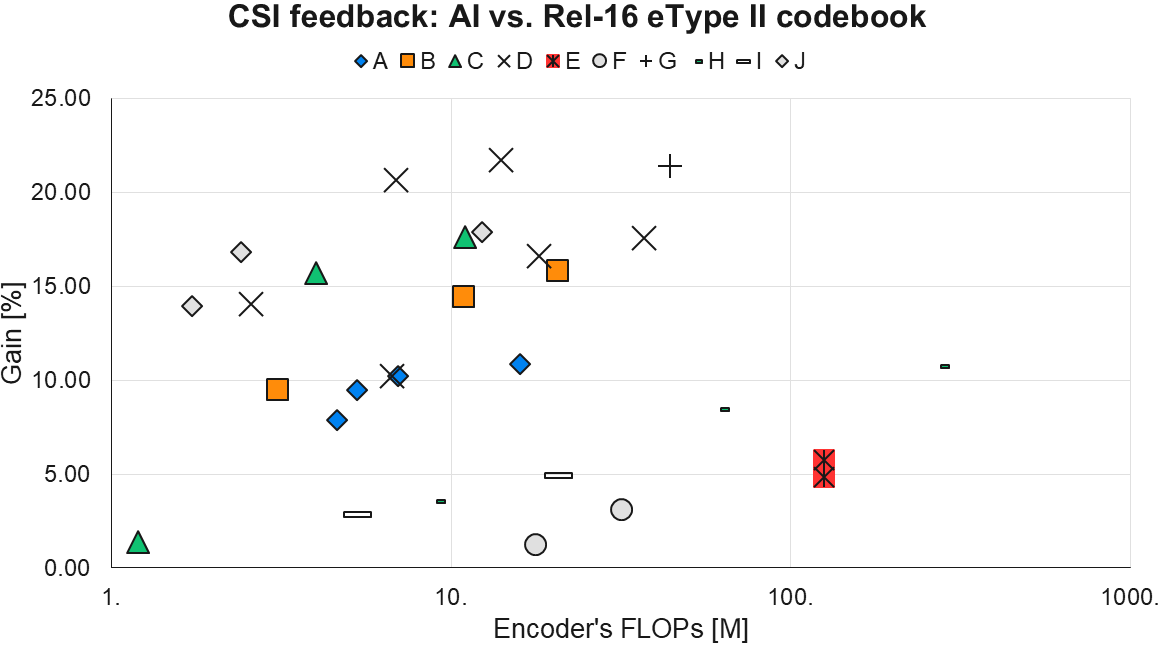}
\begingroup
\captionsetup{font=footnotesize}
\caption{Performance--complexity comparison of artificial-intelligence (AI)-based channel state information (CSI) compression with the 3rd Generation Partnership Project (3GPP) Release 16 enhanced Type II (eType II) codebook, using reported 3GPP evaluation results \cite{3gpp2024aitr}. The horizontal axis shows encoder complexity in millions of floating-point operations (FLOPs) on a logarithmic scale. The vertical axis shows the reported gain (\%) in squared generalized cosine similarity (SGCS) between target and reconstructed CSI for the first spatial layer. Labels A--J identify the ten reporting sources, distinguished by marker styles; each point represents a reported evaluation result. Higher SGCS gain indicates improved CSI reconstruction accuracy.}
\label{fig:ai_csi}
\endgroup
\vspace{-8pt}
\end{figure}

3GPP Release 19 also addressed challenges identified during the Release 18 study on CSI compression. Compressing CSI efficiently requires a two-sided model, with an encoder on the UE and a decoder on the gNB, which introduces significant complexities around inter-vendor collaboration during training. Moreover, the modest performance gains observed during Release 18 fell short of justifying the added complexity of a two-sided approach. Accordingly, 3GPP continued to study CSI compression in Release 19, striving to strike a better balance between performance improvements and complexity. Fig.~\ref{fig:ai_csi} summarizes the evaluation results from ten sources during the 3GPP Release 19 study on CSI compression \cite{3gpp2024aitr}. In this study, AI-based CSI compression was compared with the benchmark 3GPP Release 16 eType II codebook. The performance metric is SGCS between target CSI and reconstructed CSI. It can be seen from Fig.~\ref{fig:ai_csi} that the results vary across the sources, with the reported gains ranging from $1.4\%$ to $21.4\%$. Nonetheless, the results show that most of the SGCS gain is achievable using an encoder model with complexity less than 10 million floating-point operations (FLOPs). In Release 20, 3GPP aims to specify support for spatial-frequency CSI compression \cite{lin2025tale}.

The insights from 5G-Advanced serve as a foundation for more advanced AI/ML functionalities in the 6G physical layer. One potential direction is the adoption of neural receivers \cite{honkala2021deeprx}. Neural receiver architectures, which integrate neural networks at various levels of the receiver signal processing chain, have shown potential to overcome limitations of conventional methods by learning to approximate or jointly optimize multiple blocks \cite{cammerer2023neural}. In a wide sense, neural receivers encompass any design in which one or more processing blocks are enhanced or replaced by neural-network models.
AI/ML receiver design (such as neural receivers) may provide a promising path toward reducing the amount of reference-signal overhead \cite{ait2021end}. In 5G-Advanced, reference signals remain essential for channel estimation and demodulation, but they consume valuable time-frequency resources. With AI/ML-based receiver processing, the system may better exploit data-aided inference, channel correlation, and learned channel representations to maintain reliable performance with fewer reference signals. Offline learning may also support transmitter-side optimization, such as constellation shaping under representative channel and hardware conditions. However, given the standardization preference for conventional QAM-based modulation and the need for interoperability, fully learned constellations or strictly pilotless transmission are unlikely to be baseline 6G features. A more practical direction is to use AI/ML to enhance channel estimation and demodulation while reducing reference-signal density. Fully online end-to-end learning, where transmitter and receiver models continuously adapt in real time, is also unlikely in the initial 6G specification because it would require substantial standardization support for protocol extensions, conformance testing, and interoperability, among others. In short, AI/ML can make the 6G physical layer more adaptive to changing channel environments, hardware impairments, and service requirements without radically departing from established frameworks. 

More broadly, 3GPP has started to study a wide range of AI/ML use cases for 6G RAN, covering both air-interface enhancement and network-level optimization. At the physical layer, AI/ML can support reference-signal overhead reduction, AI/ML-based CSI acquisition, prediction, and compression, as well as more efficient beam management with reduced measurement and sweeping overhead. AI/ML can also be applied to predict radio measurements and measurement events, enabling the network and UE to anticipate link-quality changes before conventional threshold-based triggers are reached. Beyond the air interface, AI-based traffic prediction can support proactive scheduling, resource allocation, and congestion mitigation, while AI/ML-assisted mobility optimization can improve handover robustness, reduce radio link failures, and minimize unnecessary measurements. Network energy saving is another important use case, where AI/ML can help identify opportunities for carrier, cell, or antenna deactivation while maintaining service continuity. In addition, AI/ML can address implementation-oriented impairments, such as digital post-distortion for transmitter nonlinearity compensation and SRS power imbalance compensation for improved uplink channel sounding. These use cases suggest that AI/ML in 6G is likely to be introduced first as targeted enhancements to specific RAN functions, rather than as a full replacement of conventional protocol and signal-processing designs.

\subsubsection{Generative AI and LLM-Enabled RAN Automation}
\label{subsec:llm_ran_automation}

The preceding discussion focused on task-specific AI/ML models
embedded in the RAN air interface. Generative AI can further assist
physical-layer tasks such as synthetic data generation~\cite{Huynh2024GenAIPhysicalLayer}.
Foundation models, particularly large language models (LLMs), offer
a complementary capability for automating RAN management and
operations by providing a natural-language interface among operator
objectives, RAN data, and management functions. Potential applications include querying RAN specifications;
summarizing alarms and performance reports; correlating RAN
counters, traces, and configuration data; assisting root-cause
analysis; and generating candidate configurations or optimization
workflows~\cite{Zhou2025LLMTelecom,Long2025LLMNetOps}. For example, an LLM-based assistant could
translate an operator objective into candidate policies for
mobility management, load balancing, interference mitigation,
energy saving, or fault recovery. Such outputs should be treated
as recommendations or inputs to validated RAN control workflows,
rather than as direct and unconstrained commands to the RAN.

Realizing these benefits requires adaptation and grounding in the
wireless domain. Retrieval-augmented generation can ground an LLM
in current RAN specifications, operator policies, network topology,
configuration information, and recent telemetry. Tool use can
allow the model to invoke authorized RAN analytics, simulators,
policy engines, and management functions. Domain-specific
fine-tuning and knowledge alignment may further improve its
understanding of wireless terminology, protocols, and optimization
objectives.

At the standards level, 3GPP specifies intent-driven management
capabilities, including intent translation, feasibility evaluation,
conflict resolution, fulfillment reporting, and closed-loop
operation~\cite{3GPP_TS_28_312}.
These mechanisms can support the expression and supervision of RAN
objectives. From an industrial viewpoint, the most credible near-term role of LLMs is therefore in non-real-time and supervisory RAN management loops, including planning, policy generation, network assurance,
and optimization recommendations. Their outputs should be grounded
in up-to-date specifications and RAN data, restricted to authorized
interfaces, and subject to deterministic validation, auditability,
rollback, and human approval where appropriate.

\vspace{-0.3cm}
\subsection{Ubiquitous Coverage}
\label{subsec:ubiquitous}
Fundamentally, cellular networks address two very basic communication needs: mobility and coverage. Without coverage, the user will not benefit from any of the 6G features. Providing the required coverage is a multi-dimensional problem, involving standardization, implementation, and deployment aspects. Densifying the network (small cells, D-MIMO, etc.) is technically a nice solution but comes at a cost for the network operator, and any means to improve coverage using the existing site grid is therefore the first choice. Using large antenna arrays, as discussed in Section~\ref{sec:mimo}, is one example, and complementing the terrestrial network (TN) with satellites is another, elaborated upon in Section~\ref{subsec:ntn}.

Traditionally, mobile broadband applications have been downlink-heavy, with users consuming video content, browsing the web, and similar activities, and only to a limited degree has the uplink been in focus. However, lately the importance of uplink performance has been more and more pronounced, a change that is expected to continue in the future%
. 
There are several reasons for this increase. For example, XR applications often require an uplink video stream from the user to the cloud. The emergence of generative AI and AI agents in the handsets is another example likely to further spur an increase in uplink (UL) traffic.  

Improving the UL coverage is a challenging task. Unlike the downlink (DL), which often can be bandwidth-limited given the significantly higher transmit power, the UL is typically power-limited, and additional bandwidth may not help. Increasing the UL transmit power is often not possible, given regulatory restrictions and the fact that the UE normally is battery powered. Some improvement can be obtained by tightening the requirements defined in 3GPP RAN4 \cite{TS38101-5,TS38104}, for example, reducing the allowed back-off in the device's PA and providing the network with additional information on the applied back-offs to improve the scheduling decisions. Various forms of spectrum shaping for DFT-s-OFDM may also help, as benchmarked against other UL coverage improvement alternatives for 6G in \cite{6GWS-250004}.

One possibility that may provide large gains is so-called UL-DL decoupling. Traditionally, cellular devices camp on the best DL frequency, and the UL implicitly uses the same frequency (band), regardless of the properties of the UL frequency. However, in many situations a device in a challenging position would be better served by a UL in a lower frequency band with lower path loss (e.g., 700 MHz), with DL transmissions using a higher frequency band with wider bandwidths (e.g., 3.5 GHz).

Decoupling UL and DL is not a new idea. In \cite{Boccardi:16}, decoupling in the spatial domain in a heterogeneous network scenario is discussed. Decoupling in the frequency domain can, to some extent, be realized in later releases of the 5G standard using the carrier aggregation framework and the UL Tx switching feature \cite{Dahlman:23}. However, configuring and activating carrier aggregation takes a non-negligible amount of time in 5G. Given the typically bursty traffic mentioned in Section~\ref{sec:KPI}, the traffic might be gone before the low-band UL carrier is active. Any UL-DL decoupling mechanism in 6G therefore needs to be sufficiently fast to match the traffic characteristics.

\vspace{-0.3cm}
\subsection{Energy Efficiency}
\label{subsec:ee}
Achieving carbon neutrality is a desirable goal worldwide. The RAN is the largest consumer of energy in the network of a communications service provider. The RU contributes about 40\% of the RAN energy consumption \cite{energy}. Massive MIMO arrays use beamforming techniques, and here antenna-in-package solutions are used to realize the antenna and beamforming logic in the RU \cite{sadhu1}. Massive MIMO systems are also expected to meet high spectral efficiency and energy efficiency in 6G. This architecture is adopted even for mid-bands and becomes attractive for larger arrays. However, with increased carrier frequencies for 6G, multiplied antenna counts (see Table~\ref{tab:mimo}), and power consumed by circuitry (due to embedded signal processing, baseband processing, digital-to-analog converters, filters, etc.), developing a holistic view of energy consumption beyond just Tx power is highly desirable \cite{hoydis1,hoydis2,murdoch}. Just as is the case with airlines today, it is very likely that customers will be asking for the carbon contribution of their wireless sessions. Additionally, network management decisions (configuration, optimization) may also use energy consumption as a metric. 
For 6G to be energy efficient, the following goals are desirable:
\begin{itemize}
\item \textbf{Scaling the energy consumption with load}. The energy consumption must scale with the load. Ideally, no energy should be consumed at zero load. To achieve this, the ultra-lean design principle introduced in 5G should be further enhanced, minimizing the transmission of always-on signals. Not only should the ultra-lean principle be applied in the time domain, but the frequency and spatial domains should also be exploited. Multiple frequency bands are typically used, and by turning off some of the frequency bands at low load, additional energy can be saved. In Fig.~\ref{fig:nwee}, the base station energy consumption as a function of the SSB periodicity is plotted for an unloaded network (i.e., no user traffic) for the 3GPP power model in \cite{TR38864}. The different colors in the plot correspond to different parts in the base station. By increasing the SSB periodicity from the 20~ms used in 5G to 160~ms in 6G, it is possible to exploit deep-sleep states in the radio, which has the potential to reduce the base station power consumption by 77\%. Note that unloaded, or lightly loaded, sites are fairly common, e.g., during nighttime in an office area. Only a small number of sites operate at a very high traffic load.
\begin{figure}[t!] %
    \includegraphics[width=0.95\columnwidth]{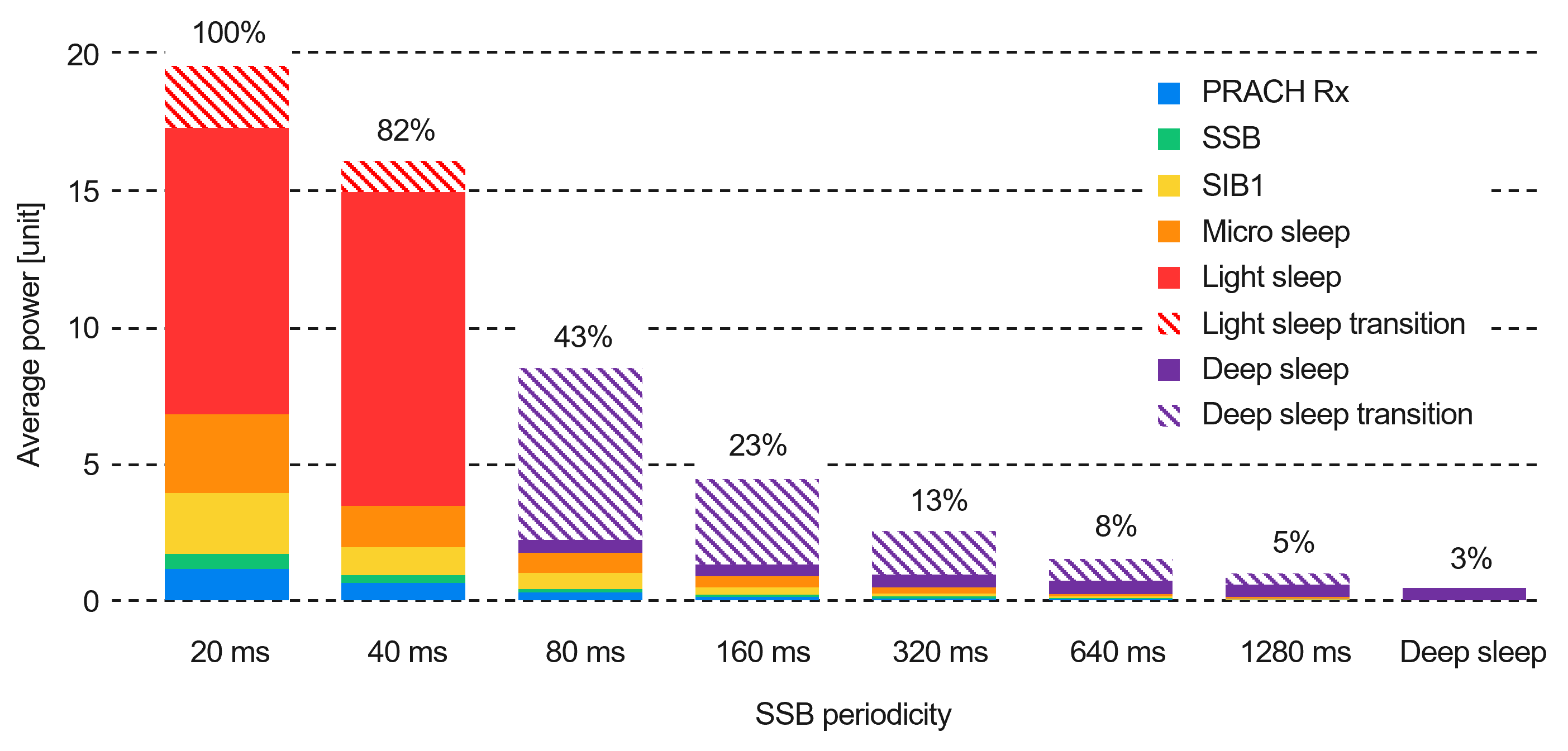}
    \begingroup
\captionsetup{font=footnotesize}
    \caption{Modeled average base station power for an unloaded network, with no user traffic, as a function of synchronization signal block (SSB) periodicity. The power model is from \cite{TR38864}, with SSB operation modeled as in 5G. Stacked components show physical random access channel (PRACH) reception, SSB and system information block 1 (SIB1) transmission, sleep states, and transitions between sleep states. The vertical axis uses the power model's units; percentages give total power relative to the 20-ms case. Increasing the periodicity to 160 ms reduces modeled power to 23\% of this baseline. The final bar shows the deep-sleep reference.}
    \label{fig:nwee}
    \endgroup
    \vspace{-10pt}
\end{figure}
\item \textbf{Requirement management} is an important aspect to consider. Modern cellular networks in general, and 6G in particular, are capable of providing very high performance in terms of data rates, latency, capacity, etc. This is visible in the KPIs typically measured and tracked, both internally within an operator and externally by third-party companies. However, not all services may require the most extreme capabilities. Paying a price in terms of energy when extreme performance is required is acceptable, but the same ``energy price'' should not apply to less demanding services.
\item \textbf{Deactivating unused Tx chains}. Reductions of transmit power may also be obtained by deactivating some active PAs \cite{deactivate} when not in use. To maintain the QoS under different traffic loads, \cite{deactivate} proposes deactivating parts of the antenna array during low load while still maintaining a high throughput, thereby achieving higher energy efficiency. This is an example of the requirement management in the previous bullet. Using detailed energy models, 
the optimal number of Tx antennas may be reduced when optimizing with respect to energy efficiency. Transmit antenna selection can also be used to improve energy efficiency \cite{zujun,liu2015massivemimoenergyefficient}.
\item \textbf{Improving PA efficiency}. PA efficiency decreases with operating frequency; see Figs.~\ref{fig:PAE1} and \ref{fig:PAE2} \cite{wang_power_amplifiers_survey}.
\begin{figure}[t!] %
    \centering
    \vspace{-5pt}
    \includegraphics[width=1.0\columnwidth]{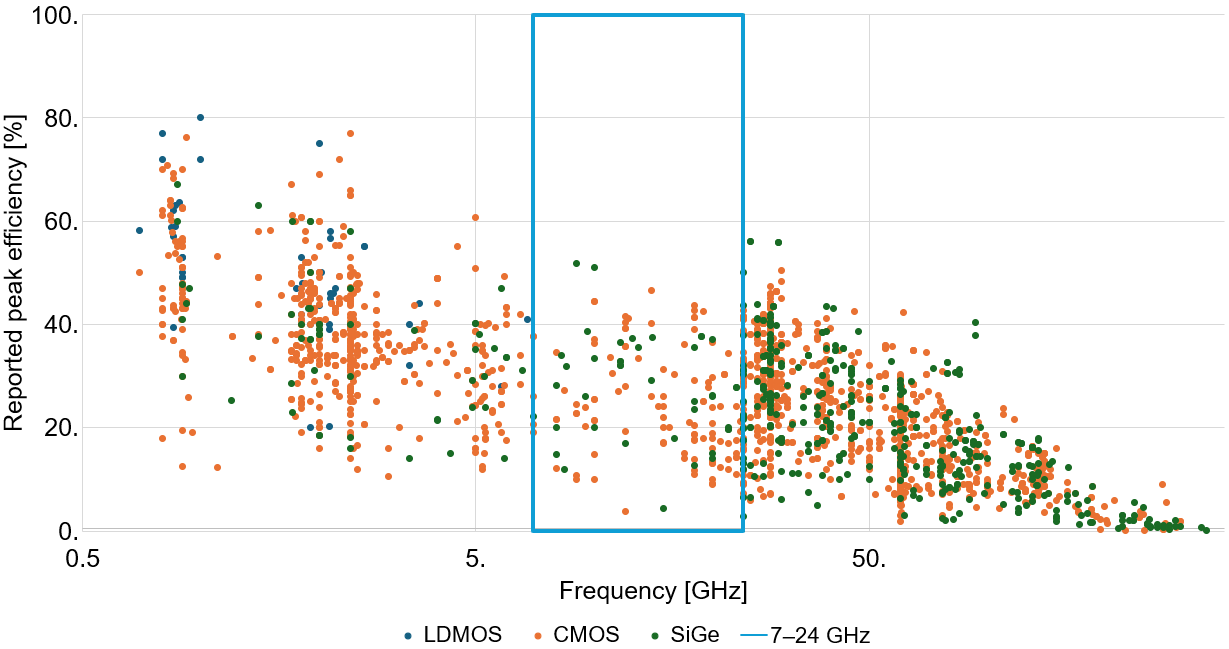}
    \begingroup
\captionsetup{font=footnotesize}
\caption{Reported peak efficiency of silicon-based power amplifiers (PAs) versus operating frequency, using the 2026 update of the ETH PA Survey \cite{wang_power_amplifiers_survey}. Each point represents a reported measurement entry. Colors distinguish laterally diffused metal-oxide-semiconductor (LDMOS), complementary metal-oxide-semiconductor (CMOS), and silicon-germanium (SiGe) technologies. Frequency is plotted on a logarithmic scale and efficiency in percent. The outlined box highlights 7--24~GHz. The scatter reflects differences in PA designs, output powers, and measurement conditions, illustrating the implementation variation relevant to radio-unit energy performance.}  
    \label{fig:PAE1}
    \endgroup
    \vspace{-10pt}
\end{figure}

\begin{figure}[t!] %
    \centering
    \vspace{-5pt}
    \includegraphics[width=1.0\columnwidth]{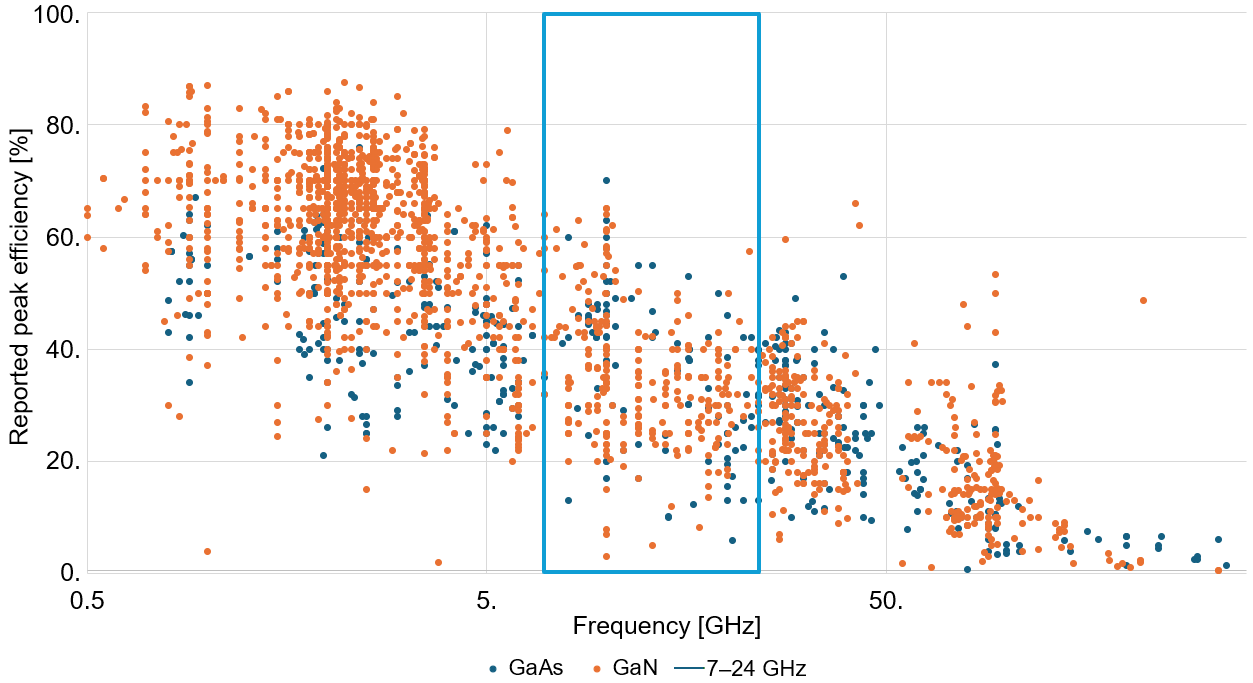}
    \begingroup
\captionsetup{font=footnotesize}
    \caption{Reported peak efficiency of compound-semiconductor power amplifiers (PAs) versus operating frequency, using the 2026 update of the ETH PA Survey \cite{wang_power_amplifiers_survey}. Each point represents a reported measurement entry, with colors distinguishing gallium arsenide (GaAs) and gallium nitride (GaN) technologies. Frequency is plotted on a logarithmic scale and efficiency in percent. The outlined box highlights 7--24~GHz. Results span different PA designs, output powers, and measurement conditions; the generally lower efficiencies at higher frequencies illustrate a challenge for limiting radio-unit power consumption.}
    \label{fig:PAE2}
    \endgroup
    \vspace{-10pt}
\end{figure}
PA architectures in commercial base stations are based on class A, class AB, Doherty, etc. Fig.~\ref{fig:PAE1} shows a scatter diagram of the PA efficiency (PAE) as a function of operating frequency for PAs made using silicon transistors (i.e., LDMOS, CMOS, and SiGe), and in Fig.~\ref{fig:PAE2} the same parameters are plotted for compound semiconductor transistors (i.e., GaAs and GaN). We note that PAE is mainly dependent on the operating frequency and not on the transistor technology. The wide spread of data is mainly due to different power levels and different amplifier architectures. Thus, one way to reduce the PA power consumption is to use advanced PAs that provide high PAE. Doherty PAs with digital predistortion achieve efficiencies in the range of 45\% at full load. Improvements to this efficiency are suggested in \cite{cune}, but the polar techniques presented there may not be applicable here. A promising approach to improve PA efficiency is to operate PAs in the nonlinear regime by limiting the back-off and using predistortion to compensate for error vector magnitude (EVM) increase due to nonlinearity.
\item \textbf{Network energy efficiency with multiple access}.
NTNs are able to cover a wide region with few satellites (e.g., GEO) that are energy-free once in orbit, but the delivered throughput is limited compared to TNs. Assuming that 6G seamlessly integrates both TN and NTN access networks, the usage of one access or another can be selected according to a global energy consumption minimization objective to provide a service \cite{22870}. Taking advantage of the different access network characteristics, their energy consumption models, and the type of service (e.g., public warning system (PWS), broadcast, messaging), the operator can turn off or limit the coverage of TN in a given zone while continuing to offer a service to the users. Data collection for energy consumption, traffic models, and an orchestration framework are essential to achieve this goal.
\end{itemize}
Improving energy consumption is also important in devices, as battery life is of paramount importance at the terminal end. Reference \cite{lozano} presents an information-theoretic approach to analyze the power consumption in a terminal's receiver radio-frequency front end (RFE). It shows that the RFE power consumption might overshadow the Tx power and that lowering it is necessary for various device types, especially smart wearables, virtual reality goggles, etc. It provides a useful discussion of tradeoffs between spectral efficiency and energy efficiency. Increasing RFE power may be worthwhile if more bits get through. Conversely, decreasing RFE power may be useful to conserve power, but the bit rate may drop. Therefore, the interplay between the two must be considered via a spectral vs. energy efficiency assessment framework.

The 6G-specific aspects of the device power consumption are in many cases features that have already been investigated, for example, in 5G-Advanced work \cite{Holma_2}, but in most cases have not been deployed, as they would have required big changes in the network-side or device-side implementations. In some cases, the features could not be used together with the earlier deployed 5G devices on the same frequency. It is therefore important to address the device energy consumption in the first 6G release with a fresh look at both features improving UE power consumption and battery life and features allowing the network to reduce the energy consumption. 

\vspace{-0.3cm}
\subsection{Non-Terrestrial Networks}
\label{subsec:ntn}

\subsubsection{Background on NTN}
\label{subsubsec:5gntn}

The integration of NTN into the terrestrial network standard started in 3GPP in June 2017 with the studies on NR support for NTN in Release 15~\cite{TR38811} and Release 16~\cite{TR38821}. The first support of NTN was integrated as part of Release 17 (2022) \cite{lin20215g} with the introduction of narrowband, wideband, and broadband communication to the 5G NR and NB-IoT technologies. %
The 5G NR satellite system architecture is composed of a space and a ground segment, with gNB (base station) functionalities on the ground in the case of a transparent architecture or on board the satellite in the case of a regenerative architecture \cite{TS38300}. For both types of configurations, the NTN control function controls the radio resources of the satellite access node (SAN) via operation and maintenance (O\&M) and provides key information about the space segment, such as the satellite ephemeris information.
The 3GPP NTN technology may be deployed for high-altitude platform stations (HAPS) (typically at $17$--$22$ km altitude), low Earth orbit (LEO) (below $2000$ km), medium Earth orbit (MEO) (orbit around $8000$ km), and geosynchronous orbit (GSO) or geostationary Earth orbit (GEO) ($35786$ km). 3GPP’s broader NTN description therefore includes HAPS, while ITU treats HAPS operating as IMT base stations (HIBS) and part of terrestrial IMT networks. Therefore, HAPS frequency bands were finally integrated as part of 3GPP TN specifications. %
For the satellites, there are those that are moving above a geographical area on Earth (non-GSO) and those that are not moving from the Earth's surface standpoint (GSO). Thus, the cells that are created by satellites are of three types: Earth-fixed, quasi-Earth-fixed, or Earth-moving, with different characteristics regarding mobility procedures, satellite radio channel dynamics, and satellite orbit types.
The frequent cell switching, due to the high speed of non-GSO satellites (greater than $7$ km per second in LEO), leads to frequent handover of all devices between the cells and the gNBs. Location- and time-based solutions have been introduced using the predictability of the satellite orbit. In addition, handover without a random access channel (RACH) procedure (i.e., RACH-less handover) and conditional handover (CHO) ensure smooth and seamless continuity with the satellite movement while reducing signaling overhead.

It is noted that the satellite channel environment is particularly challenging, with large Doppler shift and delay values compared to terrestrial communications. In Table~\ref{tab:Channel_for_Reference_Orbit_Types}, different values for the Doppler variation and delay variation \cite{TR38811, TR38821, TR38742} are presented for a minimum elevation angle of 10 degrees for both service and feeder links in an NTN architecture configuration.

\begin{table}[h!]
\centering
\vspace{-2pt}
\caption{Satellite channel for reference orbit scenarios.}
\scalebox{0.85}{\begin{tabular}{|l|l|l|l|}
\hline
\cline{1-3} \textbf{Parameters} & \textbf{LEO} & \textbf{GSO} %
\tabularnewline\hline
\midrule\hline
\textbf{Altitude [km]} & 600 & 35786 \\ \hline
\textbf{Maximum Doppler Shift [ppm]} & $\pm$24 & Negligible \\ \hline
\textbf{Doppler Variation [ppm/s]} & $\pm$0.27 & Negligible \\ \hline
\textbf{Maximum Delay [ms]} & 25.77 & 541.46 \\ \hline
\textbf{Minimum Delay [ms]} & 8 & 477.48 \\ \hline
\textbf{Maximum Delay Variation [µs/s]} & $\pm$93.0 & Negligible %
\tabularnewline\hline
\bottomrule
\end{tabular}}
\label{tab:Channel_for_Reference_Orbit_Types}
\vspace{-2pt}
\end{table}

\begin{figure*}[t!] %
\vspace{-2pt} %
    \centering
    \includegraphics[width=12cm]{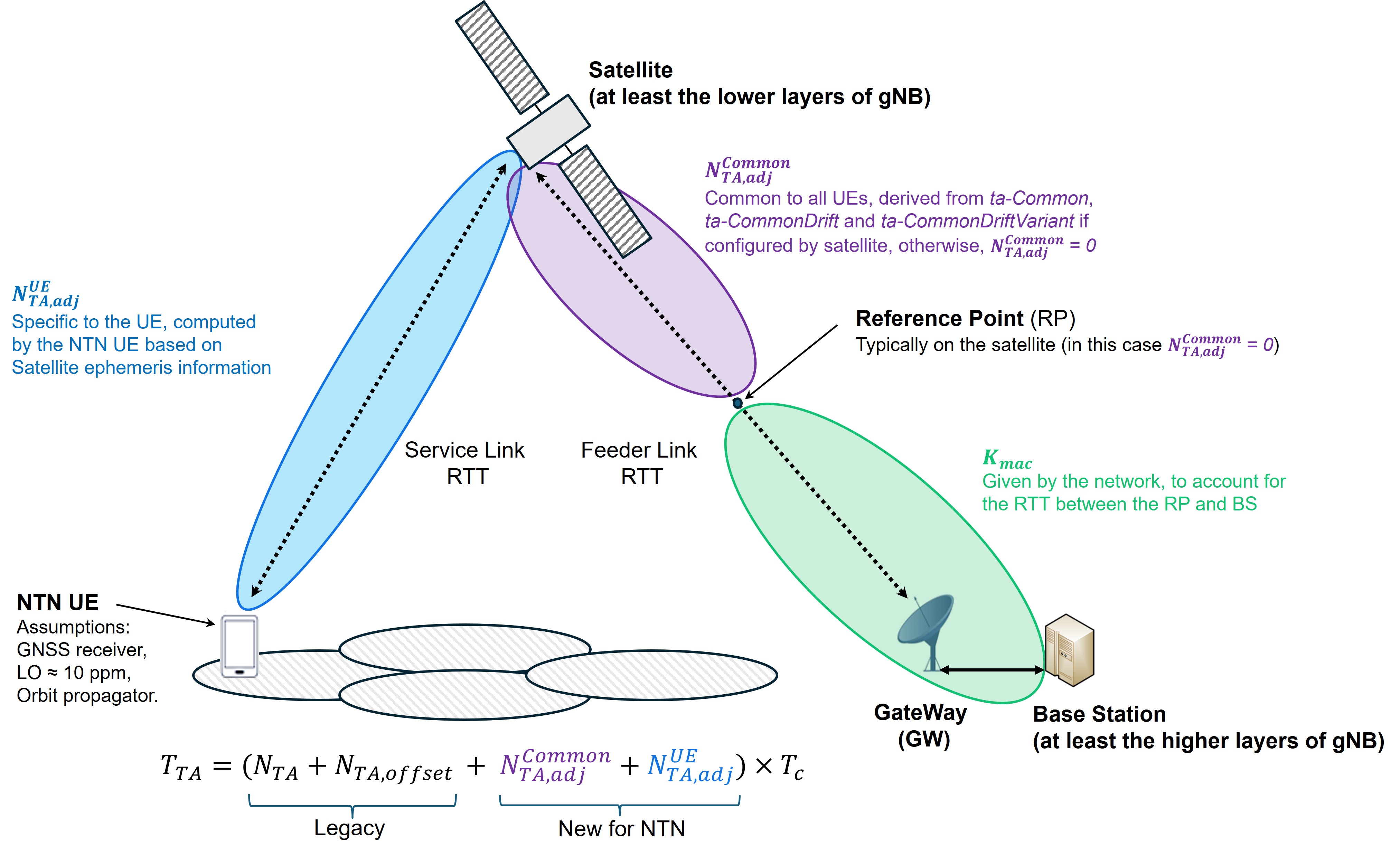}
    \begingroup
\captionsetup{font=footnotesize}
\caption{Timing-advance compensation for a global navigation satellite system (GNSS)-assisted non-terrestrial network (NTN) user equipment (UE). The service link connects the UE to the satellite, and the feeder link connects the satellite to the gateway (GW) and ground base station (BS). Blue and purple regions identify UE-specific and common timing adjustments, respectively. The UE derives service-link round-trip-time (RTT) compensation from satellite ephemeris and its position. The network configures common compensation and its drift to the reference point (RP); this term is zero when the RP is on the satellite. The green region identifies $K_{\mathrm{mac}}$, accounting for RTT between the RP and BS. Total timing advance combines the legacy command and offset with the NTN adjustments in units of $T_c$, the basic New Radio time unit. The illustrated UE assumes a GNSS receiver, a local oscillator (LO) with approximately 10-ppm accuracy, and an orbit propagator.}   
\label{fig:System_Precomp}
    \endgroup
    \vspace{-2pt} %
\end{figure*}

In NR NTN, the UE is responsible for pre-compensating the uplink Doppler and delay so that all the transmissions are aligned and synchronized at the uplink reference point. For that, the UE obtains the satellite ephemeris (i.e., satellite movement information) and the delay to a reference point/gNB (e.g., common timing advance (TA), $K_{\mathrm{mac}}$, and different offsets) from the system information broadcast in the cell, as shown in Fig.~\ref{fig:System_Precomp}.

In parallel, the UE may acquire an accurate position (depending on the NTN timing error requirement, defined for the frequency range and the numerology used) from the global navigation satellite system (GNSS). As a drawback, if the GNSS is unavailable or less accurate, the UE may be unable to connect to the network because of pre-compensation that is not sufficient to perform the initial access. Another option, adopted by some satellite network operators, is to perform the necessary compensation on board, keeping the UE unmodified with respect to legacy 4G and 5G. This solution requires more computing power on board and constrained scheduling that was not retained by the standard. In 5G-Advanced Release 20, a study followed by a normative phase is being carried out to make the NR NTN link more resilient to the GNSS system \cite{TR38742}. The solutions identified for the initial access and the connected mode shall have a limited impact on the current NR physical layer and pave the way toward GNSS-free operations for 6G.

For 6G, 3GPP has identified several use cases and enhancements for NTN, including multi-orbit operation, ubiquitous connectivity, and GNSS-free operation for the physical and access layers.

\subsubsection{Practical Challenges for 6G NTN Support in the RAN}
\label{subsubsec:6gntnrat}
This section aims at providing information with respect to the 6G satellite radio interface, distributed network functions, and multi-orbit network architecture.

\textbf{6G satellite radio interface}: The 6G RAT aims to harmonize TN and NTN radio interfaces. Most of the concepts and mechanisms introduced in 5G NR NTN may be reused in 6G NTN, with the difference that 6G will be designed for both TN and NTN, unlike 5G, where NTN integration arrived later \cite{lin2022path}. This means that constraints related to the NTN (e.g., frequency bands, Doppler, delay) will be considered in the design from the beginning. The Doppler effect introduces a large and time-variant frequency shift depending on the carrier frequency and the satellite characteristics such as altitude, orbit, and coverage. The pragmatic question to be solved by the standard and the industry is how to integrate TN and NTN when the constraints are not the same. 
To facilitate integration of TN and NTN, 6G will use a common waveform framework based on cyclic prefix OFDM (CP-OFDM) and DFT-spread-OFDM (DFT-s-OFDM). NTN-specific
adaptations, such as cyclic-prefix configuration, SSB periodicity, and timing and frequency compensation, can address satellite propagation conditions while facilitating reuse of transceiver hardware. 
Clearly, the most important challenge to enhance the NTN radio interface in 6G is GNSS-free operation. For indoor connectivity and for resilience to the degraded GNSS environment seen in parts of Europe, 3GPP technology should be less reliant on third-party infrastructure. The GNSS-resilience study in Release 20 shows that without a modification of the signals and fundamental design of the physical channel, the system cannot be GNSS-free without degradation. For 6G, two possibly complementary paths can solve this issue: designing the PRACH and physical channels to work without accurate positioning and/or introducing a positioning, navigation, and timing (PNT) service within the 3GPP technology. Some works that already study the feasibility of a positioning system for 5G show limits that can be solved in 6G. %

\textbf{Distributed network functions}: Another practical challenge for 6G NTN is to cover the different deployment scenarios of NTN for the next decade. While in NR NTN the deployment can be either transparent (the satellite is only a relay for the radio signal) or regenerative (with a full gNB on board), in 6G it is desirable to have more flexibility in the deployments. Already for NR, the ``IRIS\textsuperscript{2}'' program\footnote{``IRIS\textsuperscript{2},'' which stands for ``Infrastructure for Resilience, Interconnectivity, and Security by Satellite,'' is a satellite constellation project initiated by the European Union based on 5G NTN technology for governments and commercial users.} \cite{R2-2409400}, proposed by the European Commission, aims to use a split architecture where the base station functions are split between the space and ground segments. This design has been chosen because it represents a good tradeoff between the satellite energy consumption and the achievable capacity compared to a regenerative or a transparent architecture. The split architecture adopted by ``IRIS\textsuperscript{2}'' is standardized in TS 38.401 \cite{TS38401} but is not adapted to the NTN characteristics, making the development and the deployment more complicated.
In the 6G study, there are discussions on supporting a higher-layer split (HLS), i.e., between access-layer protocols. The difficulty for NTN is that it makes the interfaces more complex because of the dynamic nature of the satellite constellation, and it has implications for the protocols and the base station design, since the network nodes need to be reconfigured regularly, which does not occur in terrestrial deployments.
LLS options, i.e., between the physical and access layers, are also discussed in O-RAN in coordination with 3GPP. The work in \cite{6Gntn} proposes this type of architecture for 6G NTN, with the physical layer on the serving satellite and the access layer on feeder satellites, connected through high-capacity inter-satellite links (ISLs) \cite{6GntnD36}. The challenge for the LLS is to meet the delay and jitter constraints, linked to the radio framing, between the physical and access layers. Typically, the delay jitter should not exceed the length of the slot, i.e., sub-millisecond, which can be stringent in a multi-satellite deployment even though network solutions exist using buffers and time-sensitive networking techniques.
There is also a deployment scenario where the base station and core network functions are hosted in the constellation, which can be multi-orbit between non-GSO and GSO, to achieve user-to-user or user-to-application communication only through space without the need to fall back on the ground, implying dynamic routing within the constellation.

\begin{figure}[t!] %
    \centering
    \vspace{-2pt} %
    \includegraphics[width=\columnwidth]{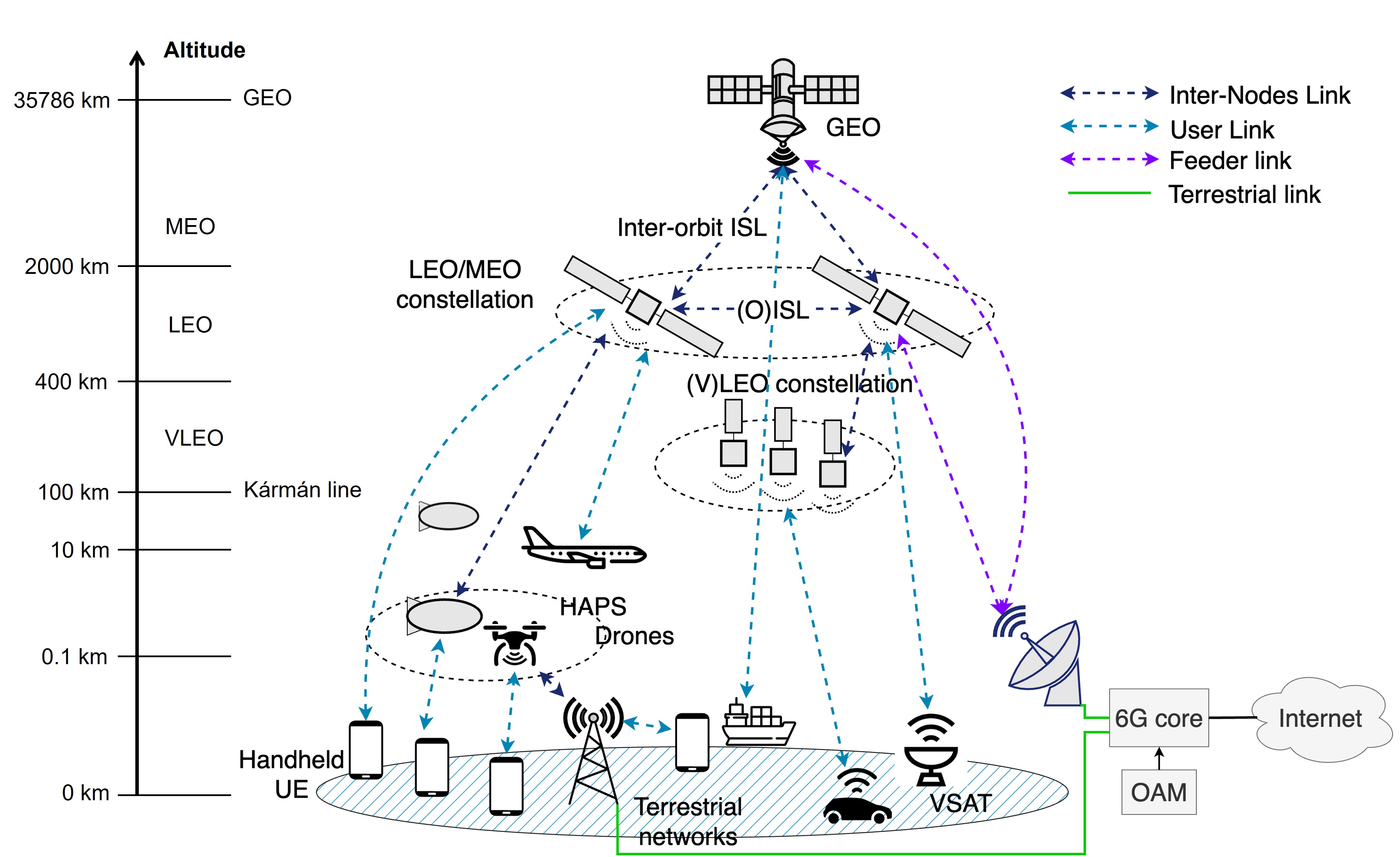}
    \begingroup
\captionsetup{font=footnotesize}
    \caption{Illustrative 6G architecture integrating terrestrial networks (TN) and a multi-orbit non-terrestrial network (NTN), from \cite{ronteix2024convergence}. Very low Earth orbit (VLEO), low Earth orbit (LEO), medium Earth orbit (MEO), and geostationary Earth orbit (GEO) satellites, high-altitude platform stations (HAPS), and drones provide complementary access for user equipment (UE), vehicles, ships, and very small aperture terminals (VSATs). Dark-blue dashed arrows show inter-node links, including inter-satellite links (ISLs) and optical ISLs (OISLs); light-blue dashed arrows show user links; purple dashed arrows show feeder links; and green solid lines show terrestrial connections. TN and NTN share a 6G core and operations, administration, and maintenance (OAM). Altitudes are shown schematically.}
    \label{fig:6GNTNarchi}
    \endgroup
    \vspace{-2pt} %
\end{figure}

\textbf{TN and NTN multi-orbit architecture}: A trend in satellite communication is the multi-orbit architecture and the TN-NTN integration~\cite{guidotti2024a}. The ``IRIS\textsuperscript{2}'' program is multi-orbit, using LEO-600, LEO-1200, and MEO-8000 shells, and Sateliot is working on the integration of the NTN and TN NB-IoT services. GEO provides wide coverage with a single satellite but at the expense of higher latency and lower throughput for a given UE antenna size. LEO provides more capacity and direct-to-device services but requires more satellites for the same coverage, with smaller, lower-capacity satellites than those in GEO. TN provides the highest capacity and lowest latency but is less resilient and needs more investment than NTN for the same coverage. Given this, the integration of the multiple layers is desirable for ubiquitous, global, and resilient connectivity in 6G.
This is illustrated in Fig.~\ref{fig:6GNTNarchi}. The 6G access network consists of a TN, three orbital layers (very low Earth orbit (VLEO), LEO, and GEO), and aerial nodes with HAPS and drones for the NTN. On the ground, different types of terminals are connected: handheld devices, boats, cars, and VSATs, each one with different connectivity needs. For the inter-node links, different frequency bands are considered according to the technology maturity, the data rate to deliver, and the channel characteristics. Typically, the optical ISL (OISL) is a mature technology that can deliver throughput up to 100 Gbit/s \cite{shang2025channel}, while ground-to-satellite laser communication for very high throughput links is at an early stage of development~\cite{6294234}.
The 6G core and associated O\&M function \cite{wang2024nonterrestrialnetworking6gevolution}, common to TN and NTN, are in charge of coordinating connectivity of the terminals in the 6G system, ensuring service continuity and resilience. The management of TN and NTN is discussed in the O-RAN Alliance \cite{polese, oranorchestration2025}. In \cite{esaai}, the European Space Agency (ESA) proposes to use AI for the NTN and TN integration, with AI in the RAN (e.g., cloud RAN), in the core for data analysis and federated learning, and in the end-to-end system for integration and optimization.

\subsubsection{Industrial Challenges for 6G NTN Implementation and Deployment}\label{subsubsec:6gntnimpl}

This section aims at providing information with respect to payload capabilities, antennas, and terminals.

\begin{figure}[t!]
    \centering

    \subfloat[Transparent satellite payload]{%
        \includegraphics[width=.95\linewidth]
        {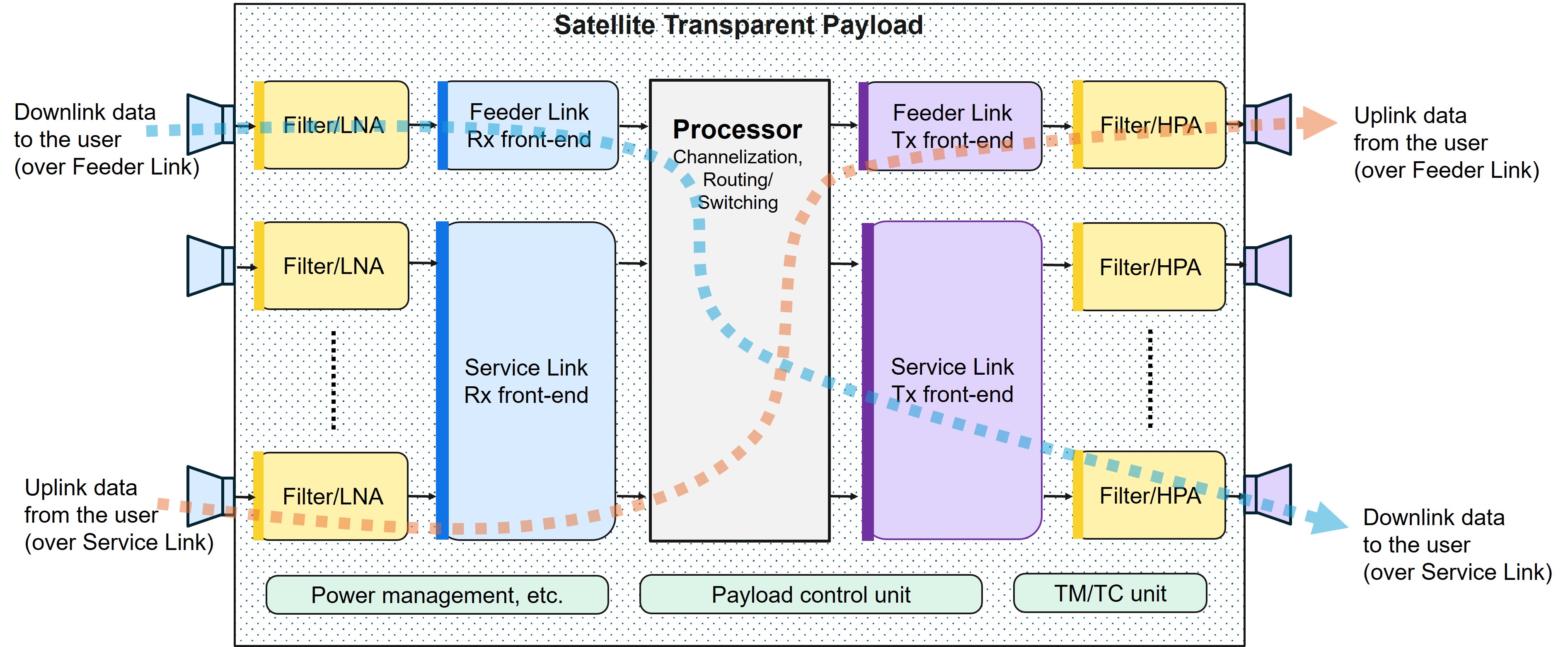}%
        \label{fig:transparentpayload}%
    }

    \par\medskip

    \subfloat[Regenerative softwarized satellite payload]{%
        \includegraphics[width=.95\linewidth]
        {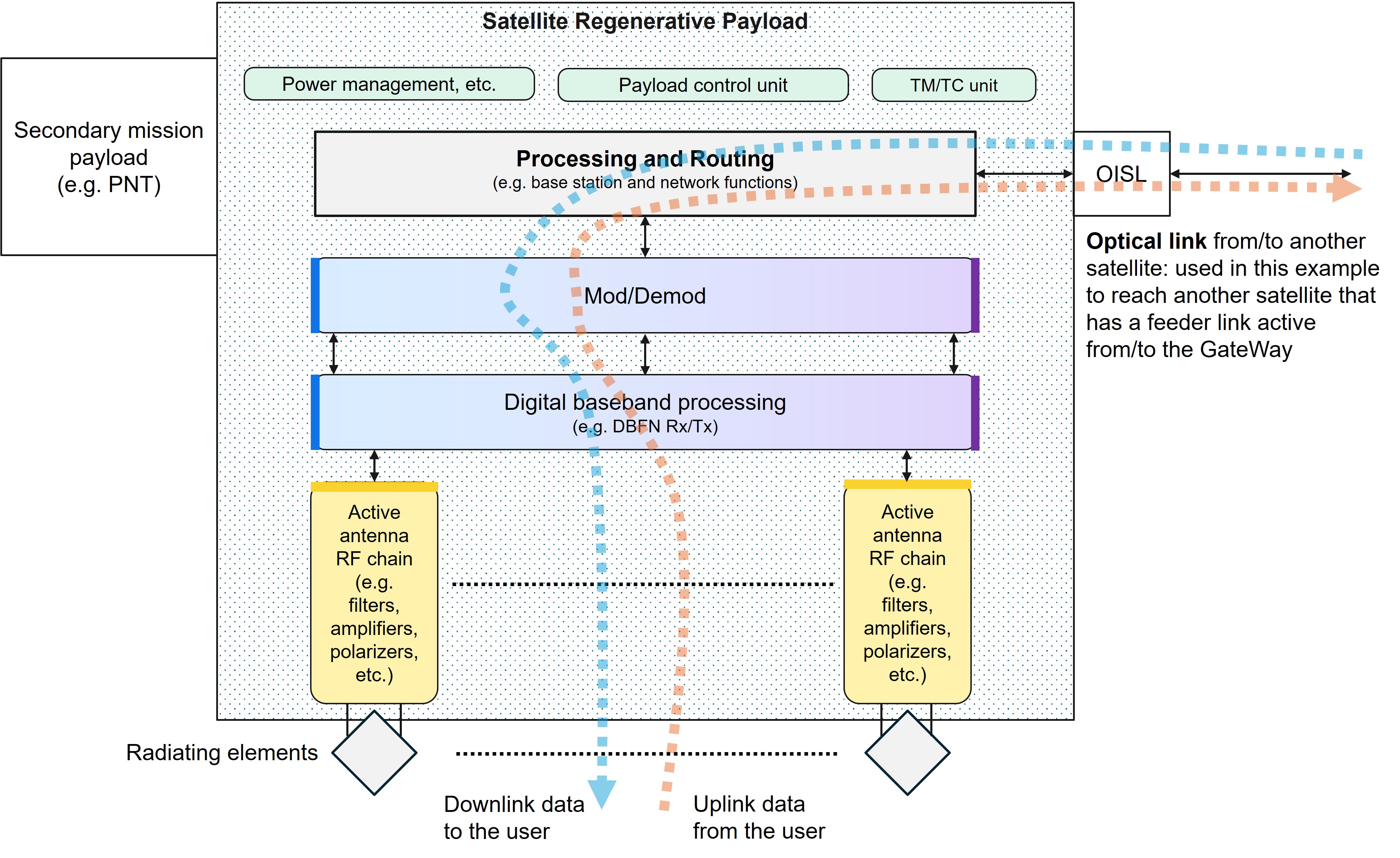}%
        \label{fig:regenerativesoftwarizedpayload}%
    }

    \begingroup
    \captionsetup{font=footnotesize}
    \caption{Examples of satellite communication payload architectures. (a) A transparent payload filters, amplifies, channelizes, and switches signals between feeder links to a ground gateway and service links to users, with radio-protocol processing on the ground. Receive (Rx) and transmit (Tx) front-ends use low-noise amplifiers (LNAs) and high-power amplifiers (HPAs), respectively. (b) A regenerative payload performs modulation/demodulation, digital baseband processing, and onboard processing and routing, including base-station and network functions. A digital beamforming network (DBFN) connects to active antenna radio-frequency (RF) chains. An optical inter-satellite link (OISL) connects to another satellite with an active feeder link to the gateway. Blue and orange dotted arrows show user downlink and uplink paths, respectively. Black dotted lines indicate omitted antenna branches. Both payloads include power management, payload control, and telemetry/telecommand (TM/TC); the secondary mission payload illustrates positioning, navigation, and timing (PNT).}
    \label{fig:satellitepayloads}
    \endgroup

    \vspace{-10pt}  %
\end{figure}

\textbf{Augmented payload capabilities}: One important constraint for the satellite capacity is the SWaP: size, weight, and power. The overall performance of the satellite is limited by the energy it can harvest from solar panels, and the weight and the size that can be launched at a reasonable cost.
Historically, satellite telecommunication payloads were analog, merely filtering, routing, and amplifying radio signals (see Fig.~\ref{fig:transparentpayload}), a technology widely used for GSO. However, progress in digital hardware and the miniaturization of electronics have enabled smaller digital payloads. Combined with the reduction in costs for launching into orbit, the number of high-capacity non-GSO constellation projects has multiplied, providing high-speed internet worldwide.
The resulting change in spaceborne capabilities is illustrated in Fig.~\ref{fig:regenerativesoftwarizedpayload}. In this example, the service link is handled by an active antenna (using dielectric resonator antenna (DRA) technology). Each radiating element is associated with an RF chain for filtering, amplifying, and converting the signals. The digital beamforming network (DBFN) \cite{esadbfn} is responsible for dynamically generating beams using the network of radiating elements (typically hundreds to thousands of radiating elements organized in tiles).
The physical-layer functions, including physical channel coding, modulation, resource element mapping, waveform generation, and fast Fourier transforms (FFTs), are implemented on the modem board. This board can process the radio signal of all cells served by the satellite. The higher layers, including the access layers and transport layers, are computed by the onboard processor in Fig.~\ref{fig:regenerativesoftwarizedpayload}, possibly supported by generic hardware (e.g., Starlink uses an ARM-based AMD Versal processor). These layers are softwarized, so the telecommunication payload is made flexible, adaptable, and scalable. It is particularly interesting considering the typical satellite lifespan of seven to ten years, as the software can be updated remotely according to standard evolutions and market changes. Additionally, a routing function for the various data and control flows between the satellite's various interfaces is used, e.g., a multiprotocol label switching (MPLS) space router. The payload is controlled from the ground by the mission control center (MCC). It is this flexibility, with generic hardware and software, that eventually enables the technically feasible transition to 5G NTN and later to the multi-orbit 6G.

\textbf{Antennas}: References \cite{TR38811, TR38821, TR38863} indicate antenna models and patterns applicable to both non-GSO and GSO satellites. While reference \cite{satpatterns} is for fixed-satellite service (FSS) antennas, it can also be used for the modeling of mobile-satellite service (MSS) antennas. %
The satellite antenna array can have different shapes such as circular, rectangular, or cross. The antenna elements are usually circularly polarized. Moreover, multi-beam antennas and phased arrays are commonly used in non-GSO satellites \cite{9852737}, as numerous high-gain beams with small Earth footprints need to be provided. However, direct radiating arrays are also used due to their wide scan range and better off-boresight performance or lower scan loss (see reference [41] in \cite{9852737}). Introducing active antennas opens the possibility of multi-beam satellite payloads, massive MIMO \cite{9110855}, multi-user precoding, and detection. 
Some NTN deployments use frequency-reuse factors of three or four
to reduce interference between adjacent satellite beams or cells,
whereas others use full frequency reuse (reuse factor one).
Precoding can help manage co-channel interference and support a
large number of beams per reuse.
See Fig. 5 in \cite{8746876} for Viasat-3, which shows that 1000 beams per reuse can be deployed using low-complexity precoding, as long as suitably dimensioned on-ground equipment is also deployed. %
The power amplifiers, for instance in Fig.~\ref{fig:satellitepayloads}, may be distributed over the antenna array, but this depends on the power output to meet the link budget and coverage required. We must also consider the power requirements of onboard equipment, which is powered by solar panels. Traditional satellite PAs are either traveling-wave tube amplifiers (TWTA) or solid-state power amplifiers (SSPA); see \cite{satpas} and references cited therein for a discussion on SSPA and TWTA technologies.

In 6G NTN, the satellite will support L-, S-, C-, X-, Ku-, Ka- and Q/V-bands, sometimes in the same payload, which poses an engineering challenge for the antenna design, as presented for instance in Table~\ref{tab:SAN_config} with key parameters for the satellite access node. The supported bands in 6G are not limited to L-, S-, Ku- and Ka-bands but can extend to C-, X- and Q/V-bands. %

 \begin{table*}[t!]
 \centering
 \caption{Targeted 6G satellite parameters with phased-array configuration for different frequency bands and orbits.}
 \scalebox{0.625}{\begin{tabular}{|l|l|l|l|l|l|l|} %
 \hline
 \cline{1-7} \textbf{} & \textbf{L-band} $\sim$1.5GHz & \textbf{S-band} $\sim$2GHz& \textbf{C-band} $\sim$4GHz& \textbf{Ku-band} $\sim$11/14GHz& \textbf{Ka-band} $\sim$20/30GHz& \textbf{Q/V-band} $\sim$37/47GHz%
 \tabularnewline\hline
 \midrule\hline

 \textbf{VLEO 300km} with antenna size, gain, &   & 1.5m, 27dBi  &   &   & 0.2/0.13m, 29dBi  &  \\  
 satellite EIRP density/beam, beam diameter, &   & 28dBW/MHz, 30km  &   &   & 5dBW/MHz, 25km &  \\  
 number of antenna elements, gain per element &   & 400, 0.92dBi  &   &   & 400, 2.98dBi &  \\ \hline 
 \textbf{LEO 600km} with antenna size, gain, & 2.5m, 30dBi & 2.5m, 32dBi &  2m, 35dBi & 0.4/0.3m, 31dBi  & 0.3/0.2m, 32.5dBi &  0.27/0.21m, 40.4dBi \\ 
 satellite EIRP density/beam, beam diameter, & 38dBW/MHz, 50km & 40dBW/MHz, 40km & 34dBW/MHz, 30km & 13dBW/MHz, 40km & 16dBW/MHz, 40km &  15dBW/MHz, 20km \\ %
 number of antenna elements, gain per element & 676, 1.14dBi &  1024, 1.28dBi & 2116, 1.74dBi &  676, 2.73dBi & 900, 2.98dBi & 3481, 5dBi \\ \hline 
 \textbf{LEO 1200km} with antenna size, gain, & 2.5m, 30dBi  &  2.5m, 32dBi &   &  0.45/0.33m, 32dBi  & 0.35/0.23m, 34dBi &  \\
 satellite EIRP density/beam, beam diameter, & 38dBW/MHz, 95km &  40dBW/MHz, 75km &   & 13dBW/MHz, 75km   & 16dBW/MHz, 60km  &  \\ %
 number of antenna elements, gain per element & 676, 1.14dBi & 1024, 1.28dBi &   & 676, 3.75dBi & 1024, 3.75dBi  &  \\ \hline 
 \textbf{MEO 8000km} with antenna size, gain, &   &   &   & 1.5/1.1m, 42.5dBi  & 1/0.67m, 43dBi &  \\ 
 satellite EIRP density/beam, beam diameter, &   &   &   & 28dBW/MHz, 140km  & 32dBW/MHz, 130km &  \\  
 number of antenna elements, gain per element &   &   &   &  1936, 9.64dBi  & 2116, 9.72dBi &  \\ \hline 
 \textbf{GSO} with antenna size, gain, &   &   & 12m, 50.5dBi &   & 6/4m, 60dBi & 2.7/2.1m, 60.4dBi \\ %
 satellite EIRP density/beam, beam diameter &   &   &  59dBW/MHz, 260km  &   & 42dBW/MHz, 100km &  45dBW/MHz, 120km \\ %
  &   &   &  Phased Array N/A &   & Phased Array N/A & Phased Array N/A  %
 \tabularnewline\hline

 \bottomrule
 \end{tabular}}
 \label{tab:SAN_config}
 \end{table*}

\begin{figure}[t!]
    \centering
    \includegraphics[width=1\columnwidth]{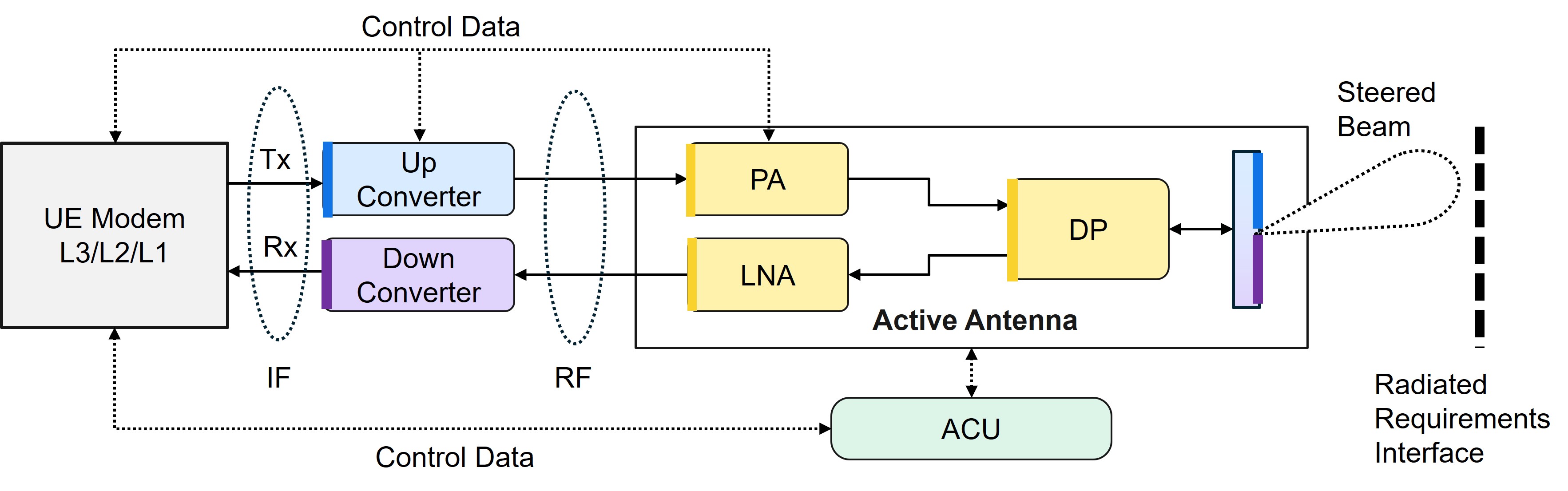} %
        \begingroup
\captionsetup{font=footnotesize}
    \caption{Reference architecture of a frequency-division duplex (FDD) very small aperture terminal (VSAT) with a steered antenna beam. The user-equipment (UE) modem implements the physical (L1), data-link (L2), and network (L3) layers. In the transmit (Tx) path, the up-converter translates the intermediate-frequency (IF) signal to radio frequency (RF), and the power amplifier (PA) amplifies it. In the receive (Rx) path, the low-noise amplifier (LNA) amplifies the received RF signal before down-conversion to IF. The duplexer (DP) connects both paths to the antenna. The antenna control unit (ACU) supports beam steering. Solid arrows show signal paths, dotted arrows show control-data connections, and the dashed vertical line marks the radiated requirements interface.}  \label{fig:VSAT_Archi}
    \endgroup
\end{figure}

\textbf{Terminals}: In addition to handheld terminals, smartphones, and IoT devices, high-power UEs (HPUEs) for S/L frequency bands and VSAT UEs in frequency bands above 10 GHz have also been introduced, as seen in Table~\ref{tab:terminals}. The latter are designed with the capability to generate at least a single beam aimed at one satellite at any given moment; however, the standards do not preclude other implementations. %
 \begin{table}[t!] %
 \centering
 \caption{Terminal types introduced in Rel-17, Rel-18, Rel-19 NR NTN \cite{TS38101-5} and NB-IoT/eMTC NTN \cite{TS36102}.}
 \scalebox{0.85}{\begin{tabular}{|l|l|l|l|}
 \hline
 \cline{1-4} \textbf{UE Type} & \textbf{NR/LTE UE} & \textbf{NR/LTE HPUE} & \textbf{NR VSAT UE}   %
 \tabularnewline\hline
 \midrule\hline
 \textbf{Tx Power} & 23 dBm & 26, 29, 31 dBm & 86.2 dBm \\ \hline 
 \textbf{Frequency} & S/L-bands & S/L-bands & Ka/Ku-bands %
 \tabularnewline\hline
 \bottomrule
 \end{tabular}}
 \label{tab:terminals}
 \end{table}
Different VSAT architectures with electronic steering (phased array), mechanical steering, or a hybrid combination are currently considered for 5G/6G NTN communications. Fig.~\ref{fig:VSAT_Archi} describes an FDD VSAT reference architecture, where UC represents the up-converter, PA the power amplifier(s), LNA the low-noise amplifier(s), DC the down-converter, DP the duplexer, and ACU the antenna control unit. The ACU controls or assists the antenna steering, where the antenna can be active, electronically or hybrid steered. RF represents the radio-frequency region, and IF the intermediate-frequency region.

\subsubsection{NTN Spectrum}
\label{subsubsec:ntn spectrum}
This section aims at providing information with respect to the satellite frequency spectrum and supplemental coverage from space.

\textbf{Satellite frequency spectrum}: Starting from Release 19, the FR1-NTN frequency range has been extended from 0.41--7.125 GHz to 0.41--14.5 GHz, while the FR2-NTN frequency range has been extended from 17.3--30 GHz to 10.7--30 GHz.
All frequency bands from \cite{TS38108, TS36108} as well as additional satellite service allocated bands below 8 GHz (e.g., C-band), above 30 GHz (e.g., Q/V-band), or near 10 GHz (e.g., X-band) may be considered for the future deployment of 6G NTN technology. Some expected 26 dBm C-band NTN UE and satellite parameters are given in Table~\ref{tab:C-band_UE_Sat} with $8\%$ active beams and $34$ dBW/MHz effective isotropic radiated power (EIRP) per beam, while targeted SINR and throughput values are indicated in Table~\ref{tab:C-band_SINR} with equations from \cite{TR38811} and \cite{TR38863} for a LEO 600 km constellation. Downlink/uplink targeted throughput values for C-band are higher than 20/2 Mbit/s, respectively, at different elevation angles.

\begin{table}[t!]
\centering
\caption{Targeted 6G UE and satellite C-band parameters.}
\scalebox{0.85}{\begin{tabular}{|l|l|l|}
\hline
\cline{1-3} \textbf{Parameter} & \textbf{FDD} & \textbf{TDD}   %
\tabularnewline\hline
\midrule\hline

\textbf{UE antenna gain [dBi]} & -2 & -2 \\ \hline
\textbf{UE Rx NF [dB]} & 7 & 5.5  \\ \hline
\textbf{Satellite ant. gain [dBi], 90°/45° elev.} & 35.3 / 33.4 & 35.8 / 33.9 \\ \hline
\textbf{Satellite Rx equivalent NF [dB]} & 2 & 2  %

\tabularnewline\hline
\bottomrule
\end{tabular}}
\label{tab:C-band_UE_Sat}
\end{table}

\begin{table}[t!]
\centering
\caption{Targeted C-band 6G throughput values.}
\scalebox{0.85}{\begin{tabular}{|l|l|l|}
\hline
\cline{1-3} \textbf{Parameter} & \textbf{FDD} & \textbf{TDD 50-50}   %
\tabularnewline\hline
\midrule\hline

\textbf{DL/UL carrier frequency [GHz]} & 3.4 / 3.9 & 3.4 / 3.4 \\ \hline
\textbf{DL/UL used bandwidth [MHz]} & 18.36 / 5.04 & 28.08 / 28.08 \\ \hline
\textbf{DL SINR [dB] 90°/45° elev.} & 10.15 / 7.51 & 11.63 / 9.00 \\ \hline 
\textbf{UL SINR [dB] 90°/45° elev.} & 4.89 / -0.23 & -1.47 / -6.02 \\ \hline
\textbf{DL peak rate [Mbit/s] 90°/45° elev.} & 32.18 / 25.08 & 27.80 / 22.20 \\ \hline
\textbf{UL peak rate [Mbit/s] 90°/45° elev.} & 5.11 / 2.43 & 5.45 / 2.26 %
\tabularnewline\hline

\bottomrule
\end{tabular}}
\label{tab:C-band_SINR}
\vspace{-15pt} %
\end{table}

\textbf{Supplemental coverage from space}: NTN could also be deployed in (terrestrial) mobile service allocated bands. This requires protecting the operations of all other radio systems as per Article 4.4 of the ITU-R Radio Regulations. In line with this, the Federal Communications Commission (FCC) announced on February 23, 2023, a future rulemaking process by which ``supplemental coverage from space'' could be provided in cellular bands by non-GSO satellite operators based on a spectrum lease agreement with mobile network operators having an exclusive terrestrial mobile license over certain geographical areas under U.S. jurisdiction \cite{fccscs}. The ITU-R is currently studying coexistence conditions (as part of WRC-27 agenda item 1.13) on possible new allocations to the MSS for direct connectivity between space stations and  IMT user equipment to complement terrestrial IMT network coverage, in accordance with Resolution 253 (WRC-23). 
This will provide a cost-effective infill service to areas where it is not economic to provide terrestrial coverage. The WRC-27 agenda item 1.13 requires that MSS systems operating in the specific bands shall not cause harmful interference to, nor claim protection from, stations operating in the mobile service. Coexistence studies to determine the conditions of coexistence and necessary regulatory conditions are currently underway in the ITU-R. Reference \cite{rappaport2025spectrumopportunitieswirelessfuture} has shown a 220~m$^2$ satellite phased array antenna at sub-GHz frequencies (providing up to 3 bit/s/Hz) with 2500 beams of $3$-degree beamwidth for direct connectivity. The area of the beam footprint is inversely proportional to the antenna area.
Other references \cite{TR38811,TR38821} for S-band satellite antenna sizes indicate equivalent diameter apertures of 2 m for LEO 600 km with 50 km beam diameter and 34 dBW/MHz EIRP density per beam, and 22 m for GEO with 250 km beam diameter and 59 dBW/MHz EIRP density.

\vspace{-0.3cm}
\subsection{New Waveforms}
\label{waveforms}

Multiple access (MA) schemes enable the efficient use of radio resources of a wireless network. 6G has triggered the need to study (or revisit) waveforms.
The 4G and 5G systems use orthogonal frequency-division multiple access (OFDMA) \cite{cimini} as a form of MA.

OFDM is the dominant modulation format (waveform) for MA \cite{Andrews_2024}. Its suitability for mobile wireless communications has been demonstrated in the landmark paper \cite{cimini}. OFDM is also the baseline waveform for assessing any improvements by using new waveforms \cite{RP-251881}. Both LTE (4G) and NR (5G) employ OFDM-based waveforms for both UL and DL across a wide range of use cases and frequency bands. In the DL, CP-OFDM is used in both LTE and NR~\cite{ali1,ali2,38211}. In the UL, LTE relies exclusively on DFT-s-OFDM, whereas NR supports both CP-OFDM and DFT-s-OFDM, enabling commonality between UL and DL waveforms, while addressing a flexible tradeoff between implementation complexity, power efficiency, and MIMO compatibility. 
OFDM's popularity in 5G is due to (1) backward compatibility with 4G and (2) its well-known information-theoretic optimality for the maximization of system capacity over frequency-selective channels \cite{TATARIA1}. 

OFDM-based waveforms have been specified in 5G for both FR1 and FR2 (up to 71 GHz). They are particularly well suited for broadband cellular systems due to their high flexibility in multiple access, excellent compatibility with MIMO and spatial multiplexing, and efficient FFT-based implementations at both the transmitter and receiver. These characteristics allow OFDM-based systems to scale efficiently with bandwidth, antenna count, and numerology, while maintaining manageable computational complexity.
From an operational perspective, the ability to reuse existing 5G site infrastructure is of high importance for operators. Consequently, 5G and 6G base stations are envisioned to be co-located, with the potential to reuse significant parts of the RF and baseband hardware across different radio access technologies (RATs). Maintaining waveform continuity through CP-OFDM and DFT-s-OFDM therefore provides substantial benefits in terms of cost efficiency, deployment simplicity, and ecosystem maturity. 3GPP has also discussed alternative waveforms, such as orthogonal time frequency space (OTFS), for 6GR \cite{saifone} but decided to reuse the existing waveforms to maintain backward compatibility. The gains of OTFS, if any, in commercial deployments are marginal. 

The UL still requires further attention, as coverage plays a key role, especially when higher frequencies are used on existing site infrastructure. It is thus important to focus on optimizing DFT-s-OFDM for an even lower peak-to-average power ratio (PAPR), which enables increasing the UE Tx power.
A low-PAPR technique already adopted in 5G NR is frequency-domain spectrum shaping (FDSS), which operates on the DFT-s-OFDM waveform by applying a shaping filter in the frequency domain, enabling effective PAPR reduction with low implementation complexity. Such techniques are currently being studied for the UL.

Different waveforms have also been discussed in 3GPP in the context of sensing, with OFDM emerging as the waveform of choice. The reasons are similar to those for communication, namely that no other waveform has shown performance benefits and that it is highly desirable to use the same waveform for sensing and communication. Note that short pulses, which are useful in some sensing scenarios, can be easily generated in the OFDM framework.

\vspace{-0.3cm}
\subsection{New Constellations and Channel Coding}
Modulation and channel coding are cornerstones of wireless communication. 

\subsubsection{New Constellations}
Conventional quadrature-amplitude modulation (QAM) up to 1024-QAM is used in 5G. The properties of QAM, which was also used in 4G, are well understood. However, it is well known from Shannon theory that there is a 1.53~dB gap to the theoretical capacity on an additive white Gaussian noise (AWGN) channel \cite{forney,pas}. The reason for this gap is the (typically) equiprobable use of the different QAM constellation points, whereas maximizing the Shannon capacity on an AWGN channel requires a Gaussian distribution of the transmitted signal. This has sparked interest in alternative modulation schemes for 6G. Two different approaches have been proposed: geometric shaping (GS) and probabilistic shaping (PS).

In GS, the constellation points are changed to mimic a Gaussian distribution \cite{forney};  this adjusts the ``shape'' of the constellation to approximate Gaussian signals, and the Euclidean distance between adjacent points can be maximized. This is the case in the digital broadcasting standard ATSC 3.0 \cite{Luke:16, Steiner:17}. 
 
In PS, a constellation point $\textbf{r}$ is chosen with the probability $p\left(\textbf{r}\right)$ given as $\sim\exp\left(-\lambda\lVert \mathbf{r} \rVert^2\right)$, where $\lambda$ is an optimization parameter of the Maxwell--Boltzmann (M-B) distribution \cite{pas}. It is noted that in PS, the geometric structure remains unchanged. The parameter $\lambda$ can be adapted to different SNRs and modulation orders, with $\lambda=0$ corresponding to regular QAM.

Although the gains from constellation shaping may look promising, they are at best 1.53 dB for the single-layer case and come at a substantial cost in complexity. Coexistence and RF requirements may also need further consideration. With this in mind, the baseline for the first 6G release is regular QAM with support for up to 1024-QAM in both uplink and downlink.

\subsubsection{Channel Coding}
Wireless channels are transmit-power-limited and prone to errors; coding gain is the principal metric to choose good codes and has been used in 3GPP. The values of some KPIs, especially those in \cite{ITURM2030},
determine the choice of suitable codes and their properties. 
Turbo codes were used in 3G and 4G due to their superior coding gain \cite{turbo}. Approaching capacity with a practical encoding/decoding complexity is a desirable goal in coding theory.

During the standardization of 5G NR, significant effort was devoted to studying and comparing the performance and efficiency among different channel codes, including (enhanced) Turbo codes, low-density parity-check (LDPC) codes, convolutional codes, and polar codes. Eventually, LDPC codes and polar codes were selected as the channel codes for the data and control channels, respectively. For the 6G system, the peak throughput target is expected to increase significantly beyond that of 5G~\cite{ITURM2030}. For enhanced mobile broadband (eMBB) use cases, a peak UE throughput of 36~Gbit/s is targeted in the first 6G release. Supporting such throughput motivates the design of new LDPC codes with improved performance at a small number of decoding iterations, thereby enabling higher area efficiency and decoder throughput than the LDPC codes used in 5G NR.

A key limitation of 5G NR LDPC codes in the small-iteration regime is the use of punctured systematic nodes. In particular, 5G NR LDPC codes employ two punctured nodes with higher degree than the remaining columns. This increases the effective average degree of the code and improves performance when decoding is performed with a large number of iterations. However, in the small-iteration regime, multiple punctured nodes may slow convergence because the decoder must first recover the punctured bits. Reducing the number of punctured nodes can therefore improve performance at low iteration counts, although it may lead to performance loss in the asymptotic regime. To address this, 3GPP will define a third base graph, BG3, to be used for LDPC coding for the highest data rates (beyond 6 Gbit/s on a single carrier).

\subsection{Duplexing Techniques}
\label{subsec:others}
The duplexing schemes agreed in 3GPP to be supported in 6G include full- and half-duplex FDD, semi-static TDD, dynamic TDD, and subband full duplex (SBFD). Full-duplex FDD and semi-static TDD are widely used in 5G (as well as in earlier generations) and are proven and well-understood technologies. Half-duplex FDD, where the UE cannot receive and transmit simultaneously (while the network has this ability), is not new either, but is an important component in low-end 6G IoT devices. 
Dynamic TDD, where slots can be dynamically used for either downlink or uplink transmission, was introduced in 5G. It is primarily useful in small-cell deployments, as the inter-site/inter-band interference in a macro deployment can be severe. It will likely be part of 6G as well, but with simplifications to the procedures determining the link direction (e.g., removing the slot-format indicator present in 5G). 

Subband full duplex, studied in 5G-Advanced \cite{TS38858}, is likely also part of 6G. In addition to problems with self-interference, it shares the same inter-site/inter-band interference aspects as dynamic TDD and is thus primarily relevant for small-cell deployments. Existing deployments are not designed to deal with the additional requirements for such an operation. If already-deployed 5G antennas are to be used with 6G deployments, then one will likely have to use the already established duplexing solutions and cannot consider SBFD, let alone a full-duplex (FD) approach. Even in pristine spectrum, such as the upper 6 GHz band, when the plan is to reuse the existing macro cell sites and share them with other operators, the use of SBFD would be a very costly solution, and the resulting increase in the antenna size (due to isolated Tx and Rx antenna sub-panels within the antenna) may not always be allowed.  

Furthermore, duplexing considerations are needed in connection with
sensing, as simultaneous transmission and echo reception in monostatic
sensing require FD capability. Bistatic sensing can avoid this
requirement by separating the sensing transmitter and receiver.
Note that the requirements may differ between FD for monostatic
sensing and FD for communication purposes; short round-trip delays
make simultaneous transmission and reception particularly relevant
when detecting objects close to the site.

\subsection{Scheduling and Radio Resource Management}
In the previous sections, several basic building blocks of the RAN, many of them studied in academia, and their applicability to 6G have been discussed. However, such technology areas are not the only ones that matter in a practical 6G system. There are many aspects such as control signaling, timing aspects, and UE measurements, just to mention some examples, where decisions in standardization can have a profound impact on the overall performance of a deployed system. 

Scheduling is a vital, time-critical, and compute-intensive part of a cellular system, controlling which UEs are to transmit/receive and on which time/frequency/spatial resources. There is a wide range of literature on different scheduling strategies. 
It is, to a large extent, an implementation aspect within the limits set by the 3GPP standard in terms of control signaling and the associated time relations. Over the years, the once relatively simple mechanisms introduced in the first release of 4G have evolved into a fairly complex framework with the addition of carrier aggregation (also across different numerologies), more advanced CSI reports, and limitations in terms of the time a UE needs to prepare an uplink transmission, all of which should work not in isolation but jointly. In particular, more asynchronous schemes for uplink control information (UCI) reporting and avoiding (time-critical) dependencies across scheduling decisions for different carriers are beneficial \cite{6GWS-250004, 6GWS-250084}, and 3GPP has agreed to study simplifications to the UCI handling.
When evaluating different schemes, it is important to consider the overall system from an end-to-end perspective, including all protocol layers and the interaction between different mechanisms with realistic traffic models, and not only focus on one technology component in isolation.

\subsection{Integrated Sensing and Communication}
\label{subsec:isac}

ISAC could extend the RAN
beyond connectivity to support detection and tracking of passive
objects, with applications such as unmanned aerial vehicle (UAV) surveillance and traffic
monitoring. Environmental information could also assist communication
functions, such as beam management and mobility. For 6G, the industrial question
is which sensing services can be provided using practical site grids
and radio hardware at acceptable incremental cost. Relevant service
measures include detection probability, false-alarm rate, position and
velocity accuracy, coverage, and update interval.

Deployment geometry and receiver design are central \cite{lin2026integrated}. Monostatic
sensing uses co-located transmission and reception, while bistatic
and multistatic sensing involve separated transmitters and receivers.
Simultaneous monostatic operation requires sufficient transmit--receive
isolation and self-interference suppression; time-separated operation
offers an alternative with different coverage and timing constraints.
Distributed sensing introduces synchronization, calibration, and
coordination requirements. Communication coverage does not imply
equivalent sensing coverage: echoes also depend on target reflectivity,
propagation geometry, blockage, and clutter. Bandwidth, observation
time, and antenna aperture determine different aspects of sensing
resolution. A common waveform framework facilitates integration,
but sensing-specific signal configurations and receiver processing
may still be needed.

Sharing spectrum, antennas, and signals does not make sensing
resource-free. The RAN scheduler must allocate transmission
opportunities, observation intervals, beams, and processing resources
to meet both communication and sensing requirements. Reusing
communication signals can reduce dedicated sensing transmissions,
but their timing and illumination may not satisfy the sensing service.
Conversely, dedicated sensing resources can reduce communication
capacity. Resource allocation should therefore account for both
traffic demand and sensing service commitments; relying only on
idle resources can make sensing availability load-dependent.
Coordination across neighboring cells may also be needed to manage
interference. Local processing can extract detections or features
before forwarding information for fusion, reducing transport
requirements compared with forwarding raw samples. Commercial
deployment also requires appropriate access control and privacy
protection for sensing information. Industrial evaluation should
therefore consider sensing performance, implementation and energy
costs, and the impact on communication services jointly.

\section{Conclusions}
\label{sec:conc}
In this paper, we have reviewed various technologies and processes that form the fabric of radio access, with particular emphasis on the physical layer for 6G. We have also considered practical implementation and deployment factors that may limit the realizability of theoretical performance gains. 6G is expected to be deployed as an overlay on existing site grids and to operate in both existing spectrum and new mid-bands. MIMO will remain an important tool for improving performance, with more antenna elements and ports in massive MIMO systems. As the number of antenna ports increases, so does the CSI acquisition and reporting overhead. AI/ML-based CSI modeling and prediction can therefore improve both single-user and multi-user performance. Uniform QAM is expected to remain the basis for modulation, complemented by enhancements to modulation and waveforms that improve uplink performance. Scheduling and other RRM-related techniques have also shown the potential for significant gains. Energy efficiency can be improved by scaling processing resources and capabilities to service requirements, since not all services require the most extreme performance.

While the individual enhancements discussed above may seem to provide only modest gains, their aggregate benefit could be significant. D-MIMO may improve spectral efficiency primarily at the cell edge, corresponding to the lower percentiles of the capacity cumulative distribution function (CDF), rather than at the peak. AI/ML can further enhance performance across different RAN functions. ISAC could extend the RAN to sensing services, subject to tradeoffs
among sensing performance, communication performance, and
implementation and energy costs. The integration of NTN into 6G can extend global coverage and improve service in underserved areas. From a deployment perspective, gains arising from changes to the fundamental RAN building blocks must be assessed together with the need for hardware compatibility and spectrum coexistence, thereby supporting a smooth and timely introduction of 6G.

As assessed in the companion article, ``Has the PHY Layer Matured?''~\cite{shafi2026has}, many core physical-layer building blocks are already mature, so isolated refinements are likely to yield only incremental gains. As we look toward 6G and beyond, adopting a truly revolutionary mindset is essential. We should reconsider core assumptions about connectivity, intelligence, and architecture while combining incremental physical-layer gains into meaningful system-level improvements. This requires looking beyond merely increasing data rates or expanding frequency bands and embracing broader architectural advances, particularly the integration of AI throughout the RAN.

\section*{Acknowledgments}
The author, Mansoor Shafi, would like to thank Dr. Harsh Tataria for many useful discussions in the preparation of this manuscript.



\begin{IEEEbiography}
[{\includegraphics[width=1in,height=1.25in,clip,keepaspectratio]{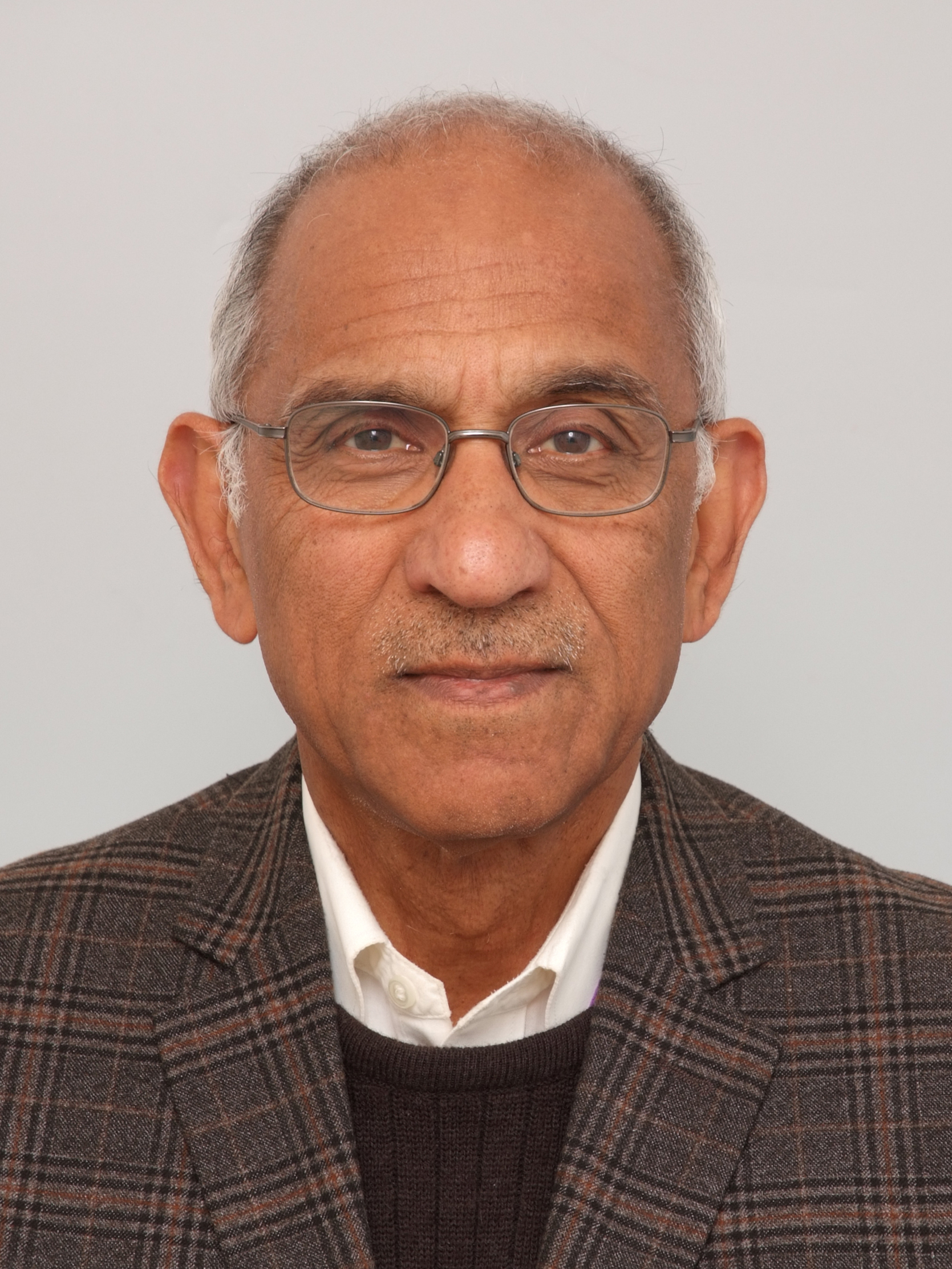}}]
{Mansoor Shafi} (S'69–M'82–SM'87–F'93–LF'16) received the B.Sc. (Eng.) and Ph.D. degrees in electrical engineering from the University of Engineering and Technology Lahore and The University of Auckland in 1970 and 1979, respectively. He is currently a Telecom Fellow Wireless at Spark NZ Ltd  and an Adjunct Professor with the School of Engineering, Victoria University of Wellington and Canterbury university , New Zealand. He is a Delegate of Spark NZ to the meetings of ITU-R and APT and has contributed to a large number of wireless communications standards in the ITU-R and 3GPP. 
He has co-shared several IEEE prize-winning papers: the IEEE Communications Society Best Tutorial Paper Award 2004 and 2021, the IEEE VTS Neal Shepherd Memorial Best Propagation Paper Award 2023, and the IEEE Donald G. Fink Award 2011. In 2013, he was awarded the Member of NZ order of Merit (MNZM)  Queen's birthday honors for contributions to wireless communications.
He has been a Co-Guest Editor for three previous JSAC editions, the Proceedings of the IEEE, and the IEEE Communications Magazine, and a Co-Chair of the ICC 2005 Wireless Communications Symposium, and has held various editorial and TPC roles in the IEEE journals and conferences.
\end{IEEEbiography}
\begin{IEEEbiography}
[{\includegraphics[width=1in,height=1.25in,clip,keepaspectratio]{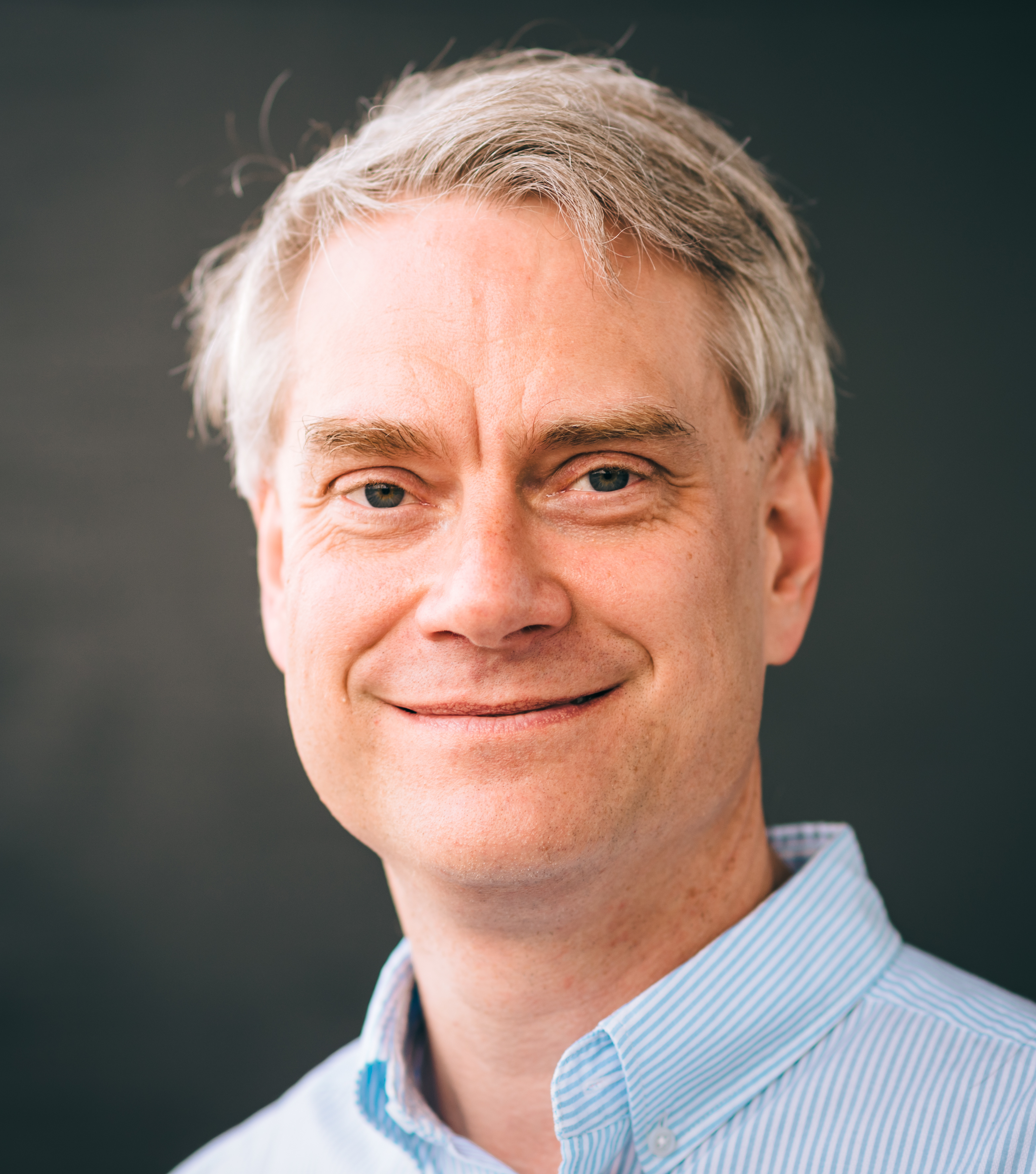}}]{Erik G. Larsson}
  is  Professor of Communication Systems at Link\"oping University (LiU) in Link\"oping, Sweden. 
He co-authored \emph{Space-Time Block Coding for Wireless Communications} (Cambridge University Press, 2003) and \emph{Fundamentals of Massive MIMO} (Cambridge University Press, 2016). He served as chair of the IEEE Signal Processing Society SPCOM technical committee (2015--2016), chair of the \emph{IEEE Wireless Communications Letters} steering committee (2014--2015), member of the \emph{IEEE Transactions on Wireless Communications} steering committee (2019--2022), General and Technical Chair of the Asilomar SSC conference (2015, 2012), technical co-chair of the IEEE Communication Theory Workshop (2019), and member of the IEEE Signal Processing Society Awards Board (2017--2019). 
He received the IEEE Signal Processing Magazine Best Column Award twice, in 2012 and 2014, the IEEE ComSoc Stephen O. Rice Prize in Communications Theory in 2015, the IEEE ComSoc Leonard G. Abraham Prize in 2017, the IEEE ComSoc Best Tutorial Paper Award in 2018, the IEEE ComSoc Fred W. Ellersick Prize in 2019, the IEEE SPS Donald G. Fink Overview Paper Award in 2023, and the IEEE ComSoc Test of Time Paper Award for Advances in Communications in 2026.
\end{IEEEbiography}
\begin{IEEEbiography}
  [{\includegraphics[width=1in,height=1.25in,clip,keepaspectratio]{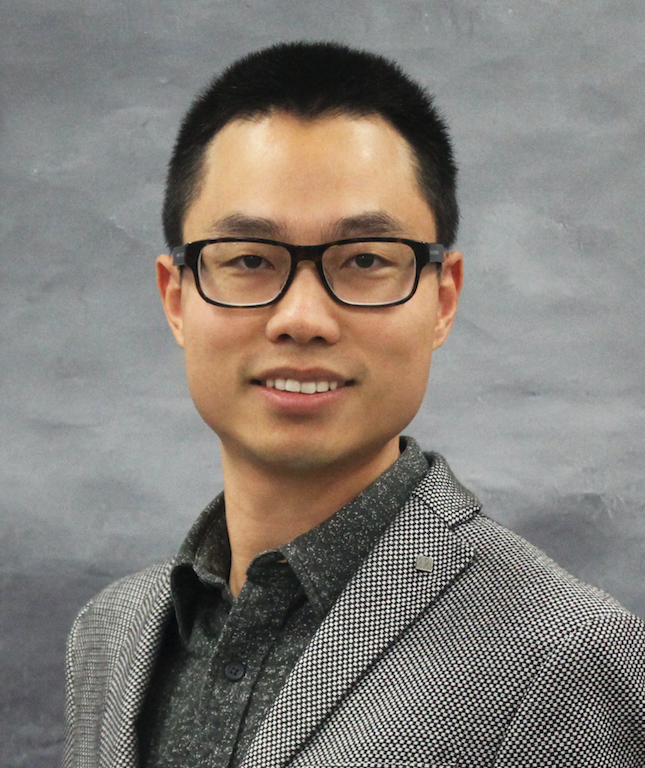}}]{Xingqin Lin} received the Ph.D. degree in electrical and computer engineering from The University of Texas at Austin and the MBA degree from the University of Illinois Urbana-Champaign. He is a Senior 3GPP Standards Engineer at NVIDIA, where he contributes to 3GPP and ATIS standardization and conducts research at the intersection of 5G/6G and AI. He was a founding co-chair of the ATIS AI Network Applications Working Group and a member of the ATIS Technology and Operations Council. He has served as an expert reviewer for the U.S. National Science Foundation and the Natural Sciences and Engineering Research Council of Canada, as well as a panelist for NASA research review programs. He received the IEEE ComSoc Fred W. Ellersick Prize and the IEEE VTS Early Career Award in 2021, the IEEE WCNC Best Paper Award in 2020, and the IEEE ComSoc Best Young Professional Award in Industry in 2020. He is a Fellow of the Institution of Engineering and Technology. He was an IEEE ComSoc Distinguished Lecturer from 2024 to 2025 and an IEEE VTS Distinguished Lecturer from 2024 to 2026.
\end{IEEEbiography}
\begin{IEEEbiography}
[{\includegraphics[width=1in,height=1.25in,clip,keepaspectratio]{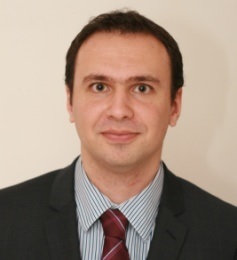}}]{Dorin Panaitopol} received the Ph.D. degree in Telecommunications and Signal Processing from both Ecole Supérieure d'Electricité (SUPELEC) and the National University of Singapore (NUS). Previously with NEC (from 2010 to 2014), Dorin joined THALES in 2014 as an R$\&$D engineer. Since 2010, Dorin has been involved in several European projects such as QoSMOS, SACRA, OneFIT, Concerto, EMPhAtiC (FP7), SHARING (CELTIC-Plus), CORRIDOR (ANR), COHERENT (H2020), HELENA (ARTES) and 6G-NTN (SNS). 
Dorin received the Best Paper Award from the European Wireless 2014 conference for the paper entitled ``On the Feasibility of Cellular Resource Reuse for Device-to-Device Communications under 3GPP Network Constraints.'' Dorin is currently actively involved in the RAN4 Working Group as moderator and rapporteur of the Technical Specification TS 38.108 for 5G NR Satellite Access Node (5G NR Non-Terrestrial Network). Since 2020, Dorin has submitted more than 600 official technical contributions at 3GPP, his work being recognized in 2024 by the Via Satellite ``2023 Satellite Technology of the Year'' Award. Moreover, Dorin is currently part of the 6G Smart Networks and Services Industry Association (6G-IA) governing board. %
\end{IEEEbiography}
\begin{IEEEbiography}
[{\includegraphics[width=1in,height=1.25in,clip,keepaspectratio]{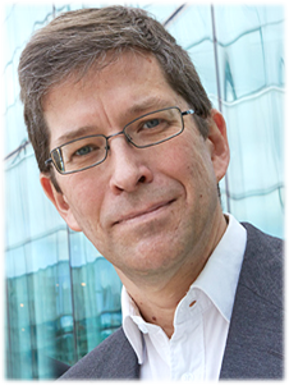}}]{Stefan Parkvall} is currently a Senior Expert at Ericsson Research working with research on 6G and future radio access. He is one of the key persons in the development of HSPA, LTE and NR radio access and has been deeply involved in 3GPP standardization for many years. Dr Parkvall is a fellow of the IEEE, served as an IEEE Distinguished lecturer 2011-2012, and is co-author of several popular books such as ``3G Evolution: HSPA and LTE for Mobile Broadband,'' ``4G: LTE/LTE-Advanced for Mobile Broadband,'' ``4G, LTE Advanced Pro and the Road to 5G,'' and ``5G NR: The Next Generation Wireless Access.'' He holds thousands of patents in the area of mobile communication. In 2005, he received the Ericsson ``Inventor of the Year'' award, in 2009 the Swedish government's Grand Technical Award for his contributions to the success of HSPA, 
in 2014 he and colleagues at Ericsson were one of three finalists for the European Inventor Award for their contributions to LTE, and in 2026 he received the IEEE J C Bose medal in wireless communication for his contributions and leadership in the area of cellular communication. Dr Parkvall received the Ph.D. degree in electrical engineering from the Royal Institute of Technology in 1996.
\end{IEEEbiography}
\begin{IEEEbiography}
[{\includegraphics[width=1in,height=1.25in,clip,keepaspectratio]{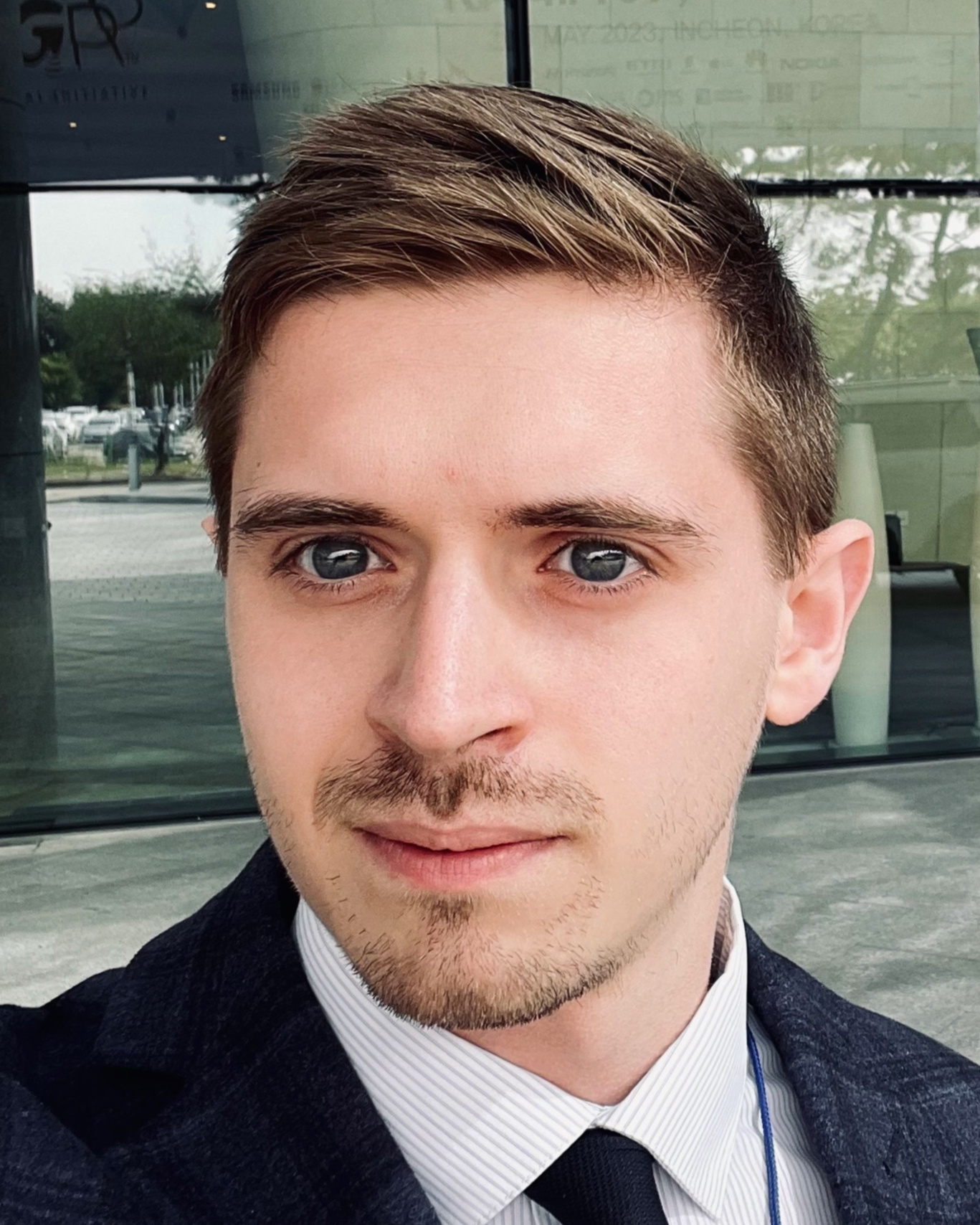}}]{Flavien Ronteix-Jacquet} is a project officer at the Joint Research Center (JRC) of the European Commission. He is in charge at the JRC of 5G NTN development for the IRIS2 program, the European Commission space-based communication infrastructure to be launched by 2030. Until December 2025, he worked at Thales Alenia Space as 5G NTN architect. He is actively involved in 3GPP standardization as a delegate for plenary and working groups, specializing in NTN topics. Dr. Ronteix-Jacquet has made significant contributions to several European projects, including 5G-Stardust and 6G NTN, and has participated in studies for the French space agency (CNES). He earned his Ph.D. degree in telecommunications from IMT Atlantique in Rennes, France, in 2022, with a thesis titled ``Reducing latency and jitter in the 5G Radio Access Network.'' His academic work includes publications and communications at international conferences such as EUCNC, WCNC, ITC, and VTC. Additionally, he holds an engineering degree in cybersecurity from INSA Centre Val de Loire and a master's equivalence from TalTech University in computer science and the IoT networks.
\end{IEEEbiography}
\begin{IEEEbiography}
[{\includegraphics[width=1in,height=1.25in,clip,keepaspectratio]{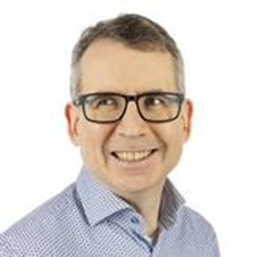}}] {Antti Toskala} joined the Nokia Research Center in 1994, where he undertook WCDMA system studies as Research Engineer and later as Senior Research Engineer and CDMA Specialist.
He chaired the UMTS physical layer expert group in ETSI SMG2 during 1998, and from 1999 until 2003 he worked in 3GPP as chairman of the TSG RAN WG1. From 2003 to 2005 he worked as Senior Standardization Manager with System Technologies, at Nokia Networks and contributed to product development as the HSDPA Chief Architect for Nokia Networks.
From 2005 onwards he worked with Nokia Networks as Senior Standardization Manager focusing on HSPA and LTE standardization, and later as Head of Radio Standardization with Nokia Siemens Networks focusing on LTE and LTE-Advanced work in 3GPP.
He has co-authored 8 books on 3G, 4G, and 5G, the latest being ``5G Technology: 3GPP New Radio.'' 
He was nominated as Nokia Fellow in 2015 and Bell Labs Fellow in 2016. Currently he is with Nokia Standards, in Espoo, Finland, heading Nokia 3GPP RAN Standardization, with technical focus on 5G-Advanced and 6G.
\end{IEEEbiography}

\end{document}